\documentclass{article}
\usepackage[utf8]{inputenc}
\usepackage{graphicx}
\usepackage{amsmath}
\usepackage{amsfonts}
\usepackage{multirow}
\usepackage{appendix}
\usepackage{caption}
\usepackage{subcaption}
\usepackage{xcolor}
\usepackage{tikz}
\usepackage{adjustbox}
\usepackage{natbib}
\usepackage{hyperref}
\usepackage{authblk}
\usetikzlibrary{matrix,decorations.pathreplacing, calc, positioning,fit}
\usepackage[a4paper,top=3cm, bottom=3.5cm, left=3cm, right=3cm]{geometry}

\title{Urban mapping in Dar es Salaam using AJIVE}
\author[1]{Rachel J. Carrington}
\author[2,4]{Ian L. Dryden}
\author[3]{Madeleine Ellis}
\author[3]{James O. Goulding}
\author[2]{Simon P. Preston}
\author[2]{David J. Sirl}

\affil[1]{School of Mathematical Sciences, University of Bath, UK}
\affil[2]{School of Mathematical Sciences, University of Nottingham, UK}
\affil[3]{N/Lab, University of Nottingham, UK}
\affil[4]{Department of Statistics, University of South Carolina, USA}

\newcommand{\boldA}{\boldsymbol{A}}
\newcommand{\boldE}{\boldsymbol{E}}
\newcommand{\boldI}{\boldsymbol{I}}
\newcommand{\boldJ}{\boldsymbol{J}}
\newcommand{\boldM}{\boldsymbol{M}}
\newcommand{\boldU}{\boldsymbol{U}}
\newcommand{\boldV}{\boldsymbol{V}}

\newcommand{\boldX}{\boldsymbol{X}}

\newcommand{\boldZ}{\boldsymbol{Z}}

\newcommand{\boldSigma}{\boldsymbol{\Sigma}}

\begin{document}

\maketitle

\begin{abstract}
Mapping deprivation in urban areas is important, for example for identifying areas of greatest need and planning interventions. Traditional ways of obtaining deprivation estimates are based on either census or household survey data, which in many areas is unavailable or difficult to collect. However, there has been a huge rise in the amount of new, non-traditional forms of data, such as satellite imagery and cell-phone call-record data, which may contain information useful for identifying deprivation. We use Angle-Based Joint and Individual Variation Explained (AJIVE) to jointly model satellite imagery data, cell-phone data, and survey data for the city of Dar es Salaam, Tanzania. We first identify interpretable low-dimensional structure from the imagery and cell-phone data, and find that we can use these to identify deprivation. We then consider what is gained from further incorporating the more traditional and costly survey data. We also introduce a scalar measure of deprivation as a response variable to be predicted, and consider various approaches to multiview regression, including using AJIVE scores as predictors.
\end{abstract}

\textbf{Keywords:} Dimension reduction; Multiview data; Urban mapping

\section{Introduction} \label{sec:intro}

Mapping levels of deprivation in urban areas is key for identifying areas of greatest need, in order to plan intervention work and decide where to best allocate resources \citep{smith2013ubiquitous,blumenstock2015predicting,steele2017mapping}. However, estimates are traditionally based on census data, which for many countries are either unavailable or outdated, or household surveys, which are expensive and time-consuming to collect \citep{smith2014poverty}. In addition, for cities that are growing rapidly in population, the urban landscape can change so rapidly that surveyed data is rendered quickly out of date. There is a need, therefore, to exploit novel sources of data that are both readily available and current.

The goal of this paper is to investigate the mapping of poverty in the city of Dar es Salaam, Tanzania, via exploitation of non-traditional types of data, including satellite imagery and phone Call Detail Record (CDR) data. Using Angle-Based Joint and Individual Variance Explained \citep{feng2018angle}, we detect sources of joint and individual variation within these distinct datasets, which allows us to understand what information is contained in both data sets, and what is unique to each. Later, we incorporate data from a survey conducted in Dar es Salaam to investigate what additional information, if any, we can gain from this.

Previous studies have shown the potential for using these sorts of data in isolation to map poverty in different locations. A number of papers \citep{xie2016transfer, babenko2017poverty, ayush2021efficient, engstrom2022poverty, nicolo2023mapping, putri2023multi} have shown the potential of using satellite imagery and remote sensing data to map poverty; \cite{hall2023review} provides a review of this area. Several studies have also used CDR data to derive socioeconomic information: for example, \cite{blumenstock2015predicting} uses the call histories of individuals to predict their socioeconomic status in Rwanda, whilst \cite{aiken2023program} uses a similar approach in Afghanistan, and \cite{smith2014poverty} uses features derived from CDR data to map deprivation in different regions of Cote d'Ivoire.

\cite{steele2017mapping} combines remote sensing and CDR data to map poverty in Bangladesh, by using features derived from both datasets to fit a model to predict poverty levels. Our approach differs from theirs in the way we use the image data (using a convolutional neural network to generate feature vectors, rather than through use of hand-crafted features such as roads and vegetation). We also aim to elicit more insight into the data by finding both joint and individual sources of variation, which the approach of \citep{steele2017mapping} does not.

The data that we consider are: (i) cell-phone call detail record (CDR) data; (ii) high-resolution satellite imagery (image data) covering most of the city; and (iii) novel survey data (see Section \ref{sec:data-survey} for details). We focus mainly on the first two, which are easily available and entail little time or expense to collect. However, later on we also incorporate the survey data to investigate what information contained within it is common to or distinct from the other two data types, reflecting on how much benefit can be obtained from its (costly) collection and incorporation into models. Our aim is to understand the relationships amongst these different data types, and the potential to exploit them to predict deprivation across different administrative divisions of Dar es Salaam.

The data are high-dimensional and of different types, and so to identify useful low-dimensional structure amongst the data --- and ultimately to map deprivation --- we employ the approach of Angle-Based Joint and Individual Variation Explained (AJIVE) \citep{feng2018angle}. The goal of AJIVE is to identify a small number of components that explain a large proportion of variation within the data, and for these components to be interpretable as reflecting individual variation (unique to each data type) and joint variation (belonging to both data types). The idea is that there is value in incorporating multiple data types into a joint ``multiview" analysis, rather than performing separate individual analyses, both in identifying a stronger signal of poverty and in understanding the information contributed by the different data types. 

In Section \ref{sec:data}, we explain some further context about Dar es Salaam and provide descriptions of available data, plus other derived measures of poverty in Dar es Salaam that we will later use for comparison. In Section \ref{sec:methods}, we summarise the AJIVE algorithm and explain how we estimate the number of components.
Section \ref{sec:results} presents results of applying AJIVE to the Dar es Salaam data. We investigate AJIVE as applied to CDR and image data, and subsequently with the addition of survey data too. We provide an interpretation of what each of the joint and individual components shows, and what this tells us about the data sources, comparing AJIVE to results from applying simpler Principal Component Analysis (PCA) to the two data sets independently, emphasising the additional insights that AJIVE provides. We also compare resulting deprivation estimates to external estimates computed via a completely different data set and method (as per \citep{seymour2022bayesian}), investigating how well deprivation statistics can be predicted utilising AJIVE outputs with regression modelling.

\section{Background \& Data} \label{sec:data}

\begin{figure}
    \centering
    \includegraphics[width=0.32\linewidth]{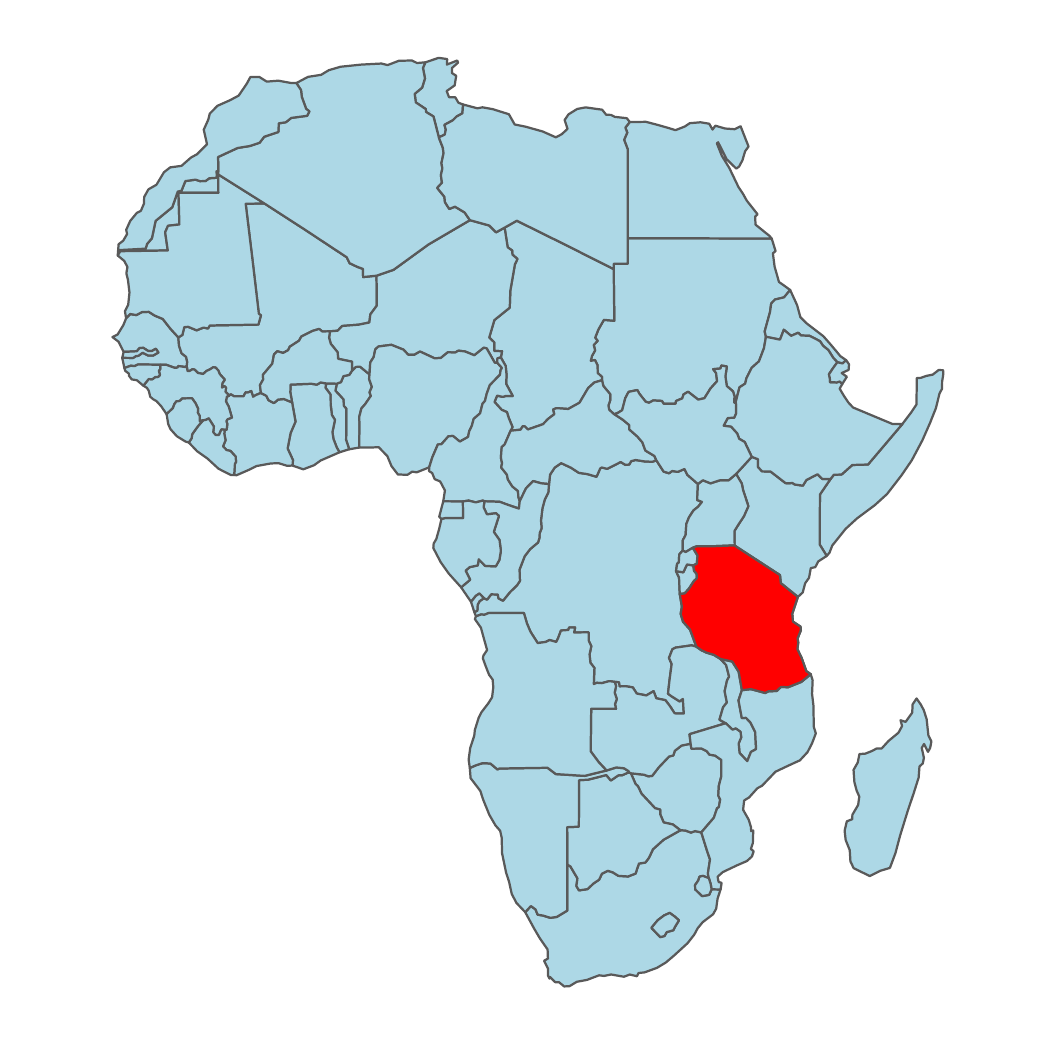}
        \put(-80,140){\small{(a)}}
    \includegraphics[width=0.32\linewidth]{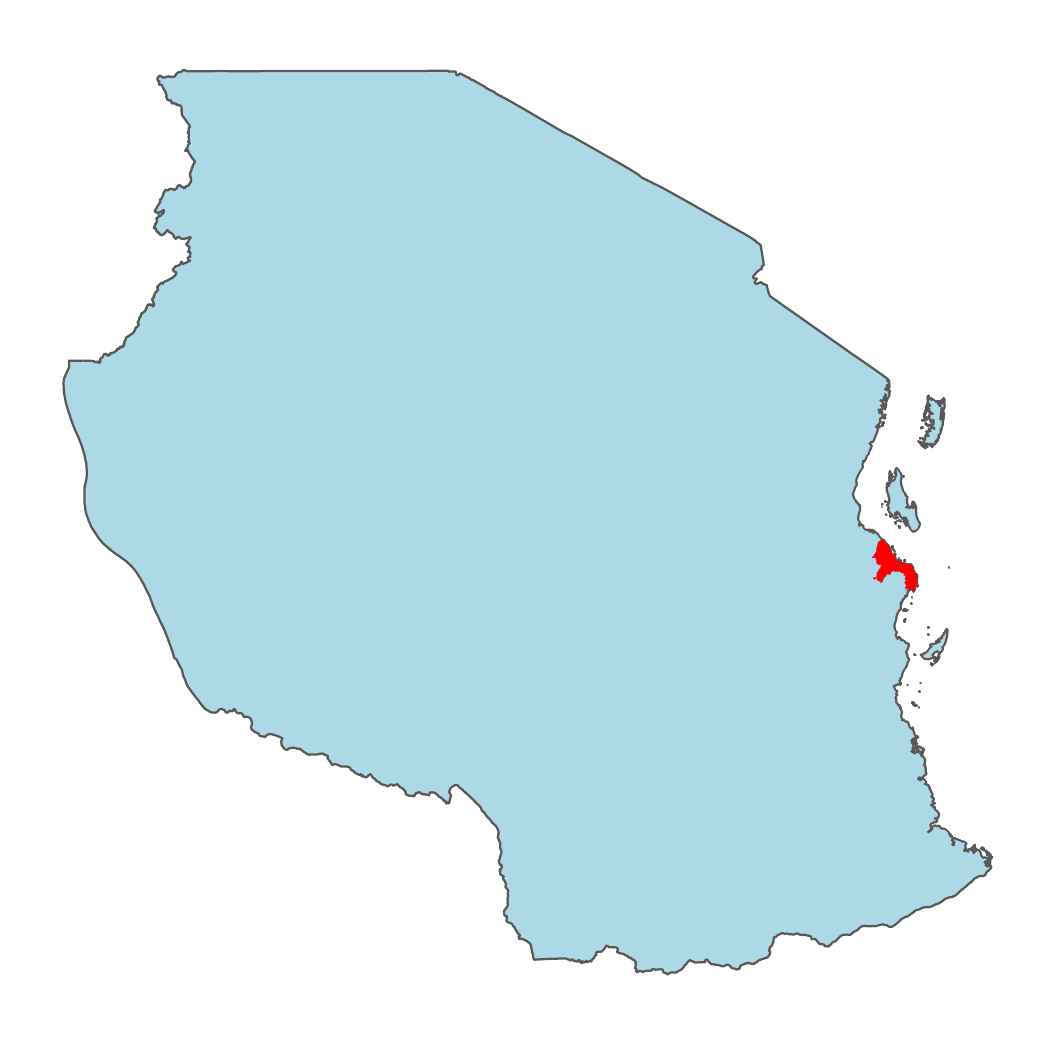}
        \put(-80,140){\small{(b)}}
    \includegraphics[width=0.32\linewidth]{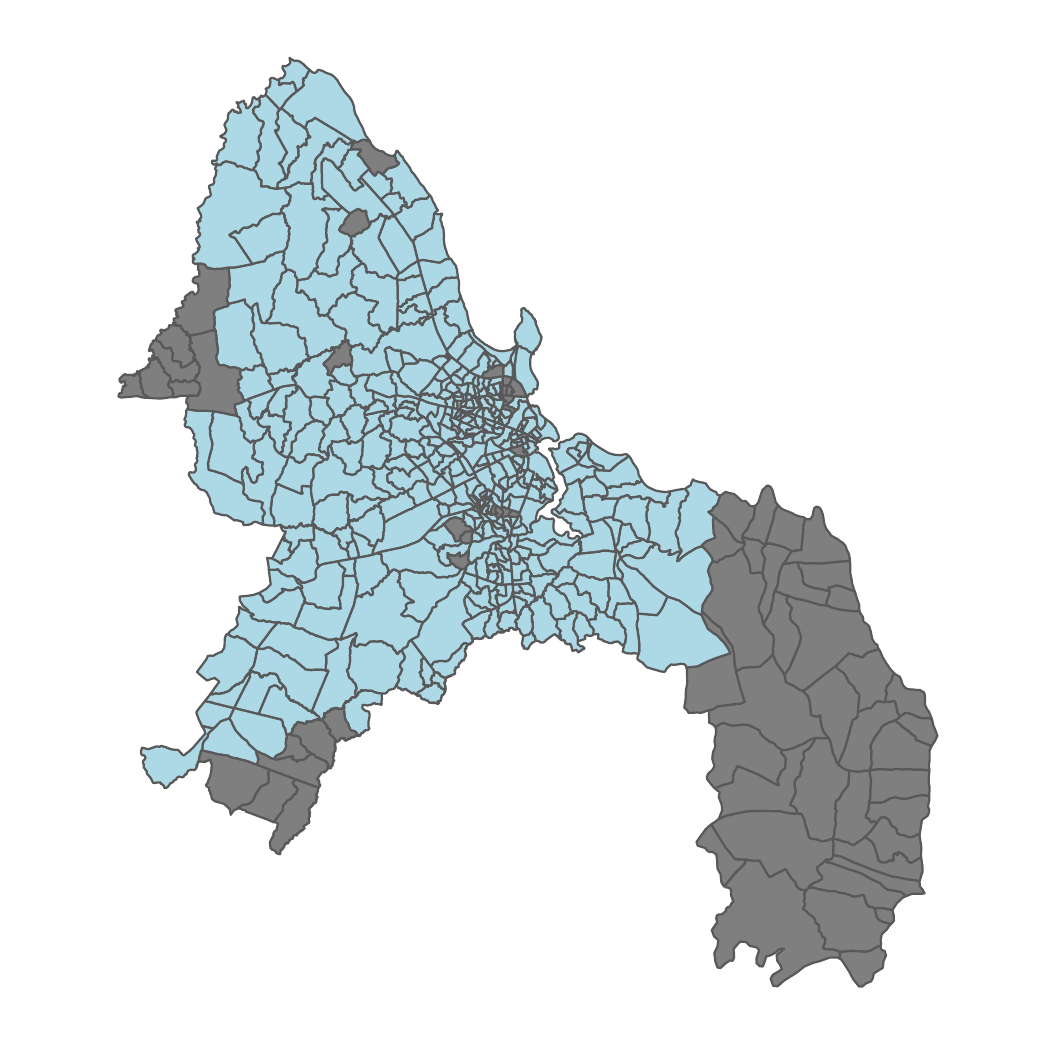}
        \put(-80,140){\small{(c)}}
    \caption{\small{(a). Location of Tanzania within Africa. (b). Location of Dar es Salaam on the east coast of Tanzania. (c). A map of Dar es Salaam showing boundaries of the subwards (administrative divisions). The blue subwards are those which are included in the analysis; grey subwards were excluded due to lack of image data.}}
    \label{fig:dar_maps}
\end{figure}

Dar es Salaam is located on the eastern coast of Tanzania. According to World Bank data, the population of Dar es Salaam in 2020 reached 6.7 million, having almost tripled from just under 2.3 million in 2000. It is the largest city in Tanzania and the fastest growing urban area in Africa, with some projections suggesting it could reach a population of over 60 million by 2050 \citep{locke2016urbanisation}. Figure \ref{fig:dar_maps} (a) and (b) show the locations of Tanzania and Dar es Salaam.

The city of Dar es Salaam is divided into 452 administrative regions called \emph{subwards}. In this paper, we treat the subward as the observational unit; in each data set each subward is represented by its own feature vector. Figure \ref{fig:dar_maps} (c) displays a map of the city, with subward boundaries shown. The image data only covers 383 subwards, so we include only these subwards in the study; these are shown in blue in Figure \ref{fig:dar_maps} (c). Throughout, we use $\boldX_1$ and $\boldX_2$ to refer to the CDR and image data respectively. Each of these matrices has $n = 383$ rows representing the subwards; the number of columns, representing feature dimensions, is $p_1 = 20$ (CDR) and $p_2 = 1536$ (image).

The survey data is denoted by $\boldX_3$. There are two subwards missing from this data set, so we leave these out when we incorporate the survey data into the analysis, giving $n = 381$. The survey data has column dimension $p_3 = 31$.

\subsection{CDR data} \label{sec:data-phone}

\begin{figure}
    \centering
    \includegraphics[width=0.6\linewidth]{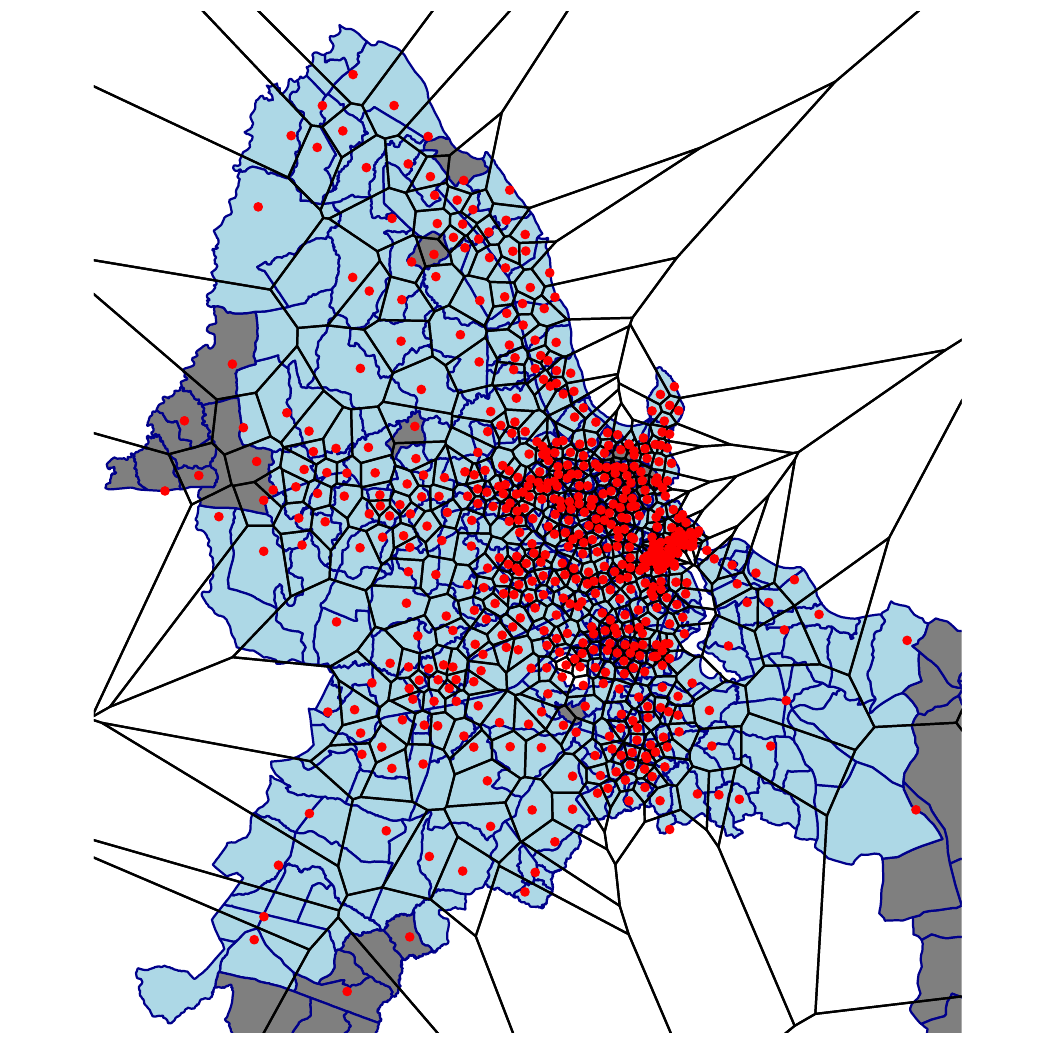}
    \caption{\small{Locations of cell towers and tower regions in Dar es Salaam. Towers are shown as red dots with region boundaries marked in red. Subwards and subward boundaries are shown in blue. As in Figure \ref{fig:dar_maps}, subwards which were excluded from the analysis due to lack of available image data are shaded in grey.}}
    \label{fig:tower-locs}
\end{figure}

The phone call detail record (CDR) data were collected from $n_T = 593$ base transceiver station (BTS) towers located over Dar es Salaam (shown in Figure \ref{fig:tower-locs}), during 122 days in 2014\footnote{For more details, see \cite{engelmann2018unbanked}.}. For each tower, information was collected about all call and SMS interactions recorded, including the time of the interaction, and the user IDs of both sender and receiver. This information was used to calculate a set of features for each user, such as the number of calls and SMS sent and received, the number of BTS towers visited, and the mean time between interactions; Table \ref{tab:phone-features} in the supplementary material gives the full list of features. There are also 5 features calculated at tower level (such as the total number of interactions recorded by each tower). To convert user-level features to tower features, each user ID was assigned a ``home" tower based on where the majority of their night-time interactions (between 10pm and 6am) took place; the tower features were calculated as the mean values across individuals who were assigned to that tower.

To calculate feature vectors for each subward, it was first necessary to determine the areas of overlap between the subwards and the regions served by each tower. The tower regions were calculated by constructing a Voronoi diagram, in which a polygon is placed around each tower in such a way that any point contained within it is closer to that tower than to any other. We refer to these polygons as the tower regions. The subward feature vectors were then calculated as weighted averages over the tower feature vectors, with weights proportional to the areas of overlap between the tower regions and subwards.

To describe this mathematically, we label the subwards $S_1, \ldots, S_n$, and the tower regions $T_1, \ldots, T_{n_T}$. Let $\boldsymbol{\Gamma}$ denote an $n \times n_T$ matrix with entries $\gamma_{ij}$, where
\begin{equation*}
    \gamma_{ij} = \mbox{area} \left( S_i \cap T_j \right), ~~~ i = 1, \ldots, n; j = 1, \ldots, n_T.
\end{equation*}
Then the feature vector corresponding to subward $S_i$ is given by
\begin{equation*}
    \boldsymbol{x}_{1,i} = \frac{\boldsymbol{\gamma}_i^\top \boldX_T}{\boldsymbol{\gamma}_i^\top \boldsymbol{1}},
\end{equation*}
where $\boldsymbol{\gamma}_1, \ldots, \boldsymbol{\gamma}_n$ are the row vectors of $\boldsymbol{\Gamma}$, and $\boldsymbol{1}$ is an $n_T \times 1$ vector with all elements equal to 1.
After calculating the matrix of subward CDR features $\boldX_1$, each column is $98\%$ winsorized (i.e. the largest and smallest 1\% are replaced with the values of the 99th and 1st percentiles) to remove the effect of large outliers, and then standardized so that each feature has mean 0 and variance 1 across subwards.

\subsection{Image data} \label{sec:data-image}
\begin{figure}
    \centering
    \includegraphics[width=0.3\linewidth]{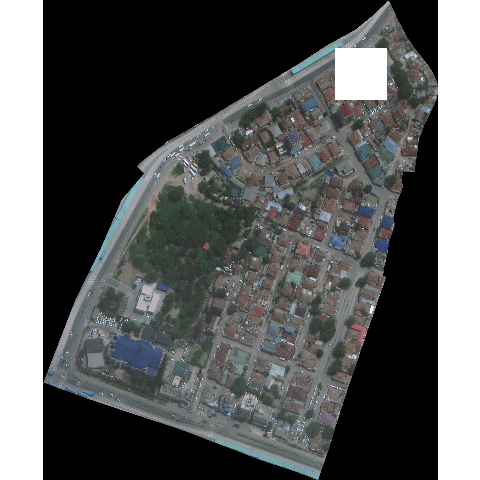}
    \caption{\small{Image for the subward of Idrisa, Dar es Salaam; the white square shows the location of one patch.}}
    \label{fig:image-eg}
\end{figure}

The image data used in our analysis consists of high-resolution, georeferenced satellite imagery of Dar es Salaam, which is made available by MAXAR (at \url{https://www.maxar.com/open-data/covid19}). We divided the images into subwards to create an image for each subward, discarding any sections that were covered by cloud. As a result, we have full or partial images for 383 subwards; the remaining 69 had to be excluded from the analysis, due either to being outside the area covered by the satellite imagery, or to the corresponding section of the image being entirely obscured by cloud.
Figure \ref{fig:image-eg} shows an example of the image data for a particular subward.

The subward images are high-dimensional and are of different sizes and shapes. To apply AJIVE or PCA, we need to construct a matrix representation of the data, where a row vector corresponds to each subward. Following in similar spirit to \citet{carmichael2021joint} we take a set of uniformly-sized patches from each subward image, and apply a pre-trained convolutional neural network (CNN) to map each patch to a lower-dimensional vector. The image feature vector for each subward is generated by taking the mean of the patch feature vectors for that subward.

Due to the irregular sizes and shapes of the subwards, using a grid of patches to cover each image was not feasible. Instead, for each subward we generated 10 random patches of dimension $200 \times 200$ pixels, sampling the top-left pixel uniformly from all possible locations (i.e. the image with the 199 right-most and bottom-most pixels excluded). Patches were allowed to overlap. Any patches containing more than $1\%$ black background pixels were replaced, to ensure that the image backgrounds were not a prominent feature in the patches.

For the CNN, we used InceptionResNet-V2 \citep{szegedy2016inception} trained on ImageNet; we chose this as it is one of the best-performing models on ImageNet. The schema is given in Figure 15 of \cite{szegedy2016inception}. A CNN takes as its input a $d_1 \times d_2 \times d_3$ dimensional array of pixel values (where $d_1 \times d_2$ are the dimensions of the image, and $d_3$ is the number of colour channels, in our case 3), where each element has an integer value between 0 and 255, and outputs a $v_1 \times v_2 \times v_3$ dimensional feature array, where $v_1 v_2 v_3 << d_1 d_2 d_3$. If we let $\mathcal{I}$ denote the space of possible images, the CNN can be defined as a mapping
\begin{equation*}
    f : \mathcal{I} \rightarrow \mathbb{R}^{v_1 \times v_2 \times v_3}.
\end{equation*}
The values of $v_1$, $v_2$, and $v_3$ are determined by the model architecture. This array is then usually fed into a final classification layer, which converts it into a vector; however, as the classification task the model was trained on is not directly relevant to our situation, we did not use this. Given the input patch dimensions and choice of model architecture, the output we obtain for each patch is a $4 \times 4 \times 1536$ dimensional array. To reduce the dimension of this further, we used maxpooling, taking the maximum value in each $4 \times 4 \times 1$ dimensional sub-array to produce a 1536-dimensional vector. We then took the mean across patches to generate a feature vector for each subward. As with the CDR features, the image features are then $98\%$ winsorized and standardized to have mean 0 and variance 1.

\subsection{Survey data} \label{sec:data-survey}

The survey data were collected in two phases in May and August 2019. In each subward, data were collected from approximately 8 respondents who were asked to answer questions on that subward. Most subwards had 8 respondents, although some had more or fewer; all had between 2 and 17.

The questions cover various aspects such as poverty and unemployment, education, and medical facilities. The data were originally collected \citep{ellis2021detection} to create a set of fine-grained, socio-demographic data about Dar es Salaam, which could be used to inform research on topics such as deprivation and forced labour risk, and to provide a ground truth against which models could be compared. Not all of the questions and responses could be used in our analysis, as some questions had categorical or write-in answers: we only included questions which had numerical or ordinal responses. Tables \ref{tab:survey-1} and \ref{tab:survey-2} in the supplementary material give the full list of survey questions in the data set used for this paper, and possible responses. For each subward we take the median responses for the respondents. The variables are then centred and scaled so that each response variable has mean 0 and variance 1.

\begin{figure}
    \centering
    \includegraphics[width=0.49\linewidth]{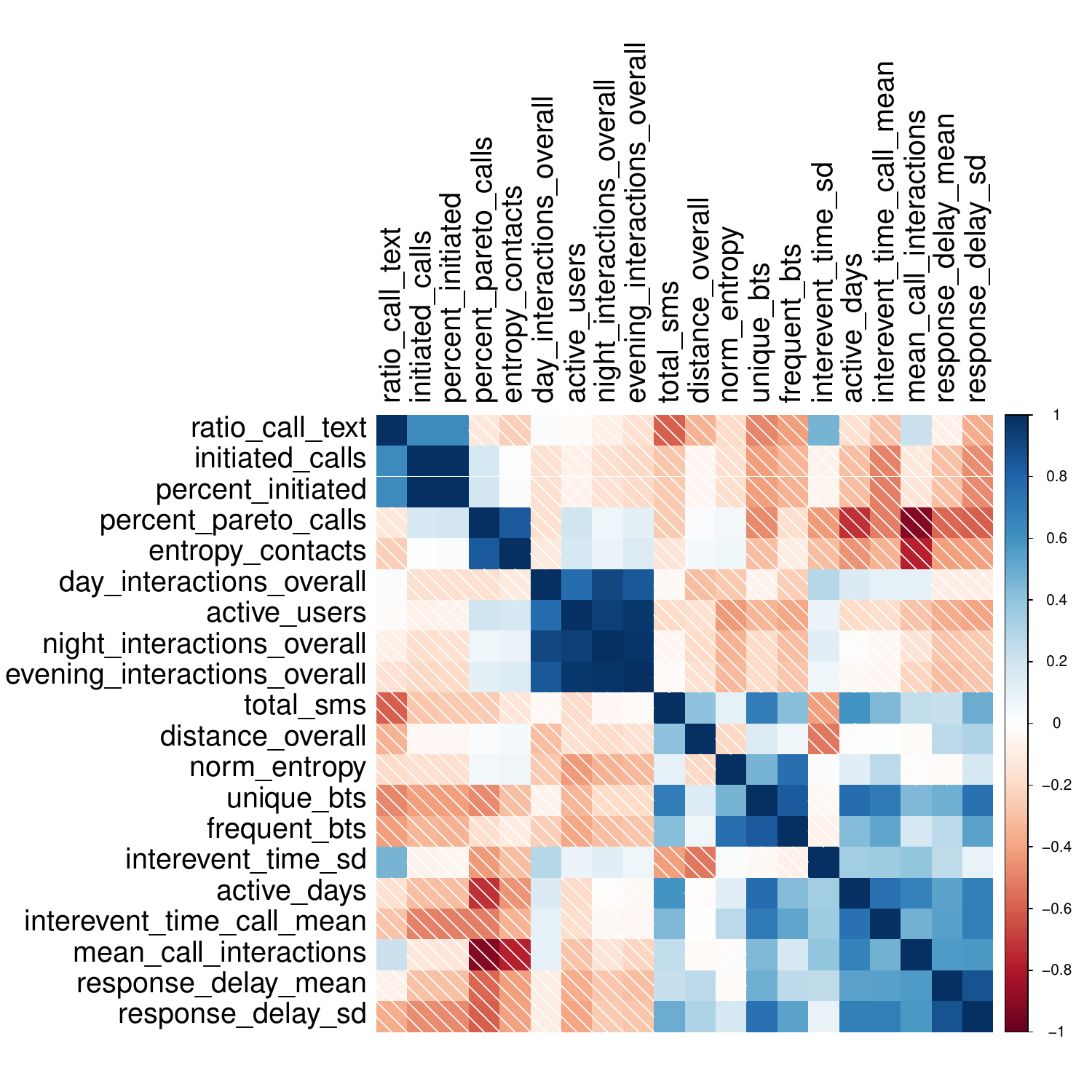}
    \includegraphics[width=0.49\linewidth]{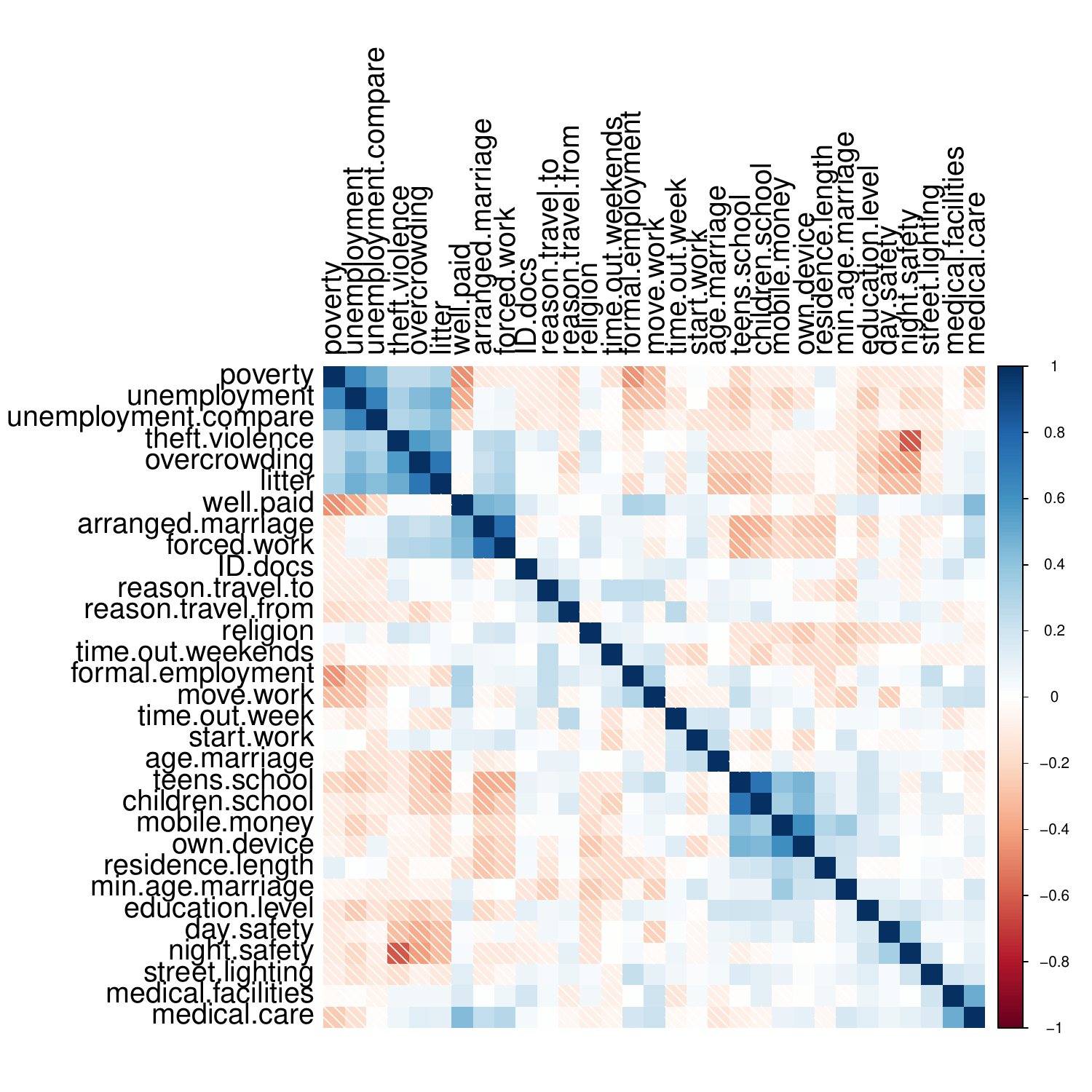}
    \caption{\small{Correlation plots of CDR variables (left) and survey variables (right). The features are ordered using complete-linkage hierarchical clustering, in order to highlight relationships between features.}}
    \label{fig:corrplots-1}
\end{figure}

\begin{figure}
    \centering
    \includegraphics[width=0.4\linewidth]{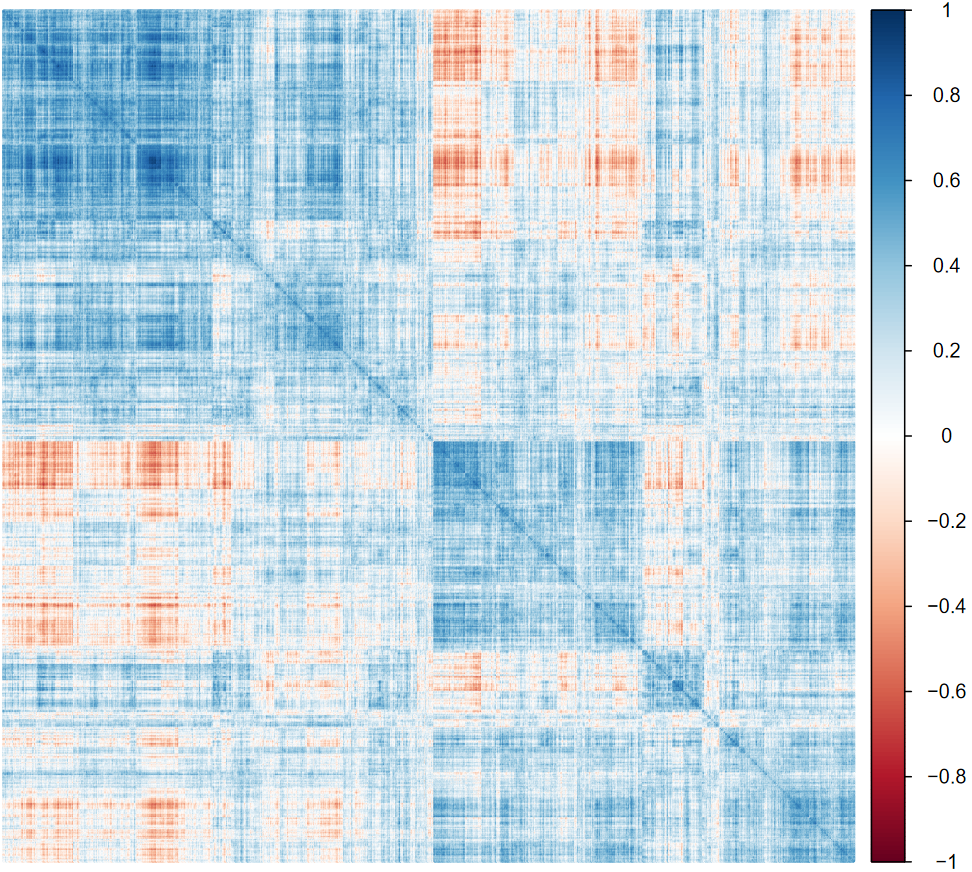}
    \caption{\small{Correlations between image features. The features are ordered using complete-linkage hierarchical clustering, in order to highlight relationships between features.}}
    \label{fig:image-corrplots}
\end{figure}

Figures \ref{fig:corrplots-1} and \ref{fig:image-corrplots} show pairwise correlations between each of the CDR, image, and survey features. In each case there are several clusters of correlated variables, which suggests that dimension reduction should be useful.

\subsection{Deprivation estimates based on comparative judgments} \label{sec:data-md}

\begin{figure}
    \centering
    \includegraphics[width=0.6\linewidth]{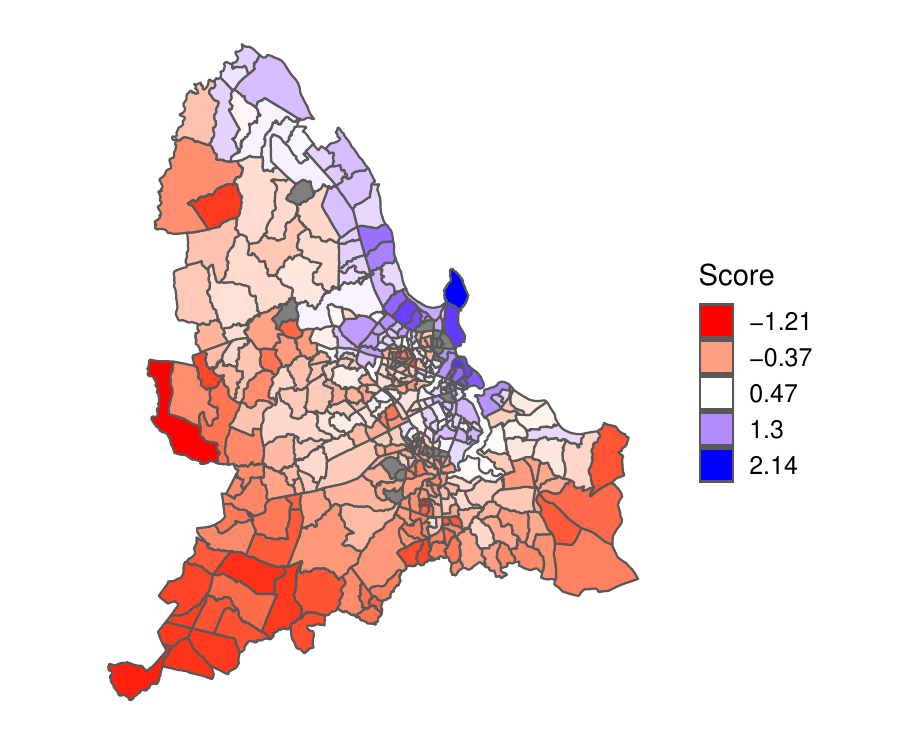}
    \caption{\small{Deprivation scores by subward, estimated from comparative judgment data using the Bayesian Spatial Bradley-Terry (BSBT) model (see \citep{seymour2022bayesian}). Subwards with higher scores (shown in blue on the figure) are those which are considered to be less deprived. Although these data cover all of Dar es Salaam, the figure only shows scores for the subwards included in our analysis.}}
    \label{fig:mean-dep}
\end{figure}

Our aim is to explore how we can use the CDR and image data to map deprivation. To provide a comparison, we use deprivation estimates for Dar es Salaam calculated from an independent data set using the Bayesian Spatial Bradley-Terry (BSBT) model \citep{seymour2022bayesian}. The input data consist of pairwise comparisons between subwards, where local participants recorded which of the two subwards they considered more deprived (ties were allowed). The BSBT model uses these to estimate a deprivation score for each subward of between -1.2 and 2.2, where higher scores indicate \emph{less} deprived subwards. The deprivation scores are available in the \texttt{R} package \texttt{BSBT}, which can be downloaded from \url{https://github.com/rowlandseymour/BSBT}. Figure \ref{fig:mean-dep} shows a plot of these for the subwards that are included in our analysis.

\section{Methodology} \label{sec:methods}

In this section we give an overview of the Angle-Based Joint and Individual Variation Explained (AJIVE) algorithm \citep{feng2018angle}, which we use to obtain the main part of our results. We also use Principal Component Analysis (PCA) for comparison; as AJIVE can be viewed as an extension of PCA to multiple data sets, we outline this first in Section \ref{sec:pca}, followed by AJIVE in Section \ref{sec:ajive}. We then, in Section \ref{sec:ranks}, discuss how to choose the ranks for low-rank approximation to our data, on which both PCA and AJIVE depend.

\subsection{Principal Components Analysis (PCA)}  \label{sec:pca}

For an $n \times p$ data matrix $\boldX$, PCA identifies the subspace of a chosen dimension, $r$, in which the data have greatest variance. We first centre $\boldsymbol{X}$ by subtracting the mean of each column, i.e. 
\begin{equation*}
    X_{ij} = X_{ij}^{original} - \frac{1}{n} \sum_{i=1}^n X^{original}_{ij} ~~~ (i = 1, \ldots, n; j = 1, \ldots, p).
\end{equation*}
We then calculate the Singular Value Decomposition (SVD):
\begin{equation*}
    \boldX = \boldU \boldSigma \boldV^\top,
\end{equation*}
where, with $m = \min \{ n, p \}$, $\boldSigma$ is an $m \times m$ diagonal matrix containing the $m$ singular values of $\boldX$ in decreasing order; and $\boldU$ and $\boldV$ are respectively $n \times m$ and $p \times m$ matrices with columns corresponding respectively to the left and right singular vectors of $\boldX$, such that $\boldU^\top \boldU = \boldV^\top \boldV = \boldI_m$ (the $m \times m$ identity matrix).

Reducing the dimension of the data to the chosen dimension $r$ $(<m)$ by PCA entails approximating $\boldX$ using the rank-$r$ truncated SVD, that is
\begin{equation*}
    \boldX \approx \boldU_r \boldSigma_r \boldV_r^\top,
\end{equation*}
where $\boldU_r$ and $\boldV_r$ are matrices containing the first $r$ columns of $\boldU$ and $\boldV$ respectively, and $\boldSigma_r$ is the $r \times r$ diagonal matrix containing the first $r$ singular values of $\boldX$. The PC scores are the rows of the $n \times r$ matrix $\boldV_r$.

\subsection{AJIVE} \label{sec:ajive}

Angle-Based Joint and Individual Variation Explained (AJIVE) \citep{feng2018angle} is a dimension reduction algorithm which can be applied to two or more data sets, where the data correspond to the same group of individuals, but are of different types. It is applicable when we believe there is some joint structure common to both data sets, but also some information which is unique to each: in contrast to methods such as Principal Component Analysis, where each data matrix is decomposed separately, or Canonical Correlation Analysis, which finds only joint components, AJIVE allows us to analyse both joint and individual variation within the data. It was developed as a more efficient version of Joint and Individual Variation Explained (JIVE) \citep{lock2013joint}.

Let $k$ be the number of data sets. AJIVE \citep{feng2018angle} aims to decompose the data $\boldX_1, \ldots, \boldX_k$ as
\[
    \left( \begin{array}{c} \boldX_1 \\ \vdots \\ \boldX_k \end{array} \right) = 
      \boldJ + \left(\begin{array}{c} \boldA_1 \\ \vdots \\ \boldA_k \end{array}\right) + \boldE
\]
where $\boldJ$ is the joint variation matrix (of rank $r_J$), $\boldA_1, \ldots, \boldA_k$ are the individual variation matrices (each of which have rank $r_i$), and $\boldE$ represents noise. For each data matrix $\boldX_i$ $(i = 1, \ldots, k)$, we can write
\begin{equation*}
    \boldX_i = \boldJ_i + \boldA_i + \boldE_i
\end{equation*}
where $\boldJ_i$ and $\boldE_i$ are the portions of $\boldJ$ and $\boldE$ corresponding to data matrix $i$.

We can decompose $\boldJ$ as
\begin{equation*}
    \boldJ = \boldU_J \boldSigma_J \boldV_J^\top,
\end{equation*}
which we obtain by taking the (exact) rank-$r_J$ SVD of $\boldJ$. $\boldU_J$ is the $n \times r_J$ matrix of joint scores (in our case, each row corresponds to a subward), and $\boldV_J$ is the $p \times r_J$ matrix of joint loadings (each row corresponds to a CDR or image feature). $\boldSigma_J$ contains the singular values of $\boldJ$ and controls the relative weights of each component. Figure \ref{fig:jc-diagram} illustrates this decomposition.

\begin{figure}
\centering
\begin{tikzpicture}
    \matrix [matrix of math nodes,left delimiter=(,right delimiter=)](U) at (3,0){
        \hspace{1cm} \\
        \hspace{1cm} \\
        \hspace{1cm} \\
        \hspace{1cm} \\
        \hspace{1cm} \\
        \hspace{1cm} \\
        \hspace{1cm} \\
        \hspace{1cm} \\
    };
     \matrix [matrix of math nodes,left delimiter=(,right delimiter=)](Sigma) at (5.2,0){
        \hspace{1cm} \\
        \hspace{1cm} \\
        \hspace{1cm} \\
        \hspace{1cm} \\
    };
     \matrix [matrix of math nodes,left delimiter=(,right delimiter=)](V) at (7.8,0){
        \hspace{2cm} \\
        \hspace{2cm} \\
        \hspace{2cm} \\
        \hspace{2cm} \\
    };
    \draw [ultra thick, purple] (2.6,1.1) to (2.6,-1.1);
    \draw [ultra thick, blue] (3.4,1.1) to (3.4,-1.1);
    \draw [ultra thick, green] (2,0.7) to (4,0.7);
    \node (s1) at (1.1,0.7)
        {\small{Subward 1}};
    \draw [ultra thick, green] (2,0) to (4,0);
    \node (s2) at (1.1,0)
        {\small{Subward 2}};
    \draw [ultra thick, green] (2,-0.7) to (4,-0.7);
    \node (s3) at (1.1,-0.7)
        {\small{Subward 3}};
    \draw [ultra thick, purple] (6.4,0.4) to (9.2,0.4);
    \draw [ultra thick, blue] (6.4,-0.4) to (9.2,-0.4);
    \draw [ultra thick, green] (6.9,0.7) to (6.9,-0.7);
    \node (l1) at (6.9,-0.9)
        {\small{F1}};
    \draw [ultra thick, green] (7.5,0.7) to (7.5,-0.7);
    \node (l2) at (7.5,-0.9)
        {\small{F2}};
    \draw [ultra thick, green] (8.1,0.7) to (8.1,-0.7);
    \node (l3) at (8.1,-0.9)
        {\small{F3}};
    \draw [ultra thick, green] (8.7,0.7) to (8.7,-0.7);
    \node (l4) at (8.7,-0.9)
        {\small{F4}};
    \node[circle, draw=purple] (s1) at (4.8,0.4){};
    \node[circle, draw=blue] (s2) at (5.6,-0.4){};
    \node (uj) at (3,1.3)
        {$\boldU_J$};
    \node (sj) at (5.2,1.3)
        {$\boldSigma_J$};
    \node (vj) at (7.8,1.3)
        {$\boldV_J^\top$};
    \node (j) at (0,1.3)
        {$\boldJ$};
    \node (eq) at (1.35,1.3)
        {$\boldsymbol{=}$};
    \node (j1) at (2.6,-1.3)
        {\small{JC1}};
    \node (j2) at (3.4,-1.3)
        {\small{JC2}};
    \node (j1a) at (9.6,0.4)
        {\small{JC1}};
    \node (j2a) at (9.6,-0.4)
        {\small{JC2}};
    \node (j1b) at (4.8,0.8)
        {\small{JC1}};
    \node (j2b) at (5.6,-0.8)
        {\small{JC2}};
    \node (x1) at (11.7,-1)
        {\footnotesize{JC = Joint component}};
    \node (x1) at (11,-1.2)
        {\footnotesize{F = Feature}};
\end{tikzpicture}
\caption{\small{Decomposition $\boldJ = \boldU_J \boldSigma_J \boldV_J^\top$ in the case where $r_J = 2$. Each of the joint components has an $n$-dimensional score vector (column of $\boldU_J$) and a $p$-dimensional loading vector (row of $\boldV_J^\top$) associated with it; in this example, $n = 3$ and $p = 4$. Each subward has an $r_J$-dimensional score vector (row of $\boldU_J$) associated with it, and each feature has a $r_J$-dimensional loading vector (column of $\boldV_J^\top$). Given that the singular values in $\boldSigma_J$ are distinct and ordered from largest to smallest, the decomposition is identifiable up to multiplication of components by -1: we can multiply any column of $\boldU_J$ and the corresponding column of $\boldV_J$ (row of $\boldV_J^\top$) by -1 without changing the value of $\boldU_J \boldSigma_J \boldV_J^\top$.}}
\label{fig:jc-diagram}
\end{figure}

Similarly, for the $i$th individual component ($i = 1, \ldots, k$) we decompose $\boldA_i$ as
\begin{equation*}
    \boldA_i = \boldU_i \boldSigma_i \boldV_i^\top,
\end{equation*}
where $\boldU_i$ is the $n \times r_i$ score matrix, $\boldV_i$ the $p_i \times r_i$ loading matrix, and $\boldSigma_i$ contains the $r_i$ non-zero singular values of $\boldA_i$.

The AJIVE algorithm is given in detail in \citep{feng2018angle}; we give an outline here. The algorithm has three main stages:

\emph{Stage 1}. We choose initial signal ranks $\tilde{r}_1, \ldots, \tilde{r}_k$ (see Section \ref{sec:ranks}) and approximate each $\boldX_i$ by its rank-$\tilde{r}_i$ truncated SVD:
\begin{equation*}
   \boldX_i \approx \tilde{\boldX}_i = \tilde{\boldU}_i \tilde{\boldSigma}_i \tilde{\boldV}_i^\top.
\end{equation*}

\emph{Stage 2}. To estimate the joint scores $\boldU_J$, we combine the score matrices into one matrix $\boldM$, from which we will extract the joint signal:
\begin{equation*}
    \boldM = \left( \tilde{\boldU}_1 ~~ \ldots ~~ \tilde{\boldU}_k \right),
\end{equation*}
and calculate its SVD:
\begin{equation*}
    \boldM = \boldU_M \boldSigma_M \boldV_M^\top.
\end{equation*}
We then choose the joint rank $r_J$, based on the singular values of $\boldM$, (again, see Section \ref{sec:ranks}), and set the estimate of the joint scores to be $\tilde{\boldU}_J = \tilde{\boldU}_M$, where $\tilde{\boldU}_M$ is a matrix containing the first $r_J$ columns of $\boldU_M$.

To estimate the joint matrix $\boldJ$, we project $\boldX = \left(\boldX_1, \boldX_2\right)$ onto $\tilde{\boldU}_J$:
\begin{equation*}
    \hat{\boldJ} = \tilde{\boldU}_J \tilde{\boldU}_J^\top \boldX,
\end{equation*}
and decompose $\hat{\boldJ}$ as $\hat{\boldJ} = \hat{\boldU}_J \hat{\boldSigma}_J \hat{\boldV}_J^\top$, by taking its rank-$r_J$ SVD (which gives an exact decomposition since $\hat{\boldJ}$ has rank $r_J$):
\begin{equation*}
    \hat{\boldJ} = \hat{\boldU}_J \hat{\boldSigma}_J \hat{\boldV}_J^\top.
\end{equation*}
(Note that although generally $\tilde{\boldU}_J \neq \hat{\boldU}_J$ (unless $r_J = 1$), we have $\tilde{\boldU}_J \tilde{\boldU}_J^\top = \hat{\boldU}_J \hat{\boldU}_J^\top$: they are both estimates of the score space of $\hat{\boldJ}$.)

\emph{Stage 3}. To estimate the individual matrices $\boldA_i$, we note that for each $i$, we require the joint scores $\boldU_i$ to be orthogonal to $\hat{\boldU}_J$. Hence, we project each $\boldX_i$ onto the orthogonal complement of $\hat{\boldJ}$:
\begin{equation*}
    \tilde{\boldX}_i = \left( \boldI_n - \tilde{\boldU}_J \tilde{\boldU}_J^\top \right) \boldX_i
\end{equation*}
where $\boldI_n$ is the $n \times n$ identity matrix.
We then take a final rank-$r_i$ SVD of this matrix $\tilde{\boldX}_i$:
\begin{equation*}
    \hat{\boldA}_i = \tilde{\boldX}_i \approx \hat{\boldU}_i \hat{\boldsymbol{\Sigma}}_i \hat{\boldV}_i^\top.
\end{equation*}

\subsection{Rank estimation} \label{sec:ranks}

To implement AJIVE we must provide estimates of the initial ranks $\tilde{r}_1, \ldots, \tilde{r}_k$, joint rank $r_J$, and individual ranks $r_1, \ldots, r_k$. We describe here how we do this. For simplicity, we will refer to the case where we have the CDR and image data, so $k = 2$: we later incorporate the survey data as well, but the methods are the same.

\begin{figure}
    \centering
    \includegraphics[width=0.32\linewidth]{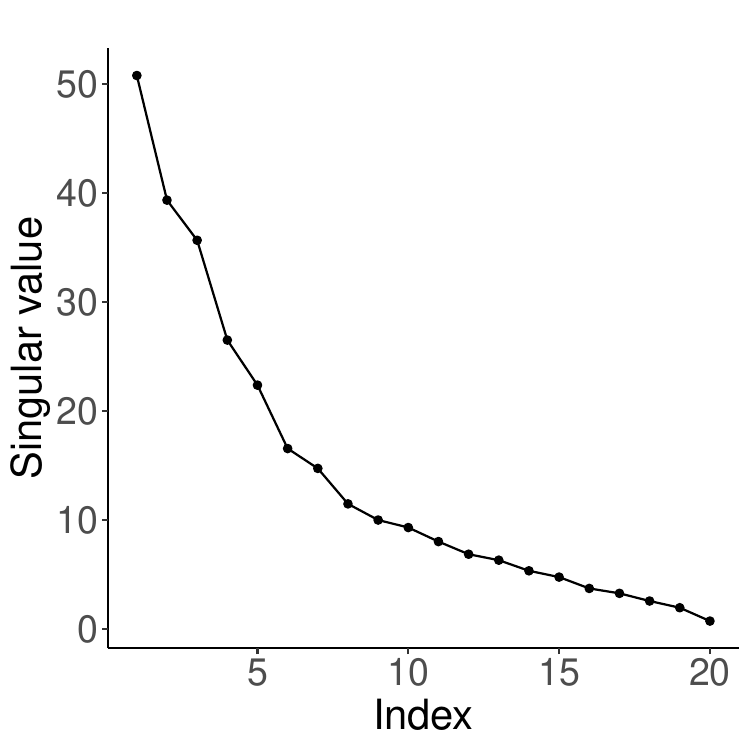}
        \put(-96,132){\footnotesize{\textbf{Scree plot: CDR}}}
        \put(-68,-7){\footnotesize{(a)}}
    \includegraphics[width=0.32\linewidth]{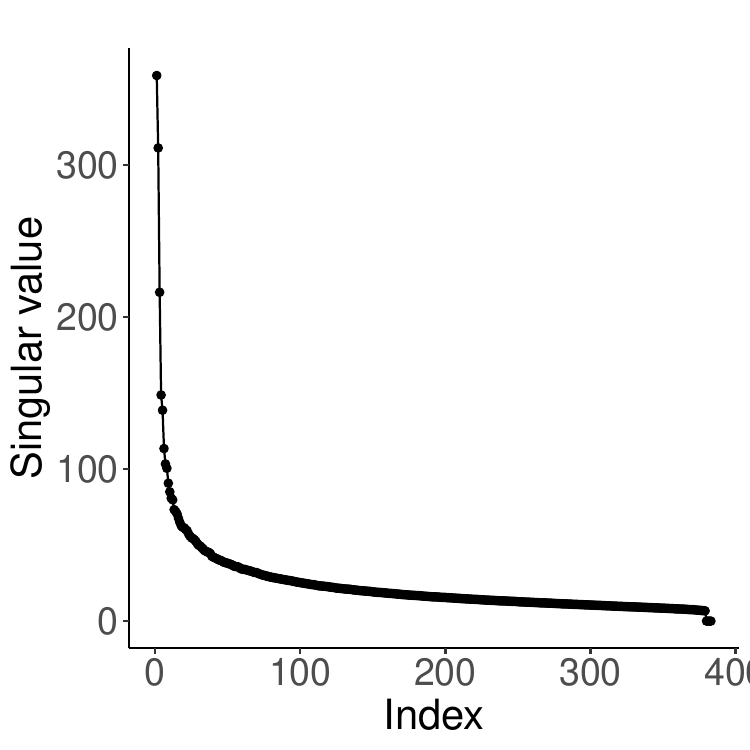}
        \put(-95,132){\footnotesize{\textbf{Scree plot: Image}}}
        \put(-68,-7){\footnotesize{(b)}}
    \includegraphics[width=0.32\linewidth]{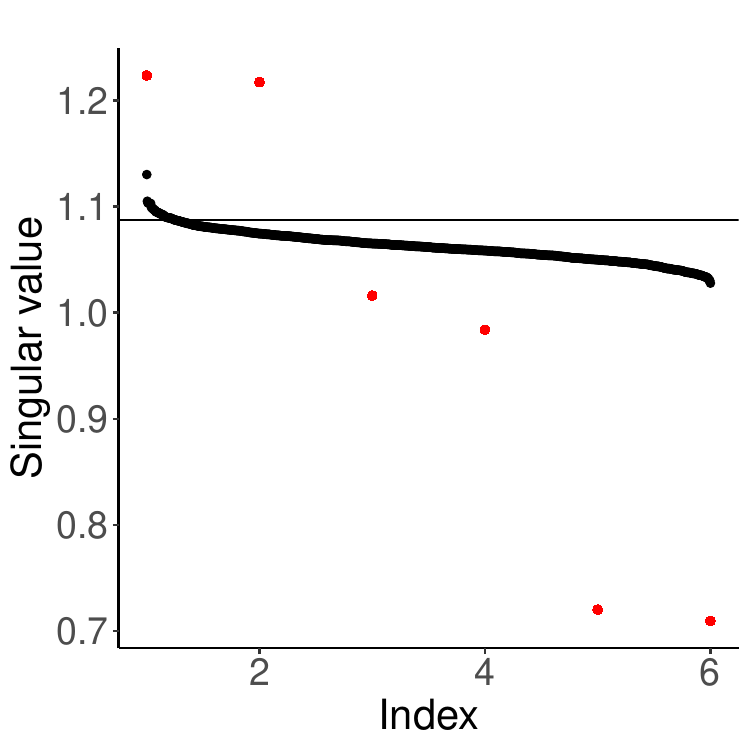}
        \put(-103,132){\footnotesize{\textbf{Joint rank estimation}}}
        \put(-68,-7){\footnotesize{(c)}}
    \caption{\small{Estimation of initial and joint ranks. (a), (b): Scree plots showing the singular values of the CDR feature matrix $\boldX_1$, and the image feature matrix $\boldX_2$. (c): Estimation of $r_J$, given initial ranks of 3 for both $\boldX_1$ and $\boldX_2$. The red points are the singular values of $\boldM$. The black points correspond to the largest singular value of $\tilde{\boldM}$ for each of 1000 random samples. The black line shows the $95\%$ threshold, above which singular values are assumed to correspond to joint components.}}
    \label{fig:scree}
\end{figure}

\emph{Initial ranks}. In selecting the initial ranks, we aim to distinguish signal from noise in each data set. We do this by inspecting scree plots of $\boldX_1$ and $\boldX_2$ (Figure \ref{fig:scree} (a)--(b)). On both plots there is a sizeable jump after the third singular value, so we take $\tilde{r}_1 = \tilde{r}_2 = 3$.

\emph{Joint rank}. \cite{feng2018angle} propose two methods for calculating the joint rank: the Wedin bound and a random direction bound. We use the random direction bound here. To recap, we have
\begin{equation*}
    \boldM = \left( \tilde{\boldU}_1 ~~ \tilde{\boldU}_2 \right),
\end{equation*}
where $\boldM$ is an $n \times \left( \tilde{r}_1 + \tilde{r}_2 \right)$ dimensional matrix.

The idea is to determine which components of $\boldM$ have sufficiently large singular values (corresponding to sufficiently small principal angles between subspaces associated with each component), to be regarded as part of the joint signal.

We generated 1000 random matrices of the same dimensions as $\boldM$, with elements sampled independently from $U(0, 1)$, and calculated the largest singular value for each. These random matrices are simulated under the assumption that $\tilde{\boldsymbol{U}}_1$ and $\tilde{\boldsymbol{U}}_2$ are have no joint components, and we expect true joint components to result in larger singular values. We therefore take the 95th percentile of these values as a lower bound for the singular values of $\boldM$ that correspond to the joint signal. Figure \ref{fig:scree} (c) illustrates this process for the CDR and image data. The red points show the ordered singular values of $\boldM$ (excluding those that are equal to 0), whilst the black points correspond to the maximum singular values of each randomly generated matrix, as described above. The horizontal black line corresponds to the threshold for the joint signal. There are two red points above the black line, so we take $r_J = 2$.

We note that this random direction bound may be too strict, as we compare all singular values of $\boldsymbol{M}$ with the largest singular values of the random matrices, which may result in too strict a bound when considering singular values of $\boldsymbol{M}$ subsequent to the largest one. As an alternative, we could calculate whether the $i$th largest singular value of $\boldsymbol{M}$ is larger than the 95th percentile of the $i$th largest singular value of the random matrices. However, in our case we find that this leads to the same result; it may be that this is usually the case in practice.

\emph{Individual ranks}. After calculating the joint components, we subtract from each $\boldX_i$ the corresponding part of the joint signal matrix, and calculate its SVD:
\begin{equation*}
    \tilde{\boldX}_i = \boldX_i - \tilde{\boldU}_J \tilde{\boldU}_J^\top \boldX_i = \tilde{\boldU}_i \tilde{\boldSigma}_i \tilde{\boldV}_i^\top.
\end{equation*}
The initial rank $\tilde{r}_i$ was selected by inspecting a scree plot, but we could equivalently have selected a threshold
\begin{equation*}
    \nu_i = \frac { \lambda^i_{\tilde{r}_i} + \lambda^i_{\tilde{r}_i + 1}}{2},
\end{equation*}
where $\lambda_{\tilde{r}_i}^i$ and $\lambda_{\tilde{r}_i + 1}^i$ are the $\tilde{r}_i$th and $(\tilde{r}_i + 1)$th largest singular values of $\boldX_i$, such that we only keep components with singular values greater than or equal to $\nu_i$. For the individual components, we use the same threshold, so we keep the individual components which have corresponding singular values greater than or equal to $\nu_i$ and discard those which are smaller. 
For the CDR and image data, this gives $r_1 = r_2 = 1$.

\section{Results} \label{sec:results}

We implement the AJIVE algorithm in \texttt{R}. We first inspect scree plots (Figure \ref{fig:scree}) for $\boldX_1$ and $\boldX_2$ to choose the initial ranks $\tilde{r}_1$ and $\tilde{r}_2$: we set $\tilde{r}_1 = \tilde{r}_2 = 3$. To select the joint rank $r_J$, we use a random direction bound (see Section \ref{sec:ranks}), with 1000 random samples, which leads to a value of $r_J = 2$. The individual ranks for both the CDR and image data ($r_1$ and $r_2$) are estimated to be 1. Throughout, we refer to the joint components as JC1 and JC2, and the individual components as IC1\textsuperscript{CDR} and IC1\textsuperscript{Image}.

\begin{figure}
\begin{tikzpicture}
    \node (scores) at (0,0)
        {\includegraphics[width=0.6\linewidth, trim={0 0cm 0 1cm}, clip]{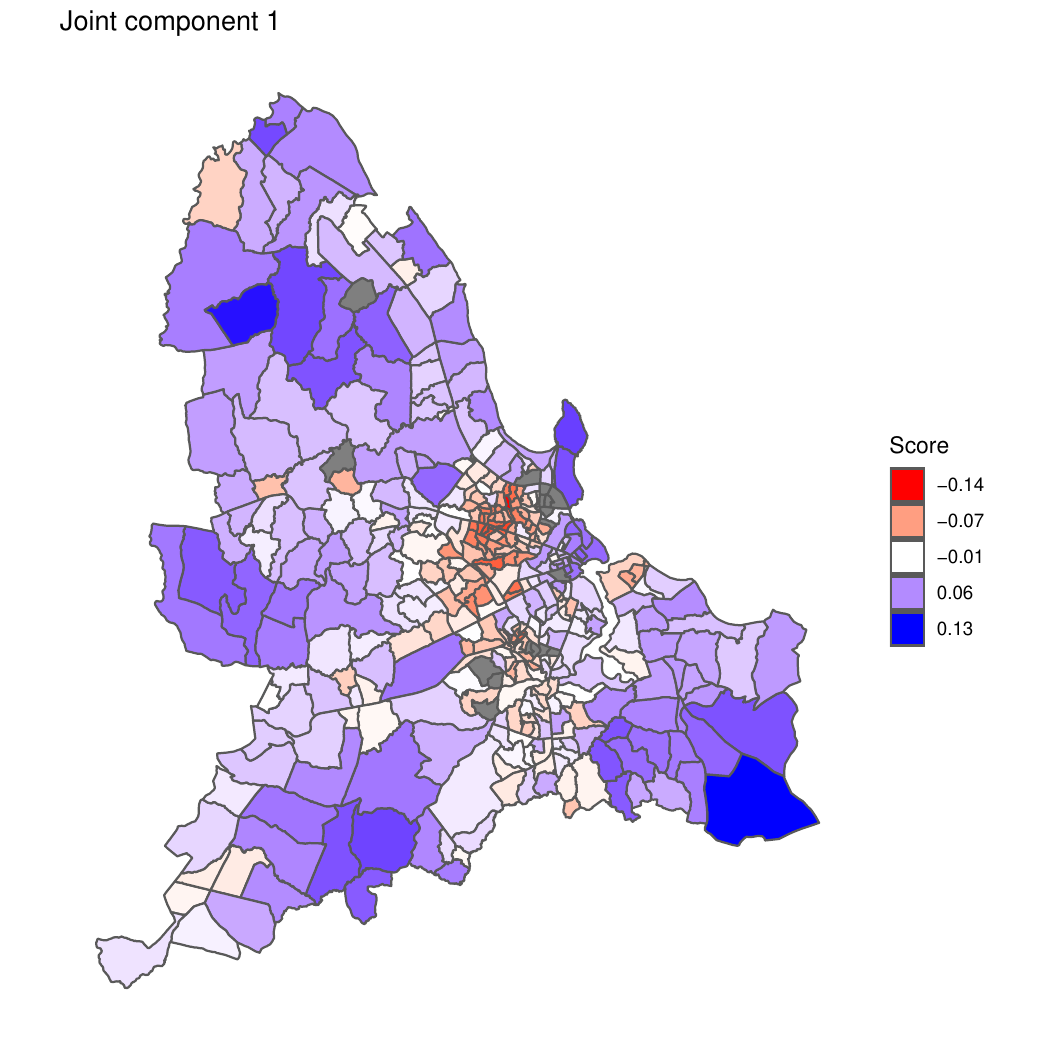}};
    \node (pos1) at (7,2.8)
        {\includegraphics[width=0.4\linewidth]{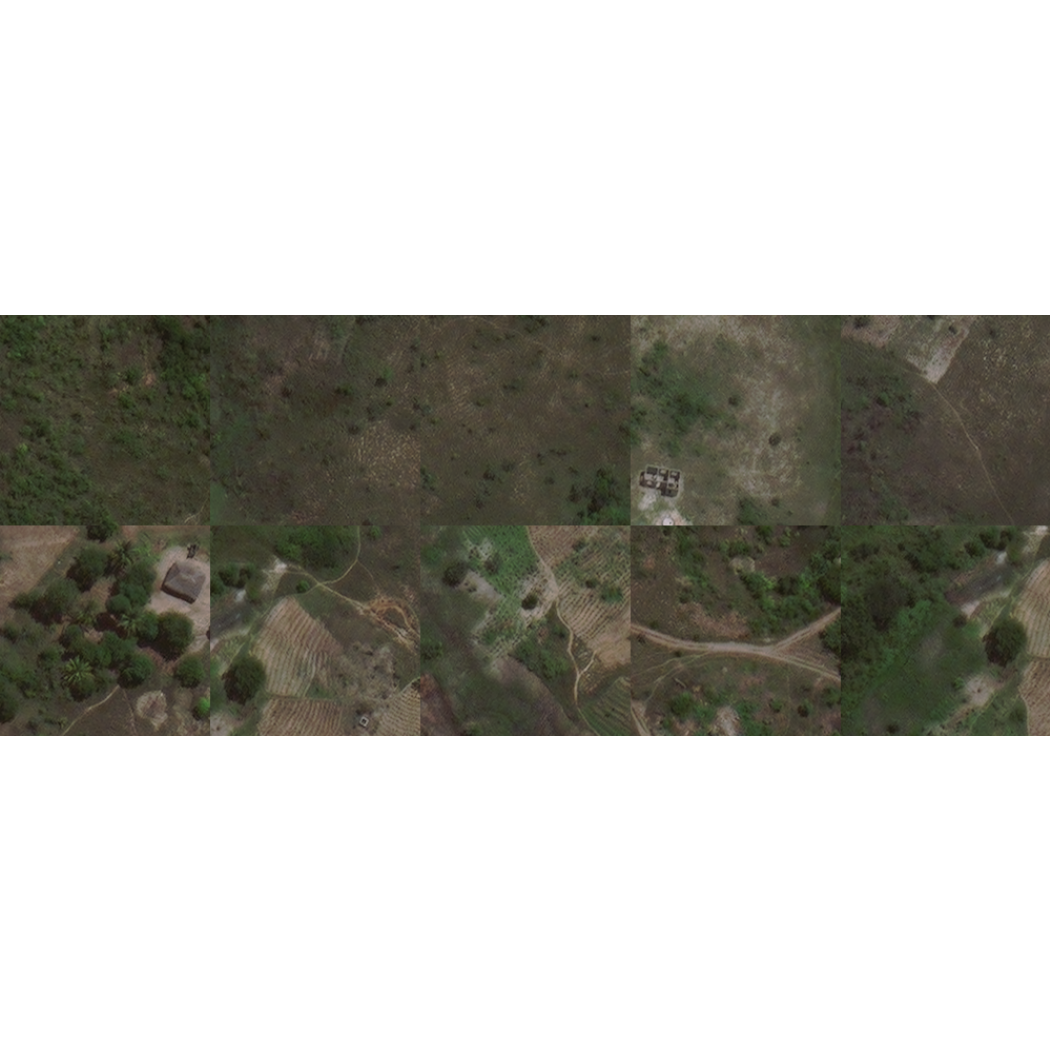}};
    \node (pos2) at (7,0)
        {\includegraphics[width=0.4\linewidth]{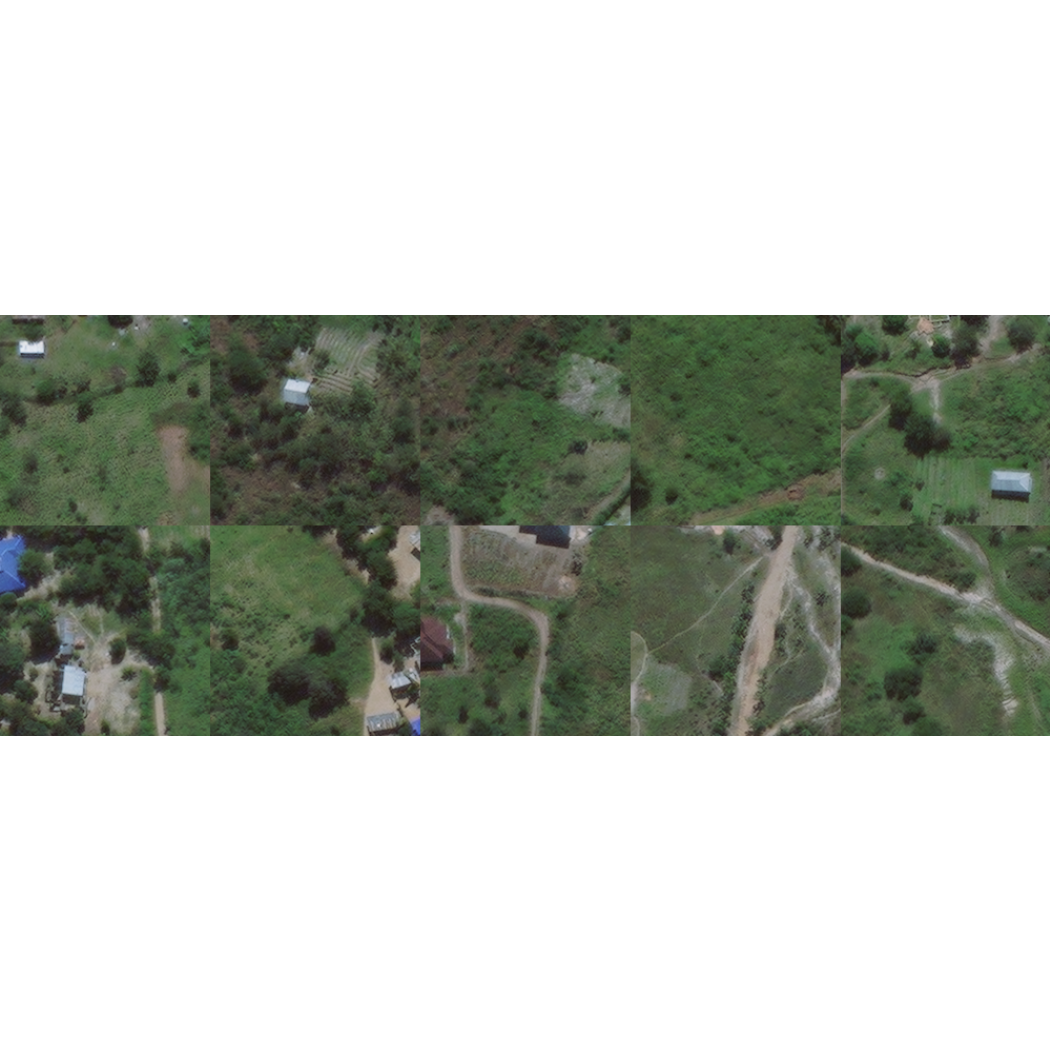}};
    \node (neg1) at (7,-3.5)
        {\includegraphics[width=0.4\linewidth]{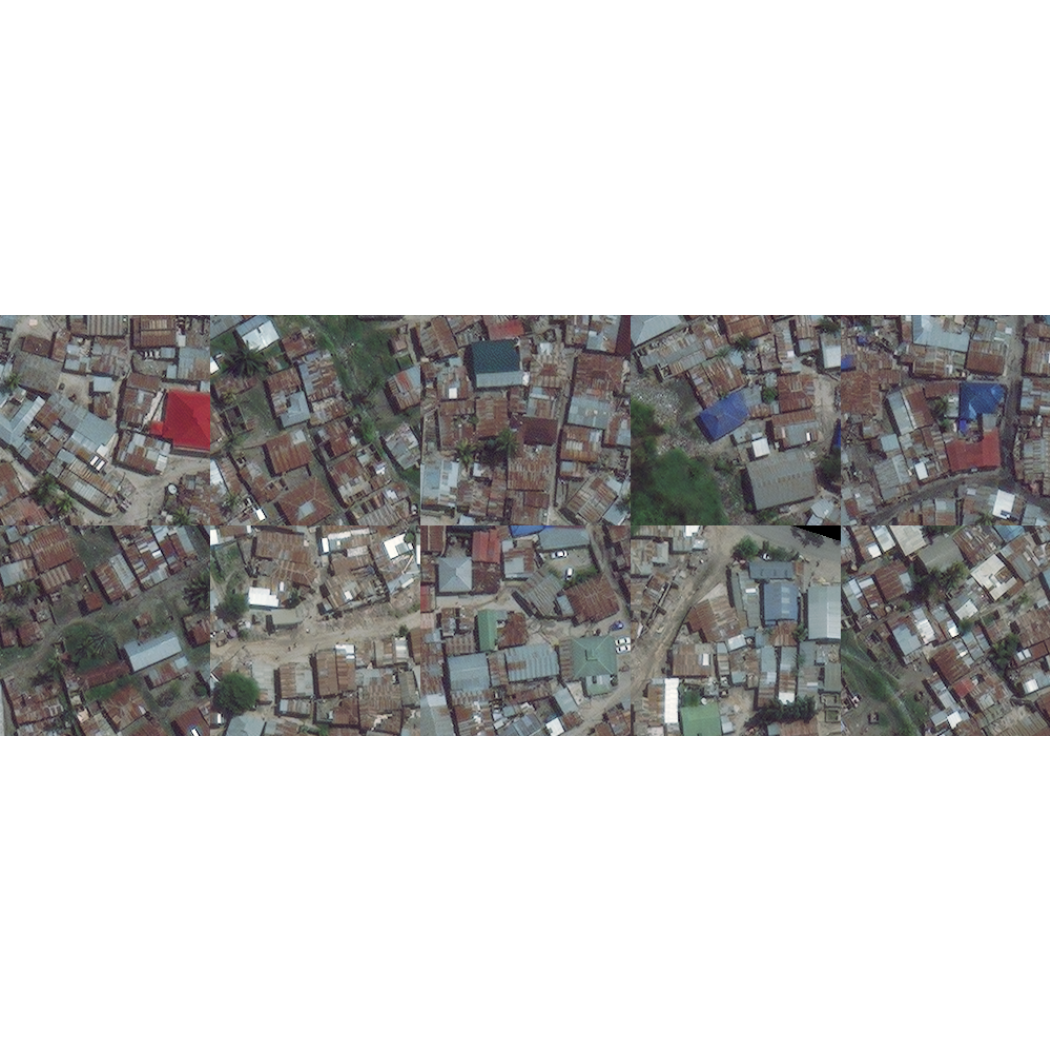}};
    \node (neg2) at (7,-6.3)
        {\includegraphics[width=0.4\linewidth]{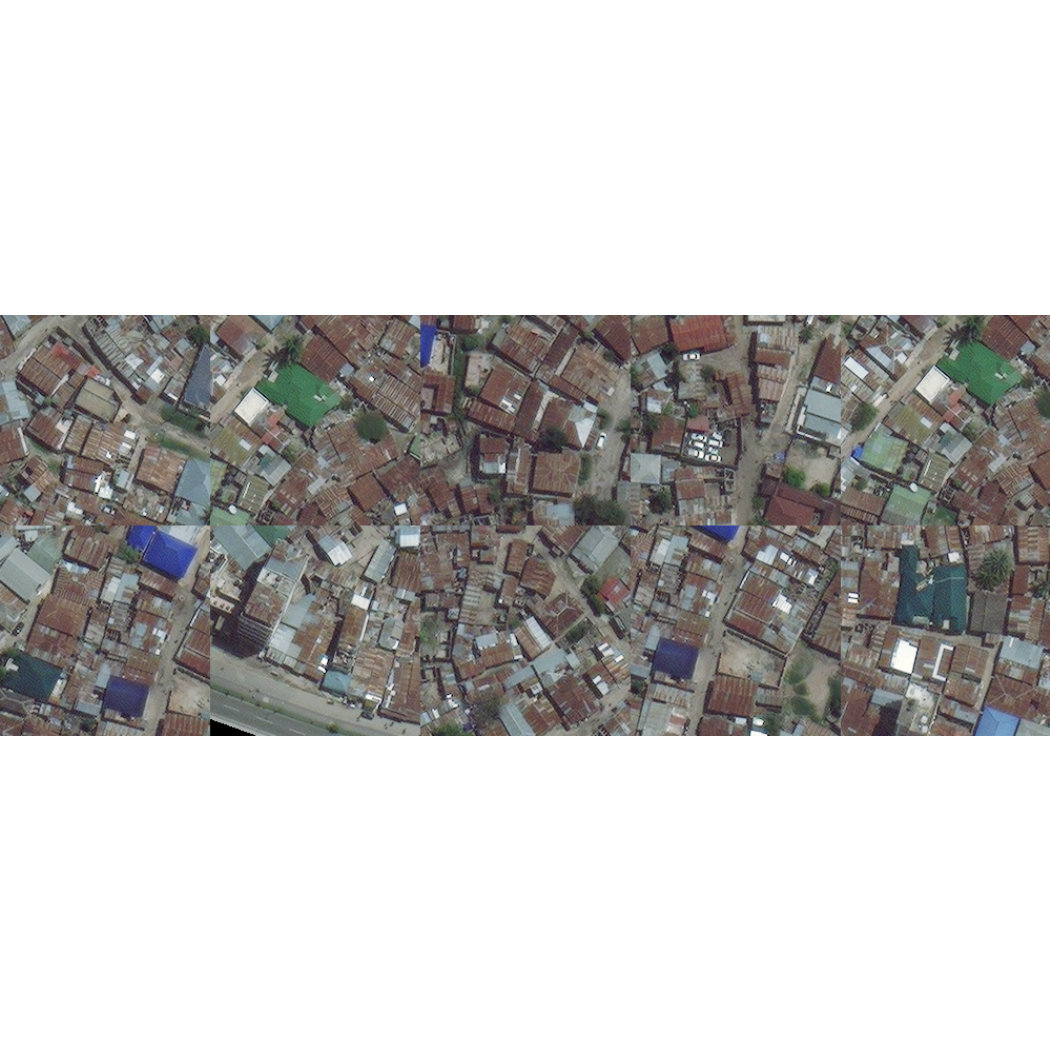}};
    \node (loadings) at (0,-7.5)
        {\includegraphics[width=0.4\linewidth]{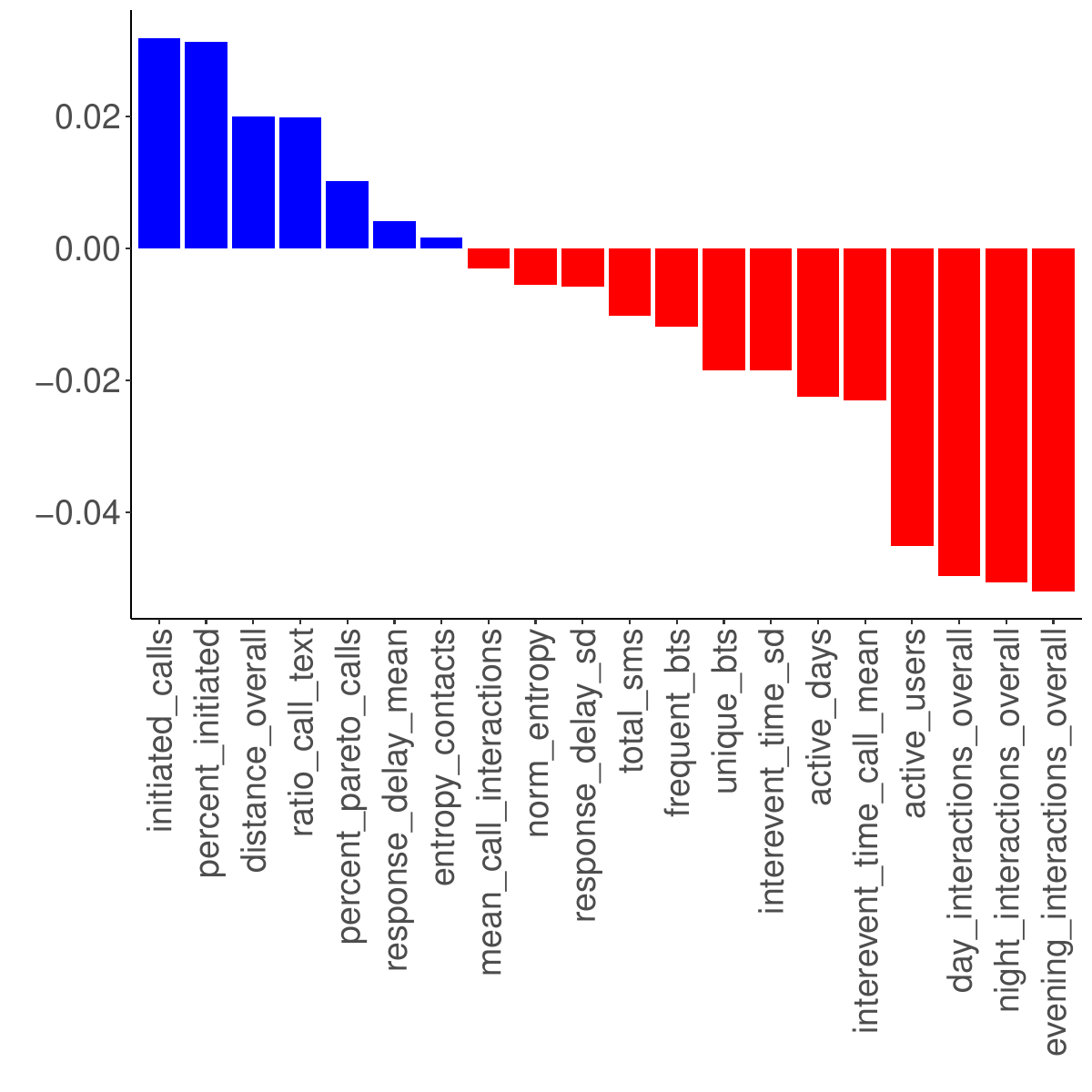}};
    \node (a) at (0,-3.7)
        {(a)};
    \node (b) at (0,-10.6)
        {(b)};
    \node (c) at (7,-8.2)
        {(c)};
    \draw [ultra thick, blue, ->] (2,-2.4) to [bend left] (4,2.8);
    \draw [ultra thick, blue, ->] (-2.4,2) to [bend left] (4,0);
    \draw [ultra thick, red, ->] (-0.17,0.48) to [bend left] (4,-3.5);
    \draw [ultra thick, red, ->] (-0.18,0.28) to [bend left] (4,-6.3);
    \node (t) at (-2.5,5)
        {\small{\textbf{Joint component 1}}};
    \node (p) at (7,4.2)
        {\footnotesize{\textbf{Patches from most positive subwards}}};
    \node (n) at (7,-2.1)
        {\footnotesize{\textbf{Patches from most negative subwards}}};
    \node (l) at (0,-4.5)
        {\footnotesize{\textbf{CDR feature loadings}}};
\end{tikzpicture}
    \caption{\small{(a) Subward scores for JC1. Red corresponds to negative scores and blue to positive scores. Grey subwards are those for which we do not have image data. (b) CDR feature loadings, sorted from most positive to most negative. (c) Patches from the two most positive and two most negative subwards in JC1, with arrows showing to which subwards they correspond. They are ordered from left to right according to how extreme are the patch scores, with more the extreme on the left.}}
    \label{fig:joint-1}
\end{figure}

\begin{figure}
\begin{tikzpicture}
    \node (scores) at (0,0)
        {\includegraphics[width=0.6\linewidth, trim={0 0cm 0 1cm}, clip]{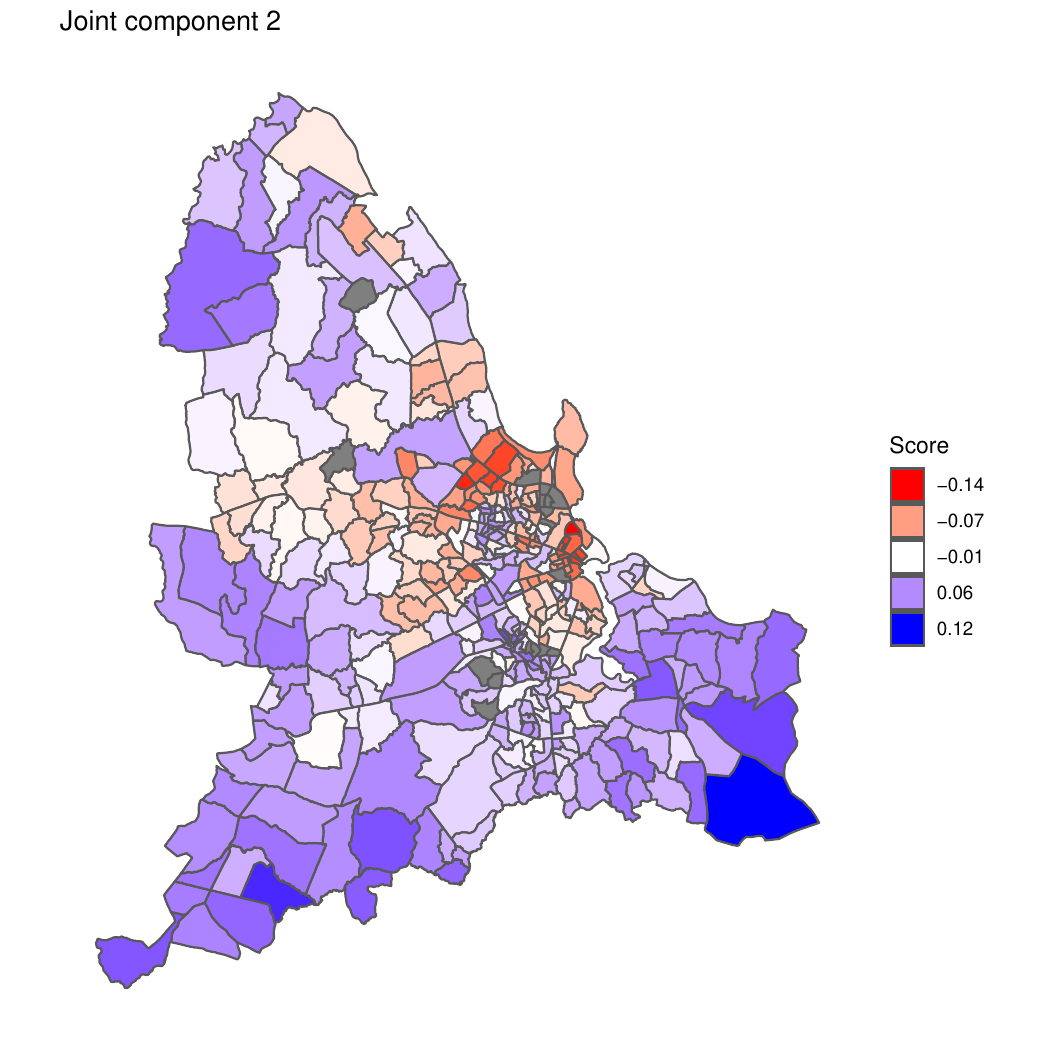}};
    \node (pos1) at (7,2.8)
        {\includegraphics[width=0.4\linewidth]{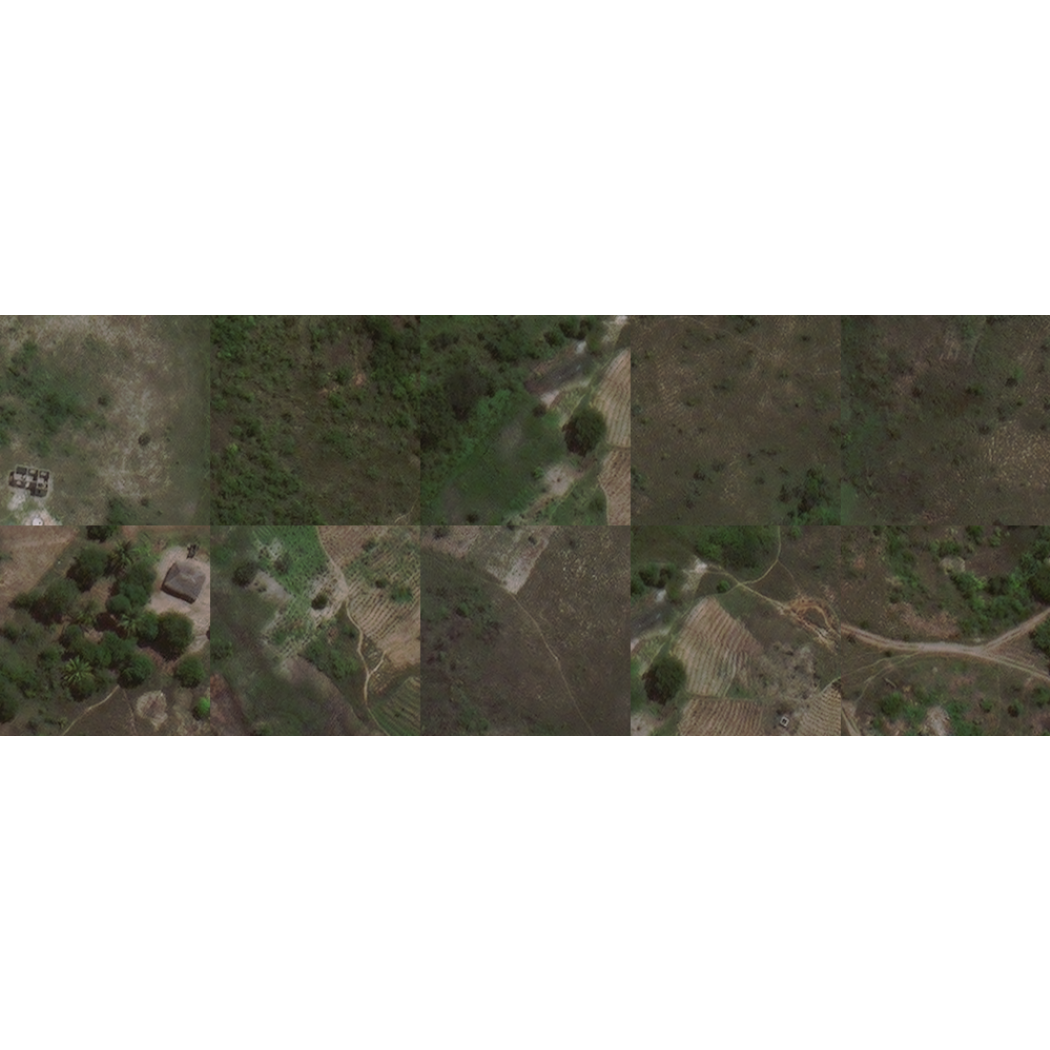}};
    \node (pos2) at (7,0)
        {\includegraphics[width=0.4\linewidth]{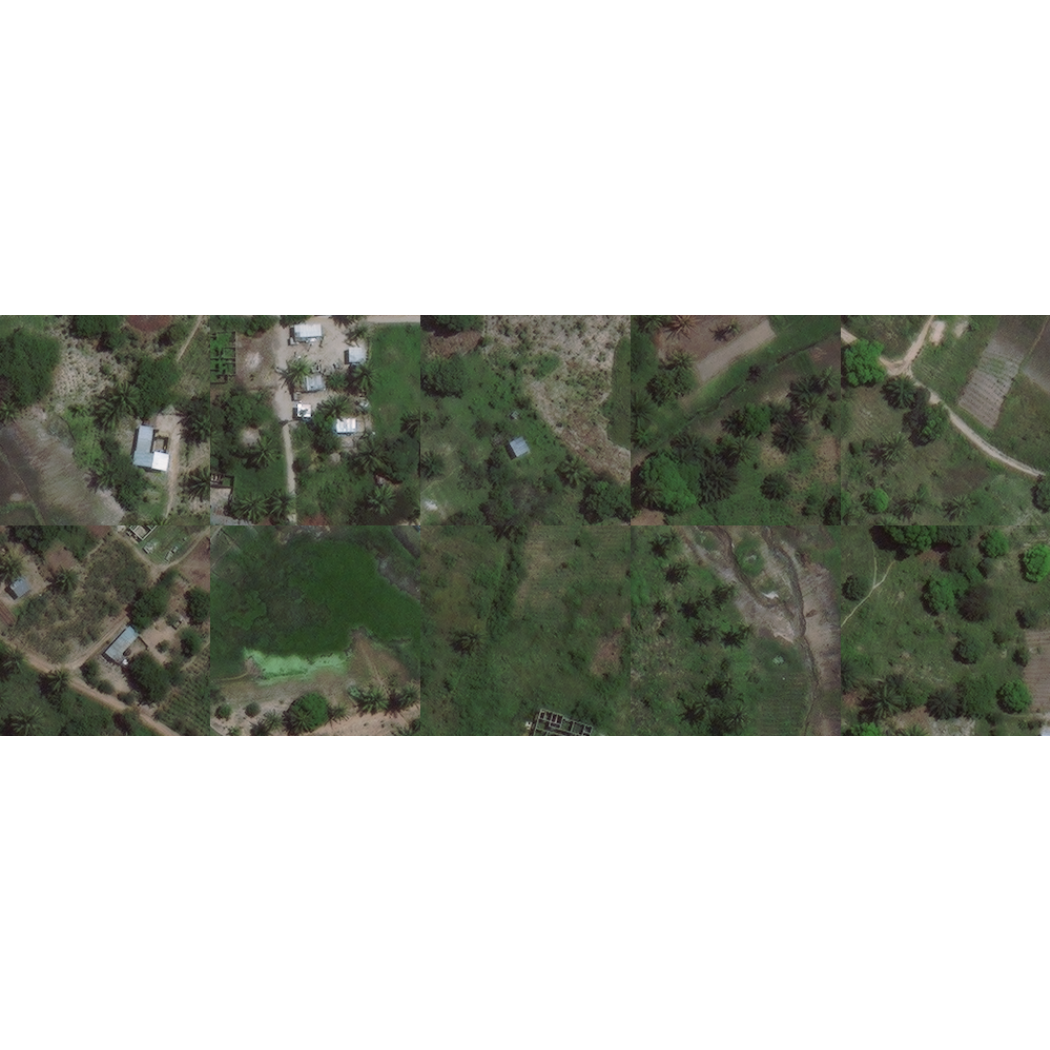}};
    \node (neg1) at (7,-3.5)
        {\includegraphics[width=0.4\linewidth]{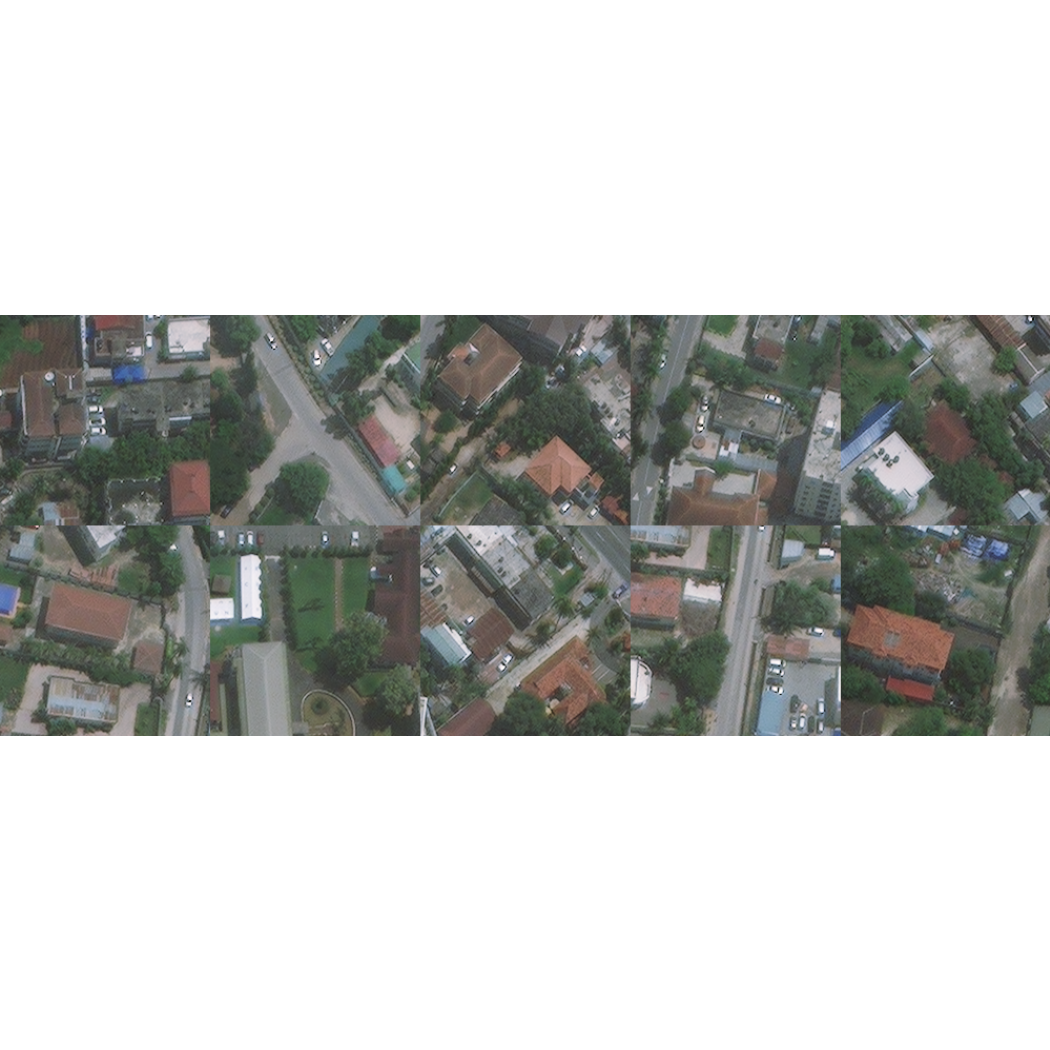}};
    \node (neg2) at (7,-6.3)
        {\includegraphics[width=0.4\linewidth]{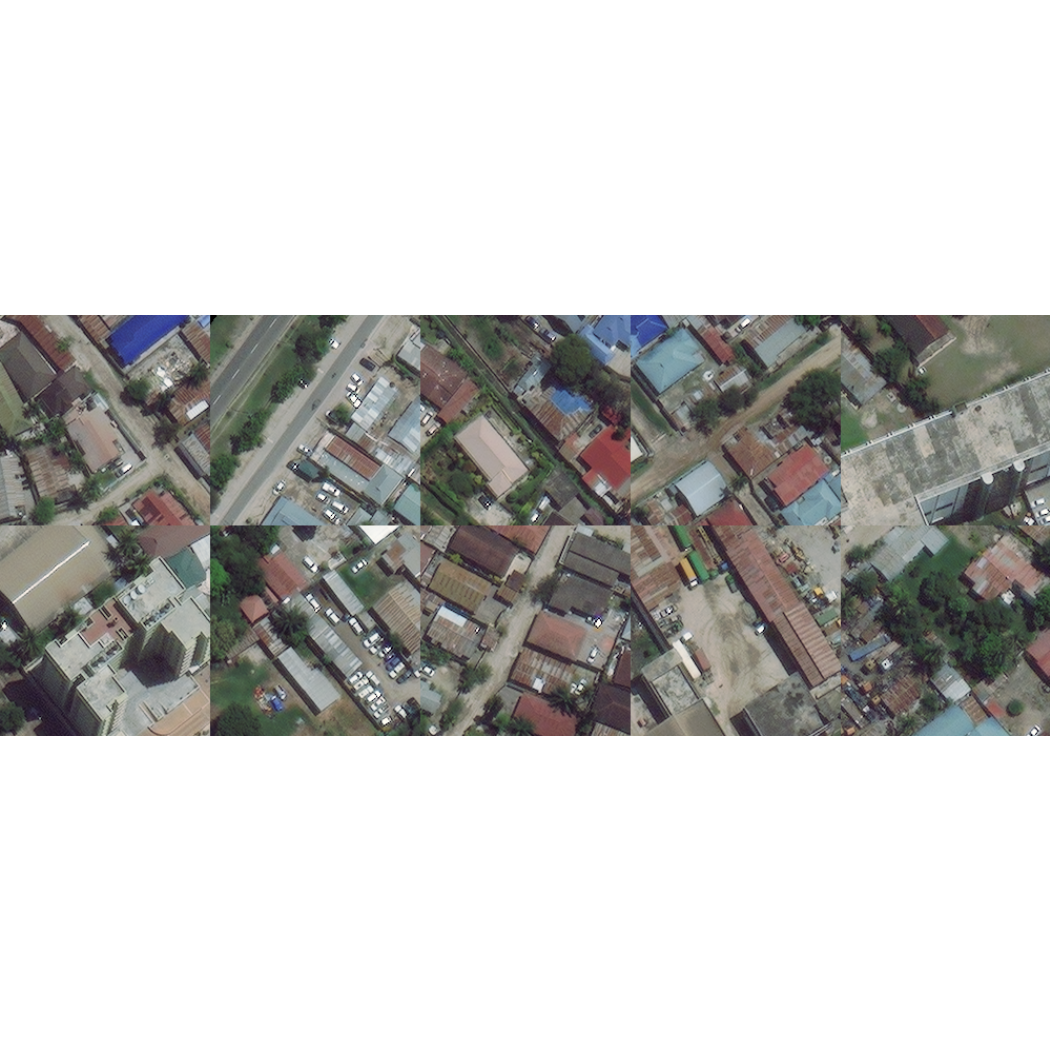}};
    \node (loadings) at (0,-7.5)
        {\includegraphics[width=0.4\linewidth]{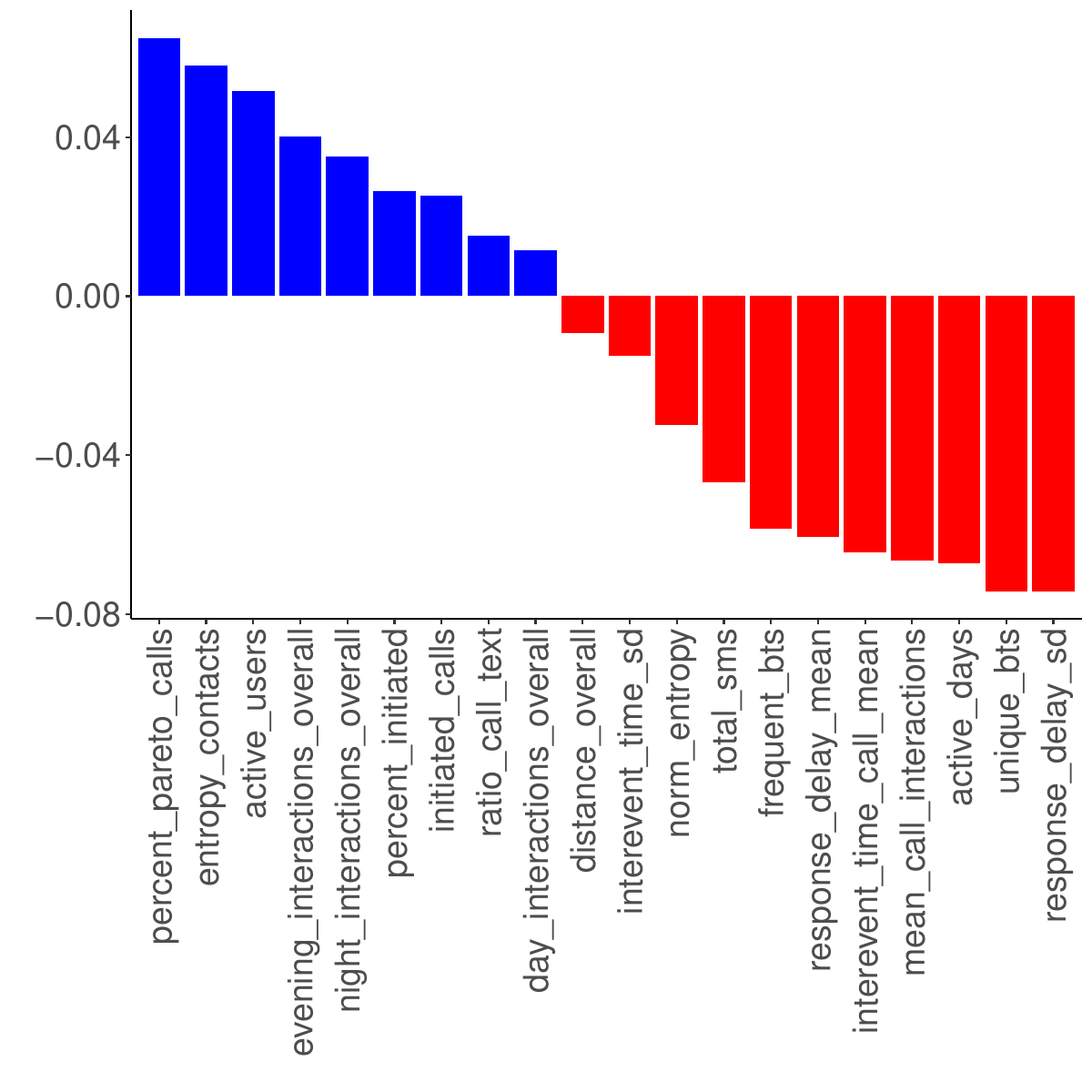}};
    \node (a) at (0,-3.7)
        {(a)};
    \node (b) at (0,-10.6)
        {(b)};
    \node (c) at (7,-8.2)
        {(c)};
    \draw [ultra thick, blue, ->] (2,-2.4) to [bend left] (4,2.8);
    \draw [ultra thick, blue, ->] (-2,-2.8) to [bend left] (4,0);
    \draw [ultra thick, red, ->] (0.35,0.22) to [bend left] (4,-3.5);
    \draw [ultra thick, red, ->] (-0.52,0.6) to [bend left] (4,-6.3);
    \node (t) at (-2.5,5)
        {\small{\textbf{Joint component 2}}};
    \node (p) at (7,4.2)
        {\footnotesize{\textbf{Patches from most positive subwards}}};
    \node (n) at (7,-2.1)
        {\footnotesize{\textbf{Patches from most negative subwards}}};
    \node (l) at (0,-4.5)
        {\footnotesize{\textbf{CDR feature loadings}}};
\end{tikzpicture}
    \caption{\small{(a) Subward scores for JC2. Red corresponds to negative scores and blue to positive scores. Grey subwards are those for which we do not have image data. (b) CDR feature loadings, sorted from most positive to most negative. (c) Patches from the two most positive and two most negative subwards in JC2 with arrows showing to which subwards they correspond.}}
    \label{fig:joint-2}
\end{figure}

\begin{figure}
\begin{tikzpicture}
    \node (scores) at (0,0)
        {\includegraphics[width=0.6\linewidth, trim={0 0cm 0 1cm}, clip]{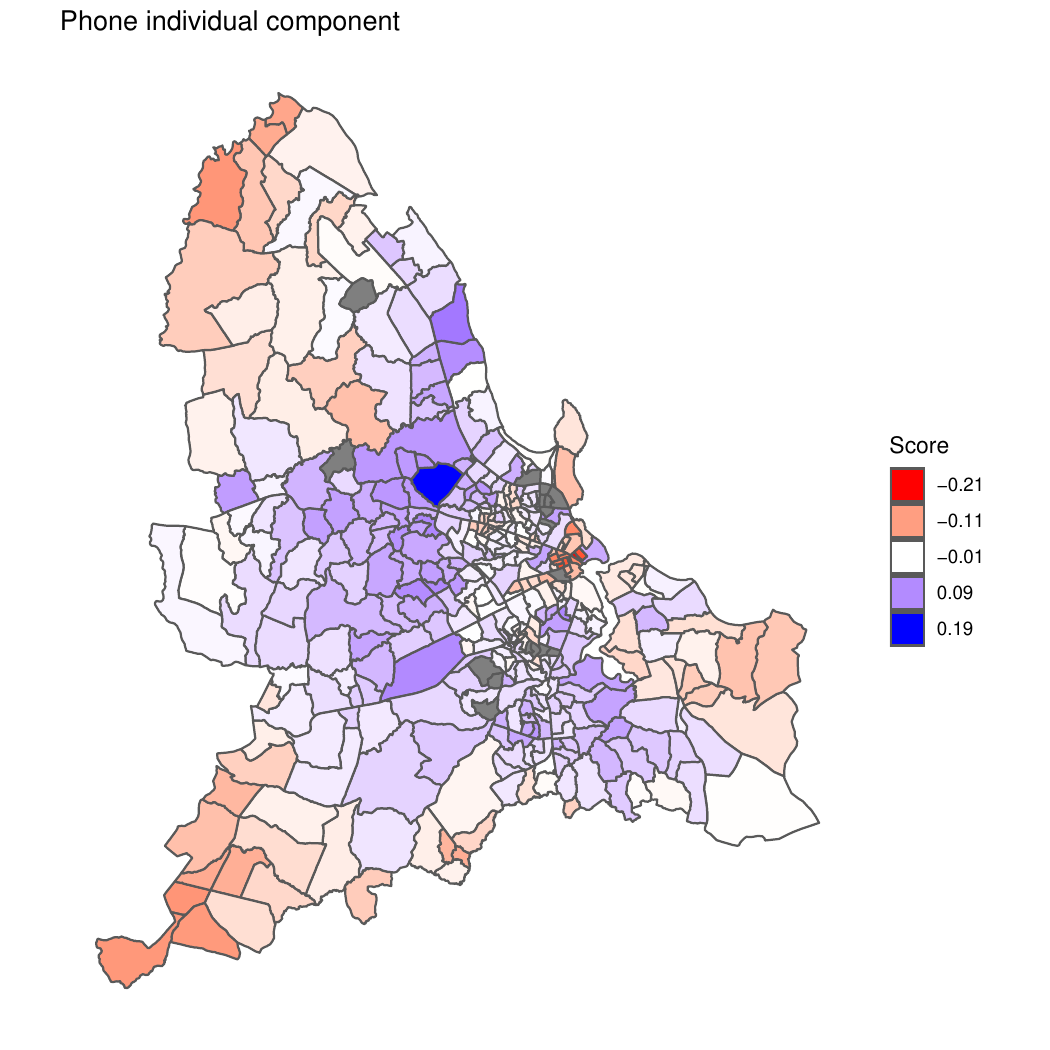}};
    \node (pos1) at (7,2.8)
        {\includegraphics[width=0.4\linewidth]{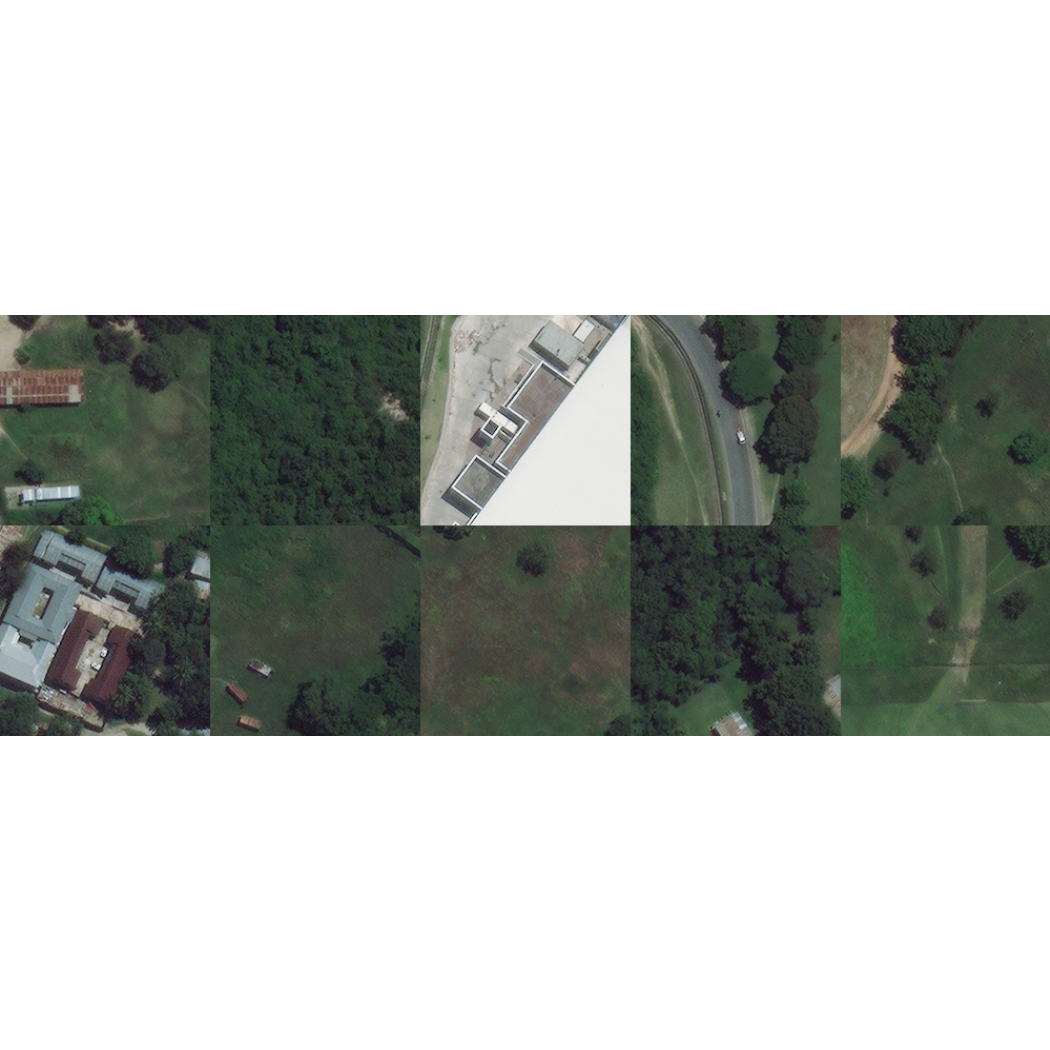}};
    \node (pos2) at (7,0)
        {\includegraphics[width=0.4\linewidth]{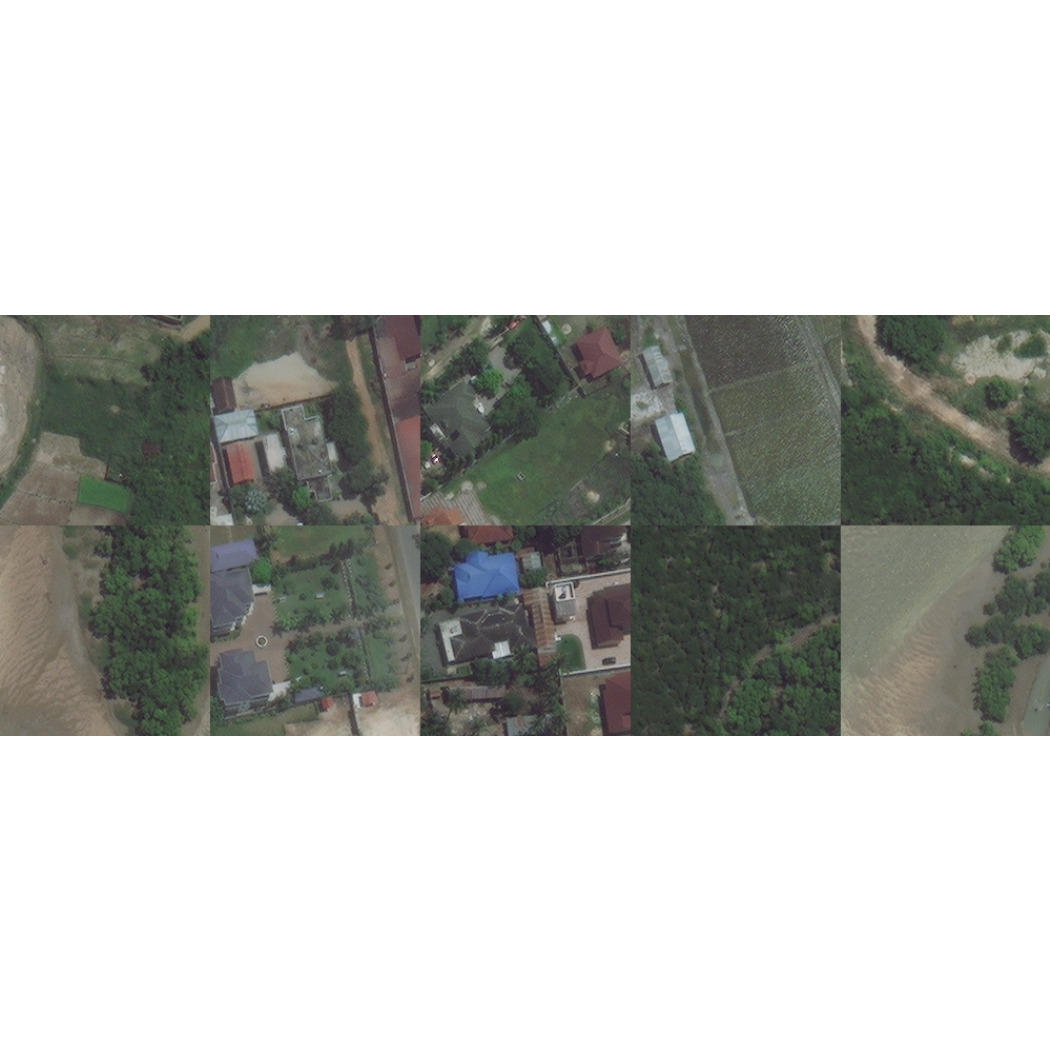}};
    \node (neg1) at (7,-3.5)
        {\includegraphics[width=0.4\linewidth]{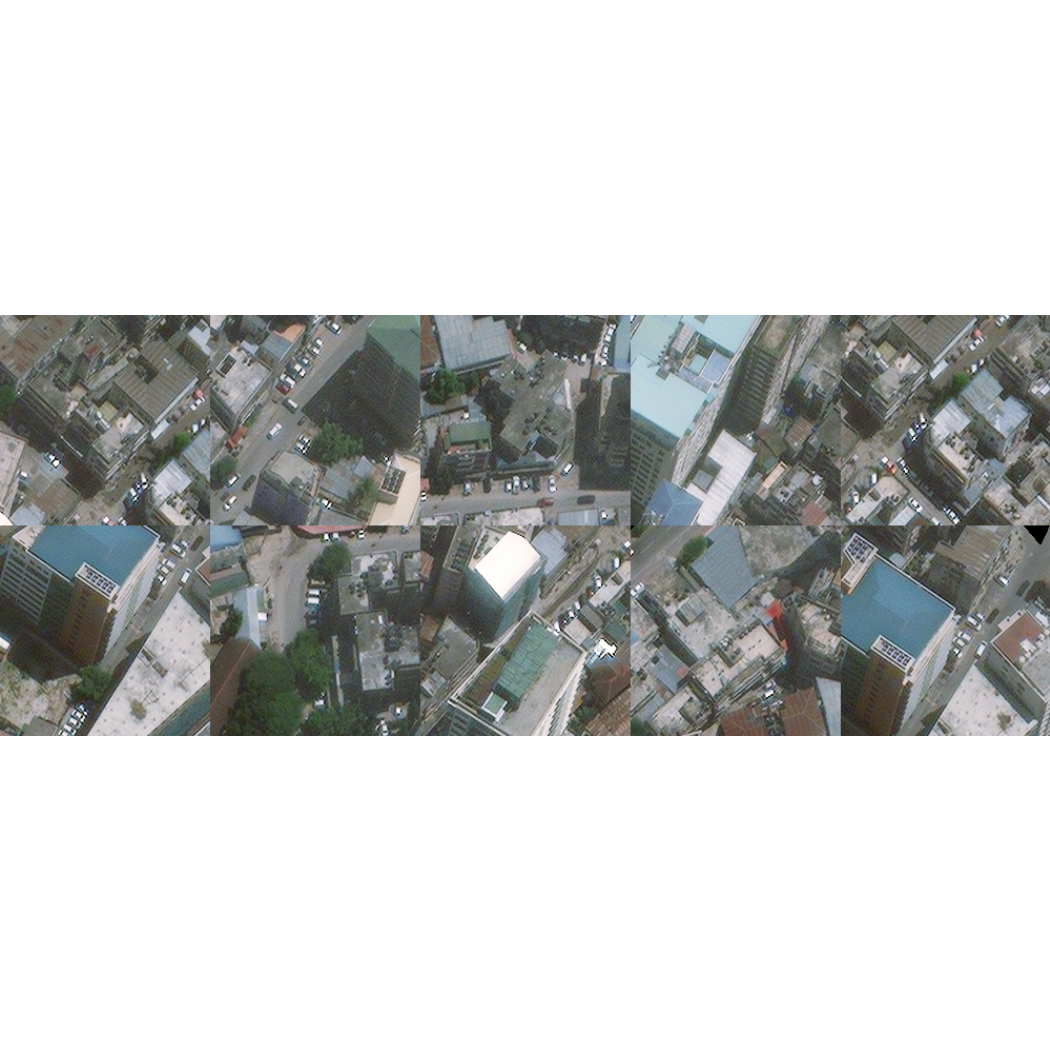}};
    \node (neg2) at (7,-6.3)
        {\includegraphics[width=0.4\linewidth]{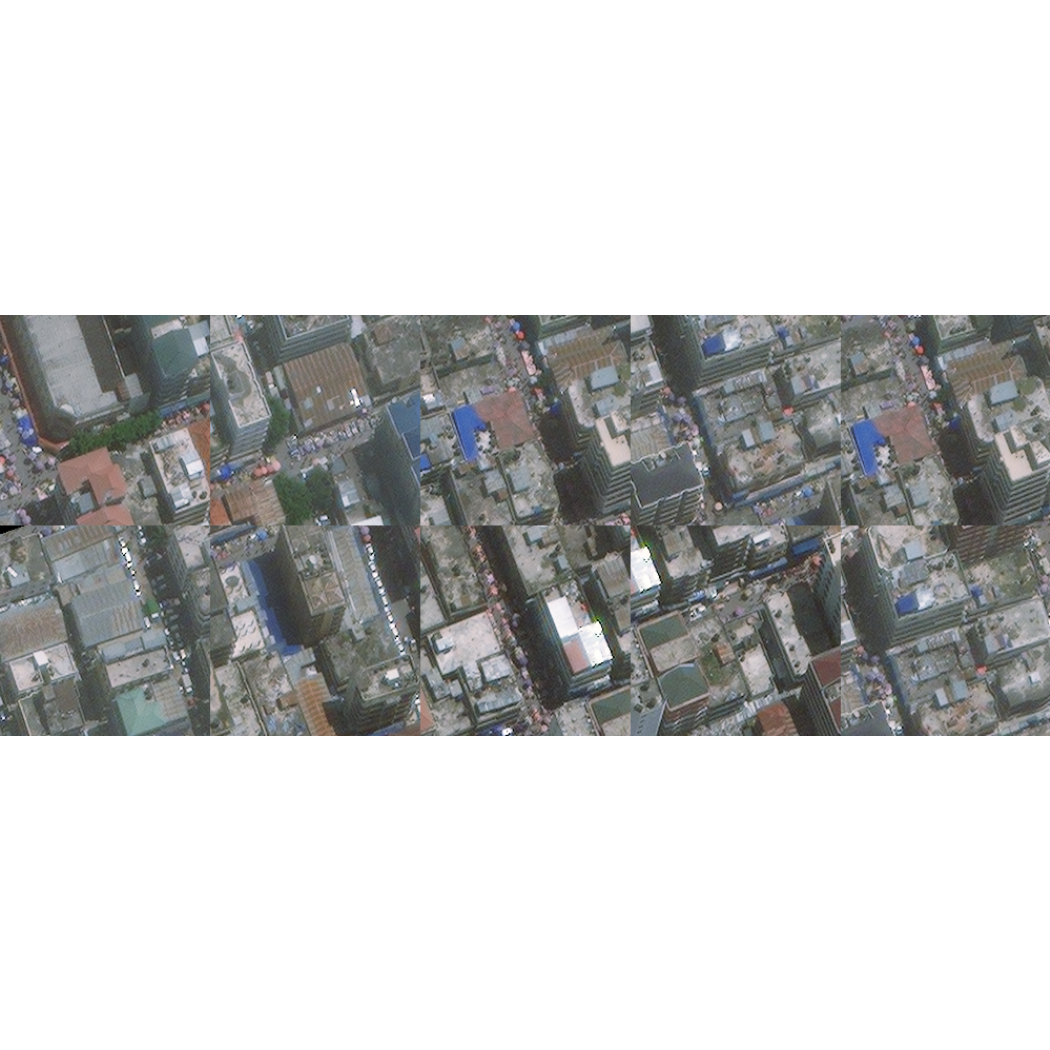}};
    \node (loadings) at (0,-7.5)
        {\includegraphics[width=0.4\linewidth]{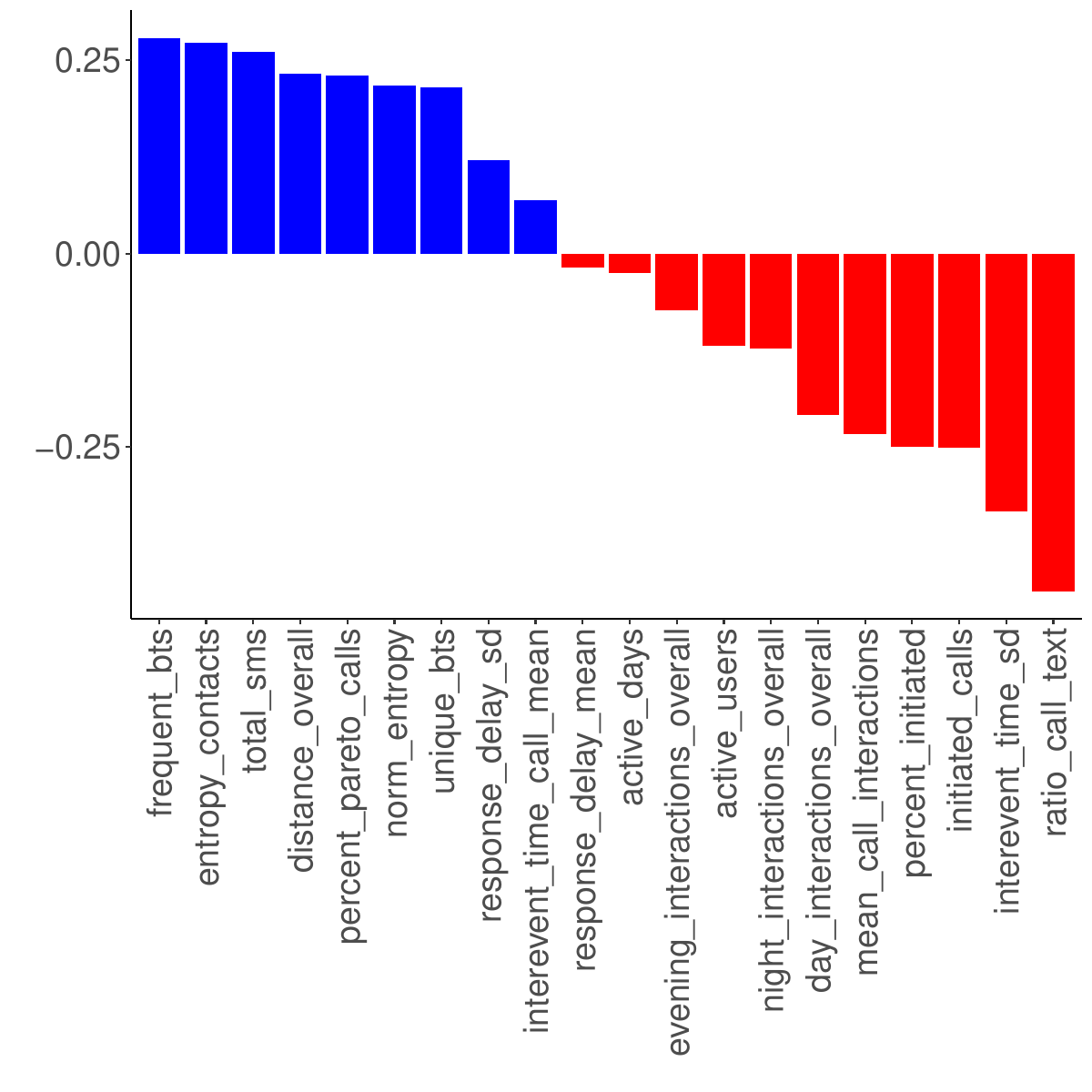}};
    \node (a) at (0,-3.7)
        {(a)};
    \node (b) at (0,-10.6)
        {(b)};
    \node (c) at (7,-8.2)
        {(c)};
    \draw [ultra thick, blue, ->] (-0.7,0.6) to [bend left] (4,2.8);
    \draw [ultra thick, blue, ->] (-0.55,2) to [bend left] (4,0);
    \draw [ultra thick, red, ->] (0.37,-0.01) to [bend left] (4,-3.5);
    \draw [ultra thick, red, ->] (0.24,-0.1) to [bend left] (4,-6.5);
    \node (t) at (-2.5,5)
        {\small{\textbf{CDR individual component}}};
    \node (p) at (7,4.2)
        {\footnotesize{\textbf{Patches from most positive subwards}}};
    \node (n) at (7,-2.1)
        {\footnotesize{\textbf{Patches from most negative subwards}}};
    \node (l) at (0,-4.5)
        {\footnotesize{\textbf{CDR feature loadings}}};
\end{tikzpicture}
    \caption{\small{(a) Subward scores for IC1\textsuperscript{CDR}. Red corresponds to negative scores and blue to positive scores. Grey subwards are those for which we do not have image data. (b) CDR feature loadings, sorted from most positive to most negative. (c) Patches from the two most positive and two most negative subwards in IC1\textsuperscript{CDR}, with arrows showing to which subwards they correspond.}}
    \label{fig:phone}
\end{figure}

We look first at each of the components separately (Figures \ref{fig:joint-1} to \ref{fig:image}). For each component, the figure shows the subward scores (columns of $\boldU_J$ or $\boldU_i$), patches from the subwards at the extreme positive and negative ends of each component, and, where applicable, the CDR feature loadings (rows of $\boldU_J$ or $\boldU_i$). (Although the relative values of scores and loadings are of interest, it is arbitrary which end is negative and which is positive.)

\subsection{Joint components} \label{sec:results-joint}

Figure \ref{fig:joint-1} illustrates the results for the first joint component (JC1). The positive end of the component shows green, rural areas; at the negative end, the patches show built-up areas with small, high-density residential buildings. We can also see from Figure \ref{fig:joint-1}(a) that most subwards with negative scores are small and located close to the city centre. The most positive CDR features are the number of calls initiated by users, the average distance between users and the people they call, and the ratio of calls to SMS. This suggests that residents in these areas are more likely to initiate calls and to make calls rather than sending SMS, compared to those at the negative end of the component. It also seems reasonable that people living in rural areas, with low population density, would tend to communicate with people further away, as there are fewer people living close by. The most negative CDR features are the number of day, evening, and night interactions, and the number of active users; these are likely to be correlated with population.

For the second joint component (JC2) (Figure \ref{fig:joint-2}), the subwards with the most positive scores again appear to be more rural areas, but the most negative subwards are in different parts of the city, notably in the coastal region near the centre. These areas correspond to the wards of Upanga -- which stands out for its post-colonial layout, hosting government institutions, diplomatic housing, and commercial zones -- and Ubungo, which also features structured residential development, commercial facilities and a key transport hub, with the patches reflecting these relatively formal, planned segments of Dar es Salaam’s urban setting. These characteristics contrast markedly with more informal, under-serviced areas of the city, indicating that this end of JC2 captures higher-order urban planning, infrastructure investment, and socio-economic status.

The most positive CDR features are the percentage of contacts that account for $80\%$ of a user's call interactions, entropy of contacts --- which are both measures of the diversity of contacts with whom a user interacts --- and the number of active users. Looking at the CDR features with the most negative loadings, we infer that at this end of the component, users move around more (to different towers), make more calls, and use their phones on more days.

\subsection{Individual components} \label{sec:results-ind}

For the CDR individual component (IC1\textsuperscript{CDR}) (Figure \ref{fig:phone}), the most positive CDR features are those related to entropy of locations and contacts, and the total number of SMS sent. At the negative end are the ratio of calls to texts, and the standard deviation of interevent times. This end of the component also seems to contain built-up areas with taller buildings, although it is less clear what the patches from the positive end show. (This is perhaps not surprising, as this component does not directly use information from the patches.)

The most positive subward seems to have a particularly extreme value: we can see that it is much brighter than all the other subwards in Figure \ref{fig:phone} (a). This subward contains the University of Dar es Salaam, as well as a large shopping centre (part of the roof of this is shown in one of the patches in Figure \ref{fig:phone} (c)); it seems reasonable to assume that these would have an impact on patterns of phone usage, as people's activities in this subward will likely be different to other areas of the city.

\begin{figure}
\begin{tikzpicture}
    \node (scores) at (0,0)
        {\includegraphics[width=0.6\linewidth, trim={0 0cm 0 1cm}, clip]{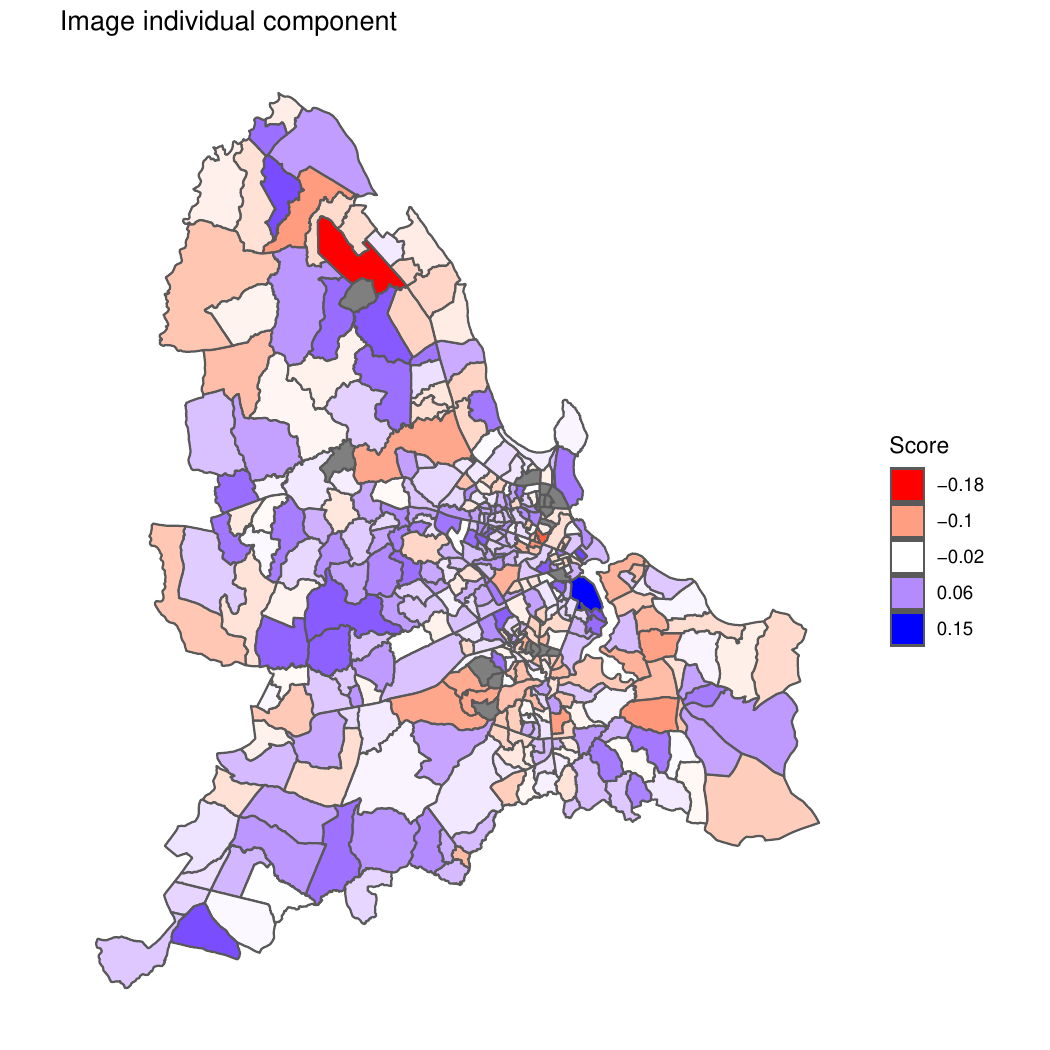}};
    \node (pos1) at (7,2.8)
        {\includegraphics[width=0.4\linewidth]{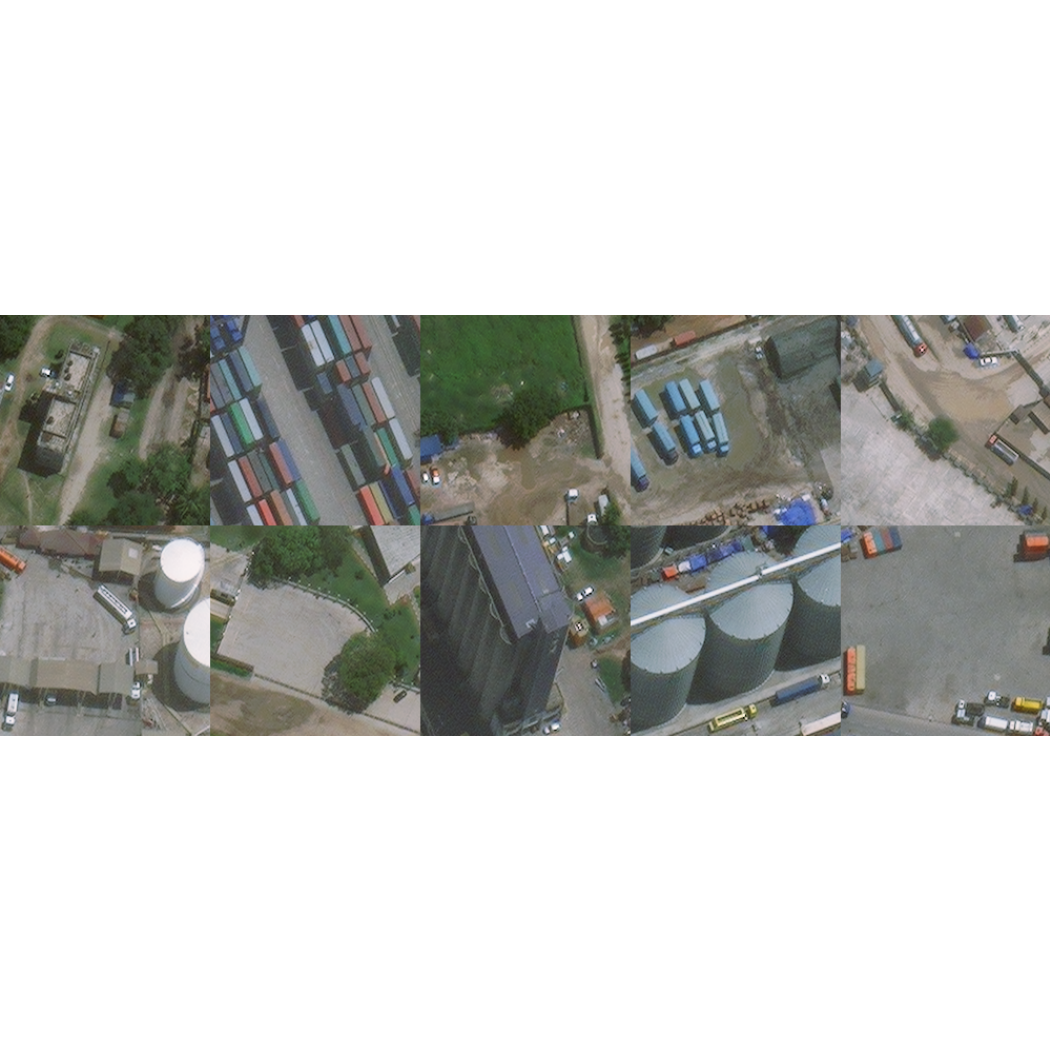}};
    \node (pos2) at (7,0)
        {\includegraphics[width=0.4\linewidth]{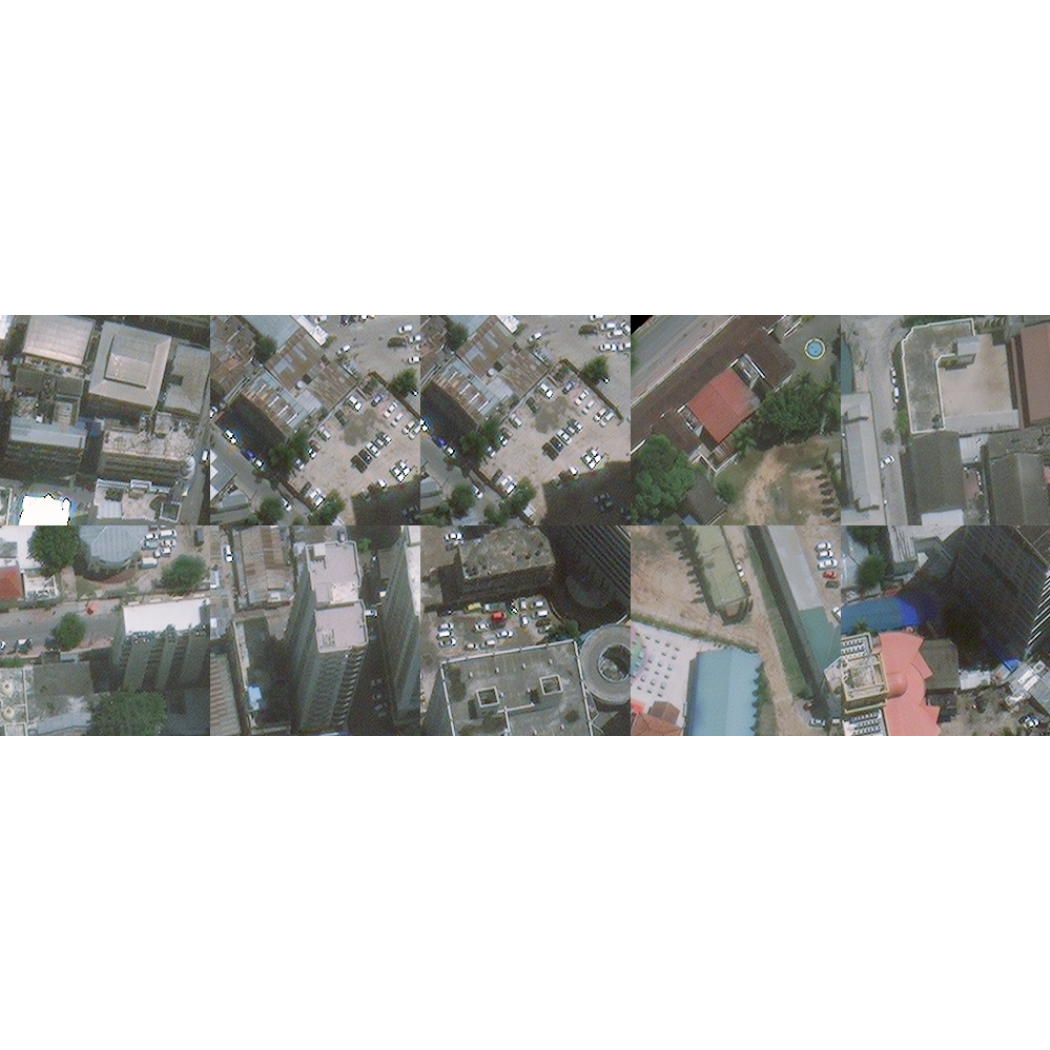}};
    \node (neg1) at (7,-3.5)
        {\includegraphics[width=0.4\linewidth]{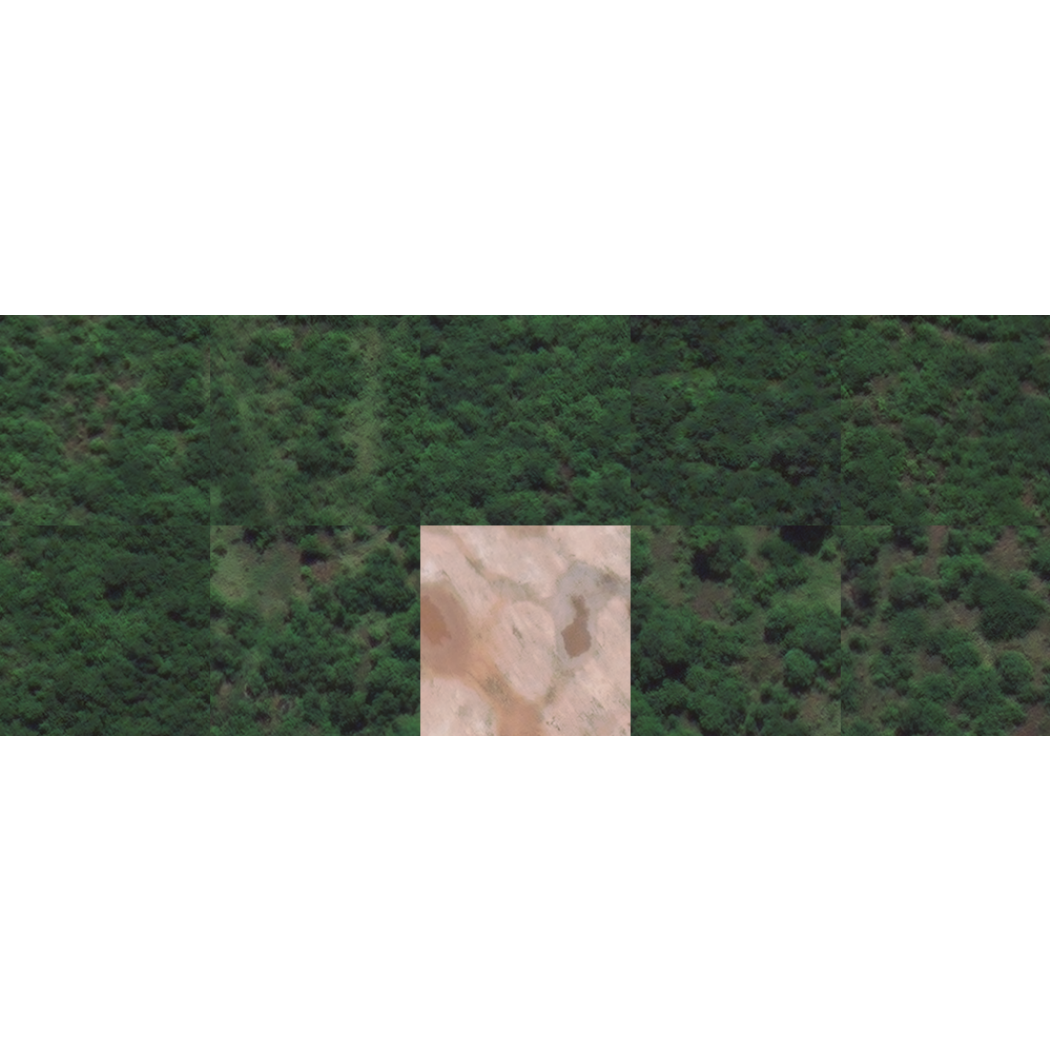}};
    \node (neg2) at (7,-6.3)
        {\includegraphics[width=0.4\linewidth]{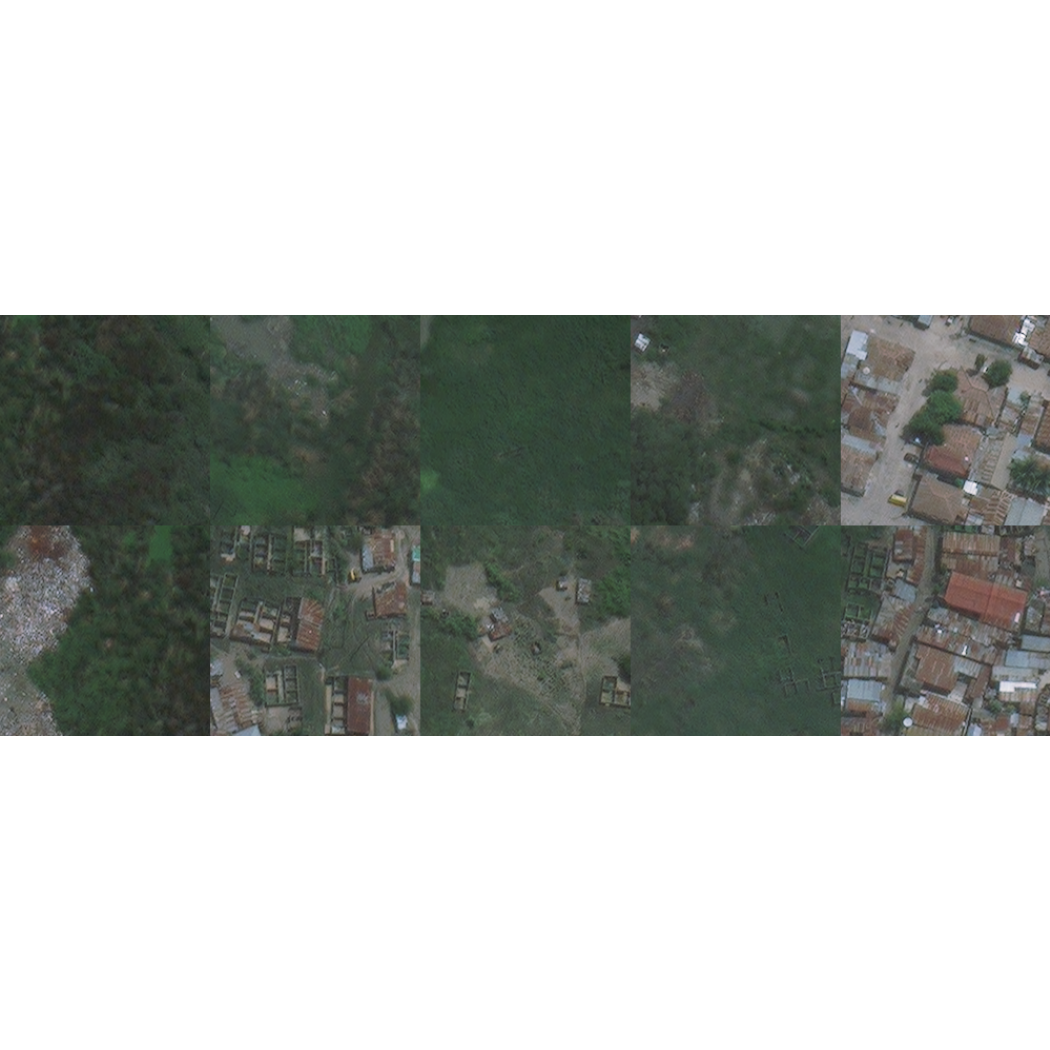}};
    \node (a) at (0,-3.7)
        {(a)};
    \node (b) at (7,-8.2)
        {(b)};
    \draw [ultra thick, blue, ->] (0.5,-0.5) to [bend left] (4,2.8);
    \draw [ultra thick, blue, ->] (0.5,0) to [bend left] (4,0);
    \draw [ultra thick, red, ->] (-1.5,2.45) to [bend left] (4,-3.5);
    \draw [ultra thick, red, ->] (0.15,0.12) to [bend left] (4,-6.3);
    \node (t) at (-2.5,5)
        {\small{\textbf{Image individual component}}};
    \node (p) at (7,4.2)
        {\footnotesize{\textbf{Patches from most positive subwards}}};
    \node (n) at (7,-2.1)
        {\footnotesize{\textbf{Patches from most negative subwards}}};
\end{tikzpicture}
    \caption{\small{(a) Subward scores for IC1\textsuperscript{Image}. Red corresponds to negative scores and blue to positive scores. Grey subwards are those for which we do not have image data. (b) Patches from the two most positive and two most negative subwards in IC1\textsuperscript{Image}, with arrows showing to which subwards they correspond.}}
    \label{fig:image}
\end{figure}

For the image individual component (IC1\textsuperscript{Image}) (Figure \ref{fig:image}), it looks like the positive end corresponds to industrial and commercial areas, whereas the negative end is mostly rural. There are no CDR feature loadings to display for this component. As for IC1\textsuperscript{CDR}, there appears to be one extreme subward, this time at the negative end of the component: one of the patches from this component (Figure \ref{fig:image} (b)) looks unusual, so this may be responsible. However, when we re-implement AJIVE with this subward removed, there is not a noticeable difference in the results for the other subwards, so it does not seem to be having too much influence on the results. (This is also the case for the outlier subward in IC1\textsuperscript{CDR}.

\begin{table}
    \centering
    \begin{tabular}{ccc}
      \hline
        & \textbf{Positive} & \textbf{Negative} \\
      \hline
        Joint component 1 & Rural & High-density, slum housing \\
            & More initiation of calls & Total CDR interactions \\
      \hline
        Joint component 2 & Rural or semi-rural & Industrial/commercial \\
            & Diversity of contacts & \\
      \hline
        CDR individual component & Low-density housing & Industrial/commercial \\
            & High entropy CDR data & High calls to text ratio \\
      \hline
        Image individual component & Industrial/commercial & Rural \\
      \hline
    \end{tabular}
    \caption{\small{Qualitative interpretation of the AJIVE joint and individual components shown in Figures \ref{fig:joint-1}, \ref{fig:joint-2}, \ref{fig:phone}, and \ref{fig:image}, summarising the discussion in Sections \ref{sec:results-joint} and \ref{sec:results-ind}}}
    \label{tab:results}
\end{table}

The scores vectors for IC1\textsuperscript{CDR} and IC1\textsuperscript{Image} have virtually no correlation ($\rho = 0.03$), so we can surmise that all joint variation is adequately captured by JC1 and JC2. We can therefore assume that the information captured in IC1\textsuperscript{CDR} and IC1\textsuperscript{Image} is unique to the CDR and image data sets respectively.

Table \ref{tab:results} briefly summarises our observations in this section.

\begin{figure}
    \centering
    \includegraphics[width=0.8\linewidth]{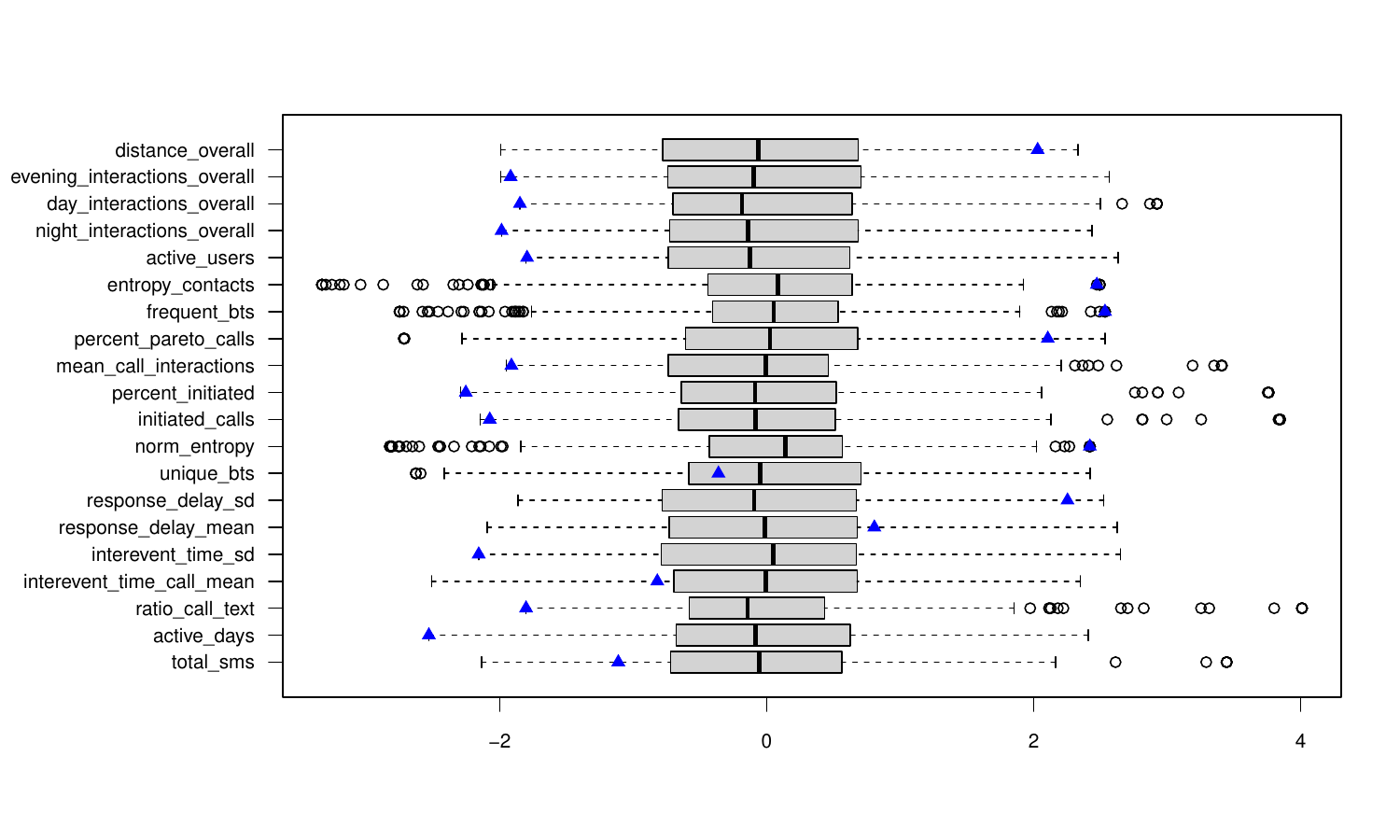}
    \caption{\small{Boxplots of CDR features, with blue triangles showing the values for the subward of Chuo Kikuu.}}
    \label{fig:phone-outlier}
\end{figure}

\subsection{Comparison with Principal Components Analysis} \label{sec:results-pca}

An important question to consider is whether using AJIVE provides a significant advantage over using Principal Components Analysis (PCA). PCA can be used either on $\boldX_1$ and $\boldX_2$ separately, or on the concatenated data matrix $\boldX = \left( \boldX_1, \boldX_2 \right)$. We consider both of these options here, and compare their output to that of AJIVE.

As a reminder, from Section \ref{sec:pca}, using rank-$r$ PCA we represent an $n \times p$ data matrix $\boldX$ as
\begin{equation*}
    \boldX = \boldU_r \boldSigma_r \boldV_r^\top,
\end{equation*}
where $\boldU_r$ and $\boldV_r$ are $n \times r$ and $p \times r$ matrices containing the first $r$ left and right singular vectors of $\boldX$ respectively, and $\boldSigma_r$ is an $r \times r$ diagonal matrix containing the first $r$ singular values of $\boldX$ along its diagonal. As with AJIVE, we can interpret the columns of $\boldU_r$ and $\boldV_r$ as score and loading vectors respectively. The singular values in $\boldSigma_r$ control the relative importance of each component.

\begin{figure}
    \centering
    \includegraphics[width=0.7\linewidth]{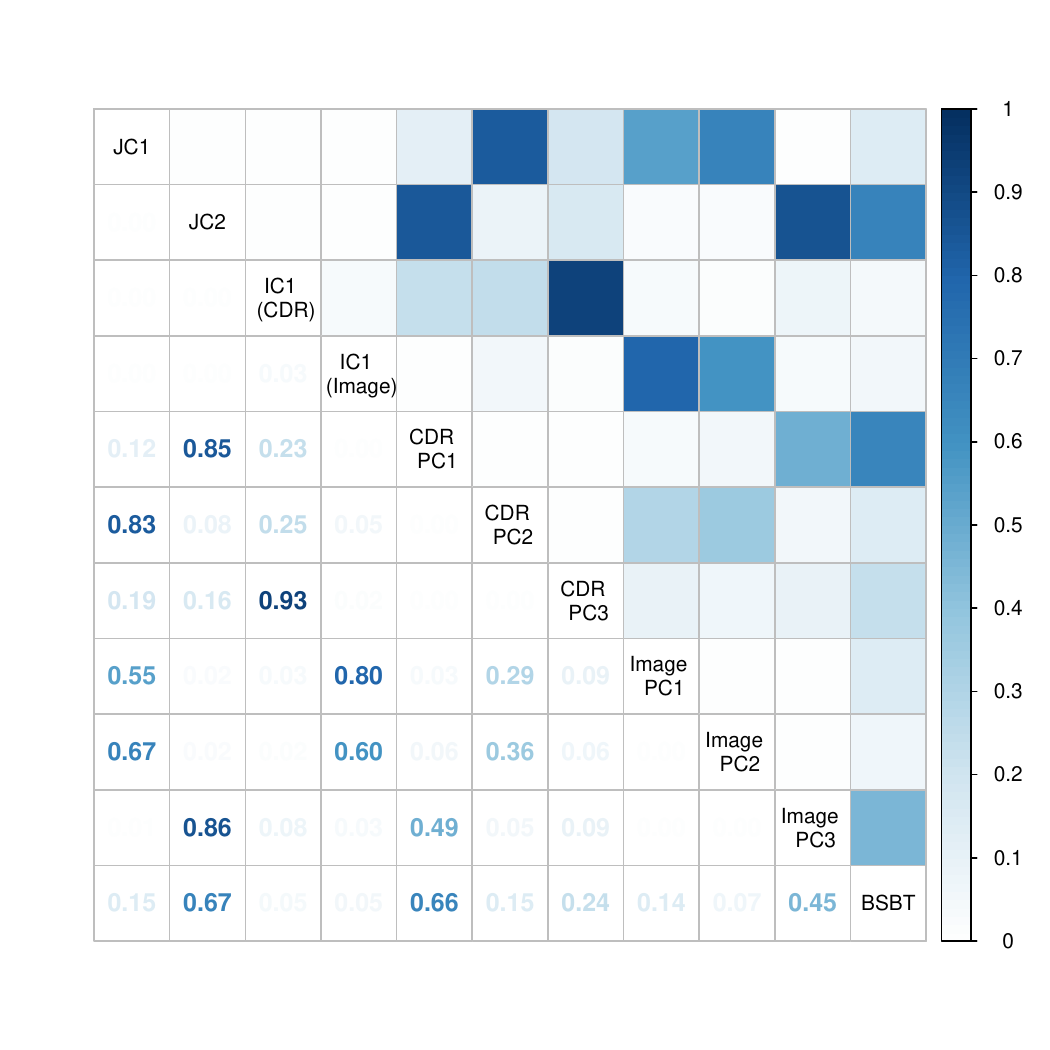}
    \caption{\small{Absolute values of correlations between AJIVE and PC scores, when both are applied to the CDR and image data. The signs of the score vectors are arbitrary, so the directions of the correlations between them are not important. Deprivation scores estimated using the Bayesian Spatial Bradley-Terry model (see Section \ref{sec:data-md}) are also included for comparison.}}
    \label{fig:svd-corrplot}
\end{figure}

To compare AJIVE with separate PC analyses of $\boldX_1$ and $\boldX_2$, we computed the first three principal components (referred to as PC1, PC2, and PC3) for each data set. (Each data set has three AJIVE components related to it --- two joint and one individual --- so this seems a fair comparison.) Figure \ref{fig:svd-corrplot} shows correlations between scores for the AJIVE and PC components (as well as with independently calculated deprivation estimates, which we will explore in Section \ref{sec:results-md}). Note that JC1 and JC2 are forced to be orthogonal to each other, as well as to the individual components, and the PC components for each data matrix are also forced to be orthogonal.

We see that JC1 is strongly correlated with PC2\textsuperscript{CDR} and is also correlated with both PC1\textsuperscript{Image} and PC2\textsuperscript{Image}, whilst JC2 is correlated with PC1\textsuperscript{CDR} and PC3\textsuperscript{Image}. IC1\textsuperscript{CDR} is strongly correlated with PC3\textsuperscript{CDR}, and IC1\textsuperscript{Image} is correlated with PC1\textsuperscript{Image} and PC2\textsuperscript{Image}. IC1\textsuperscript{CDR} appears to be uncorrelated with the image PC scores, and the same applies to IC1\textsuperscript{Image} and the CDR PC scores. AJIVE appears to give similar overall results to doing separate PCAs of the CDR and image data, but the division of components is different. Hence, AJIVE identifies which parts of the components are common to both data sets and which are unique to each data set.

When doing PCA on the concatenated data matrix $\boldX = \left( \boldX_1 \hspace{0.1cm} \boldX_2 \right)$, we find that the first few PCs of $\boldX$ are almost identical to the corresponding PCs of $\boldX_2$. This is not surprising as $\boldX_2$ has a much higher dimensionality than $\boldX_1$ ($ p_1 = 20, p_2 = 1536 $). However, this shows that applying PCA to the concatenated data matrix $\boldX$ is not useful for our data, as we essentially lose the information from $\boldX_1$. This would also apply to other situations where the number of dimensions differs greatly between data matrices.

\subsection{Comparison with deprivation estimates} \label{sec:results-md}

A major aim of this paper is to investigate the extent to which AJIVE can be used to predict deprivation. We are interested in whether our methods and data sources are able to achieve this in situations where alternative data sources are unavailable; however, to assess how well this approach works, we here use deprivation estimates created from a completely distinct data set (details were given in Section \ref{sec:data-md}).

Figure \ref{fig:svd-corrplot} shows correlations between the deprivation estimates and AJIVE and PC scores. There is a fairly strong correlation between deprivation and JC2 ($\rho = 0.67$). The sign of the correlation (not shown in the figure) is negative, meaning that subwards at the positive end of JC2 tend to have lower deprivation scores (corresponding to more deprived areas; see Section \ref{sec:data-md}), whilst subwards at the negative end tend to be less deprived. Hence, this component seems to be picking up joint variation in the data which is relevant for determining deprivation. In particular, this indicates that deprivation information is present in both $\boldX_1$ and $\boldX_2$. For each of the other AJIVE components there is no or very little correlation with deprivation.

\subsection{Incorporating the survey data} \label{sec:results-survey}

So far, we have looked at the application of AJIVE and PCA to the CDR and image data sets. We now repeat the preceding analysis, but with addition of the survey data introduced in Section \ref{sec:data-survey}. Recall that the survey data are slower and more costly to collect than the CDR and image data, so it is natural to ask: what extra information does the survey data provide, and do any improvements warrant the cost of collecting the data in the first place?

\begin{figure}
    \centering
    \includegraphics[width=0.33\linewidth]{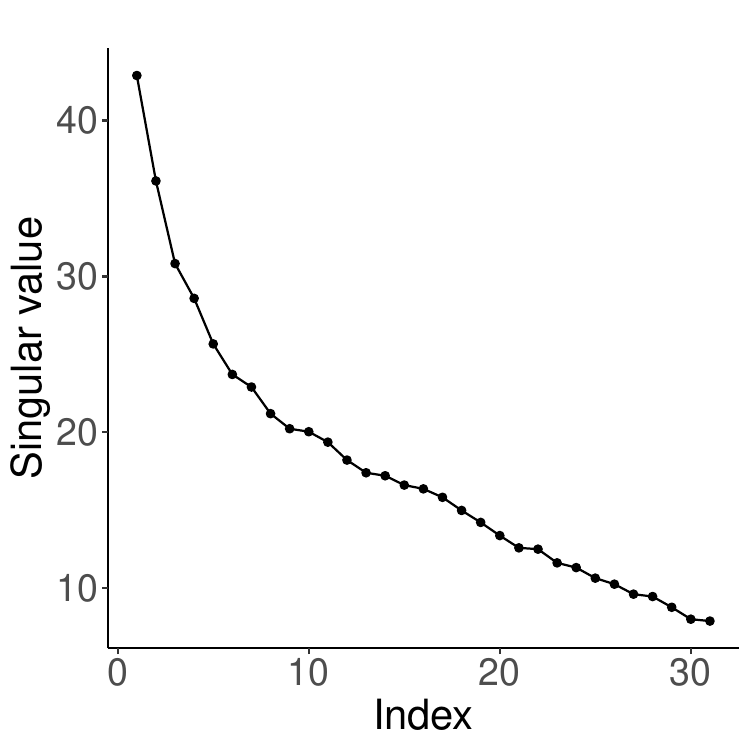}
        \put(-66,-7){\footnotesize{(a)}}
    \includegraphics[width=0.33\linewidth]{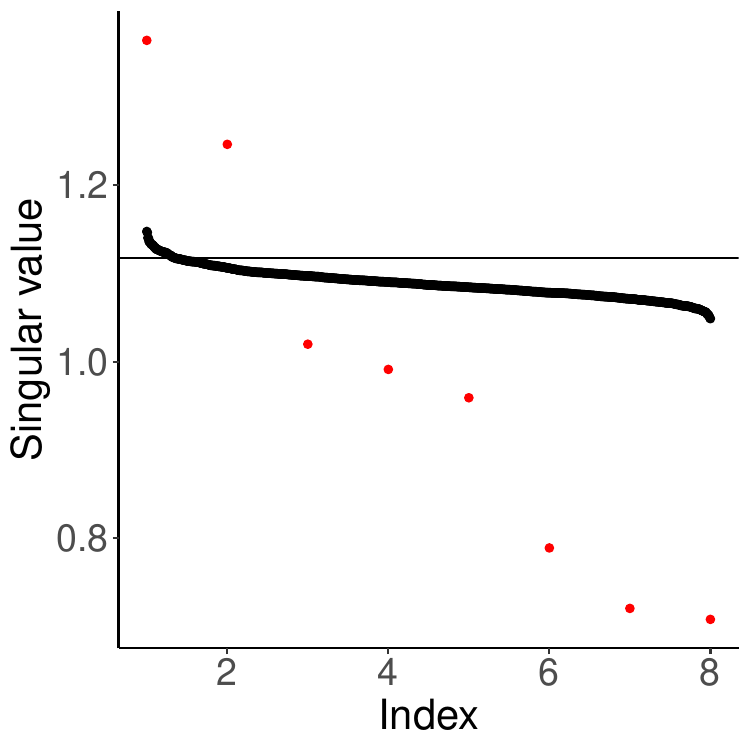}
        \put(-64,-7){\footnotesize{(b)}}
    \caption{\small{Rank estimation for survey data. (a). Scree plot. (b). Estimation of joint rank using the random samples method, with initial ranks of (3, 3, 2).}}
    \label{fig:scree-survey}
\end{figure}

The scree plot (Figure \ref{fig:scree-survey}) suggests an initial rank of 2 for the survey data. Applying the random direction bound to estimate the joint rank (as in Section \ref{sec:ranks}, but using all three data matrices) returns a joint rank $r_J = 2$. We also obtain individual rank estimates of $r_1 = r_2 = r_3 = 1$.

\begin{figure}
\begin{tikzpicture}
    \node (scores) at (0,0)
        {\includegraphics[width=0.6\linewidth, trim={0 0cm 0 1cm}, clip]{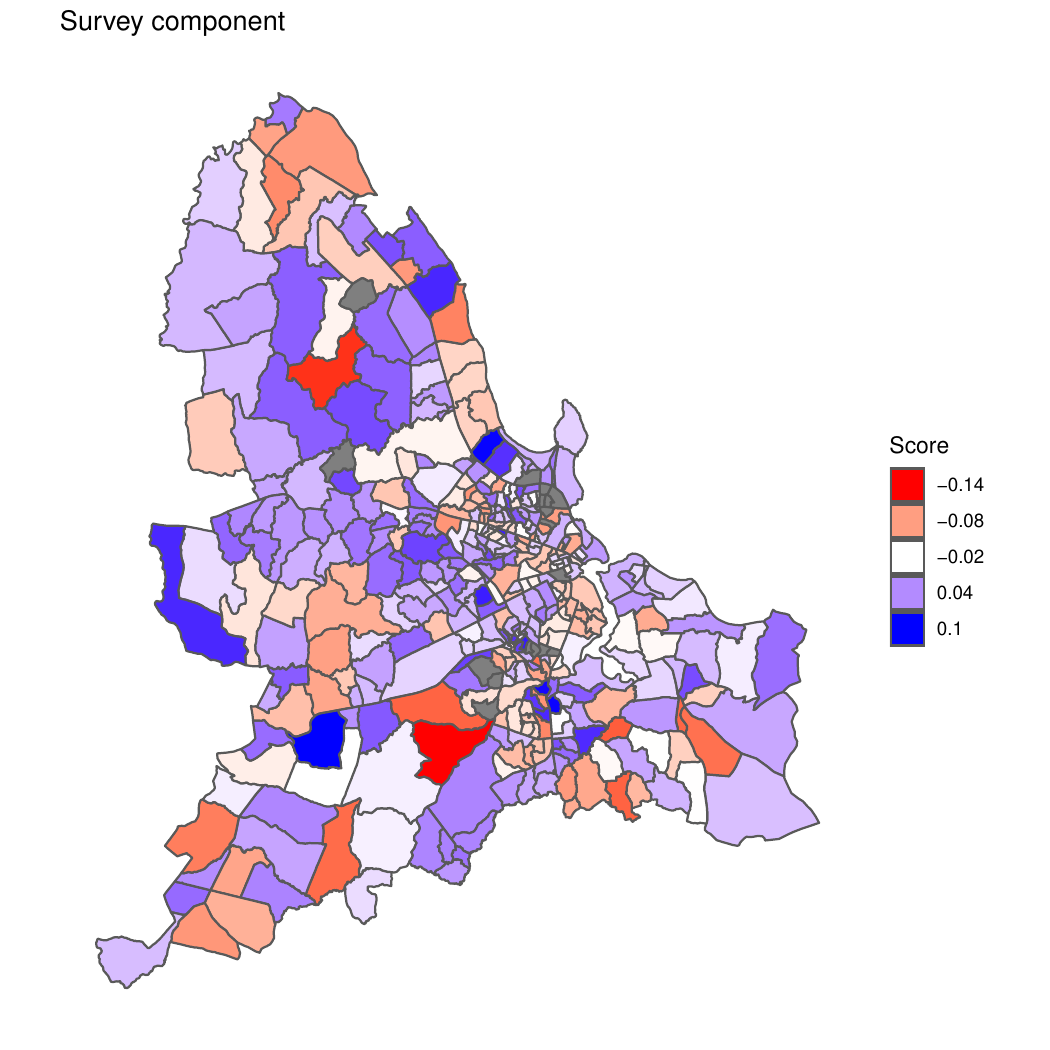}};
    \node (pos1) at (7,2.8)
        {\includegraphics[width=0.4\linewidth]{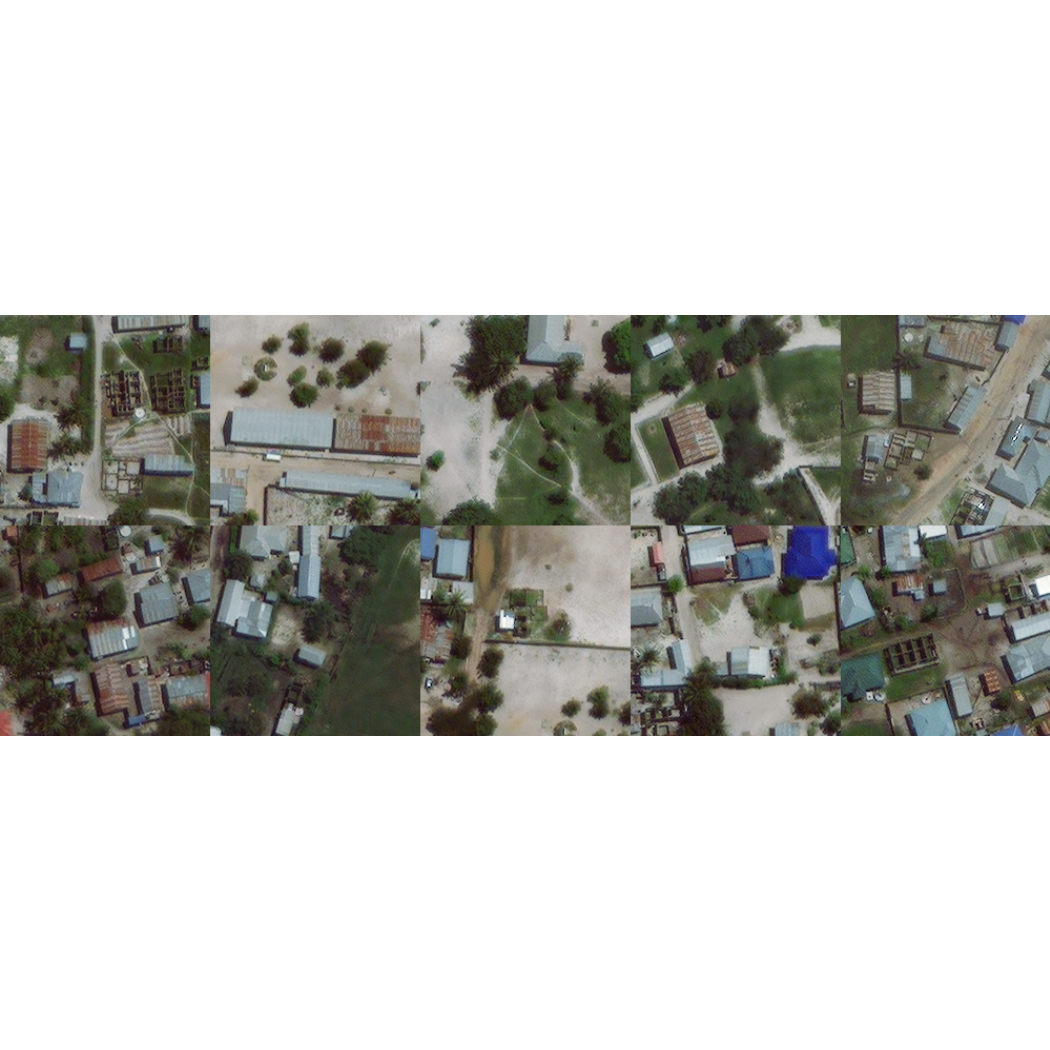}};
    \node (pos2) at (7,0)
        {\includegraphics[width=0.4\linewidth]{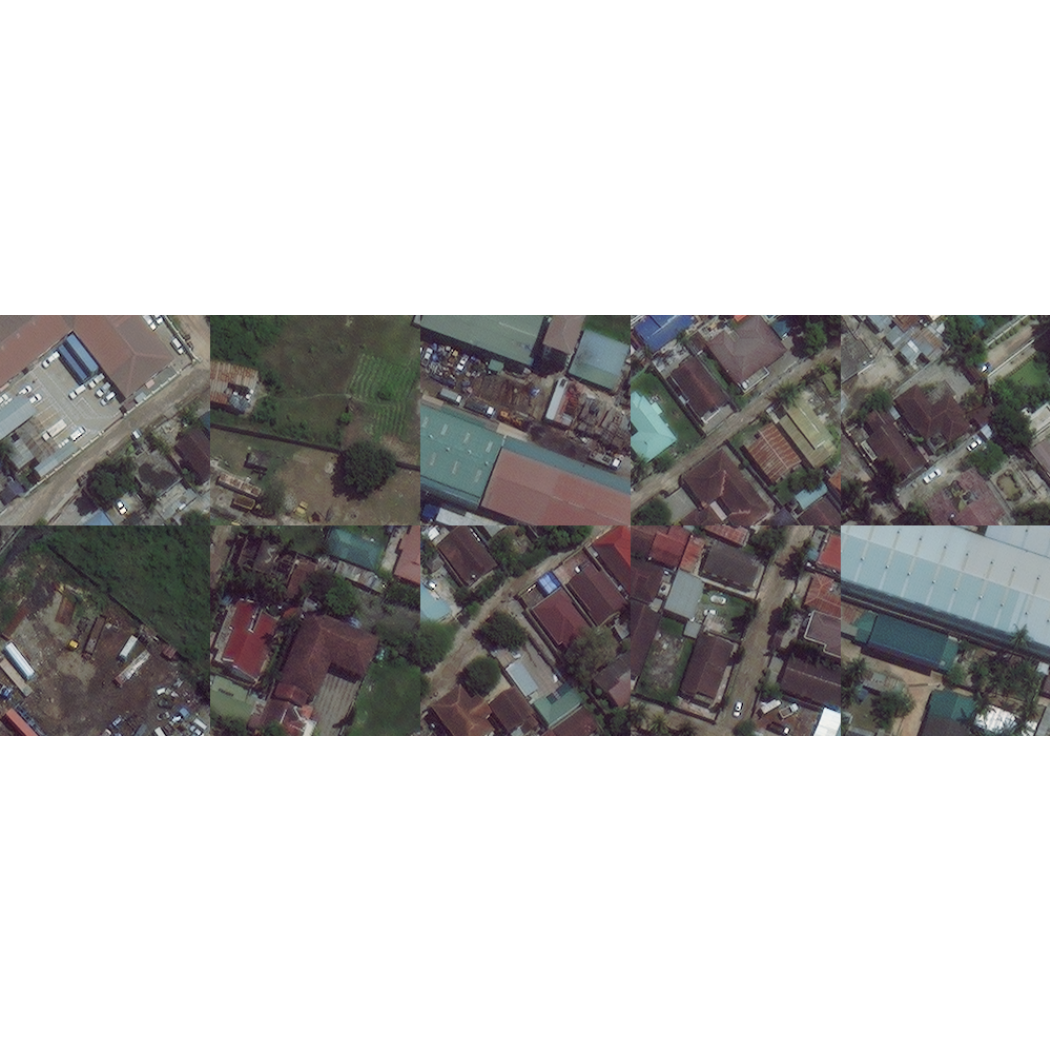}};
    \node (neg1) at (7,-3.5)
        {\includegraphics[width=0.4\linewidth]{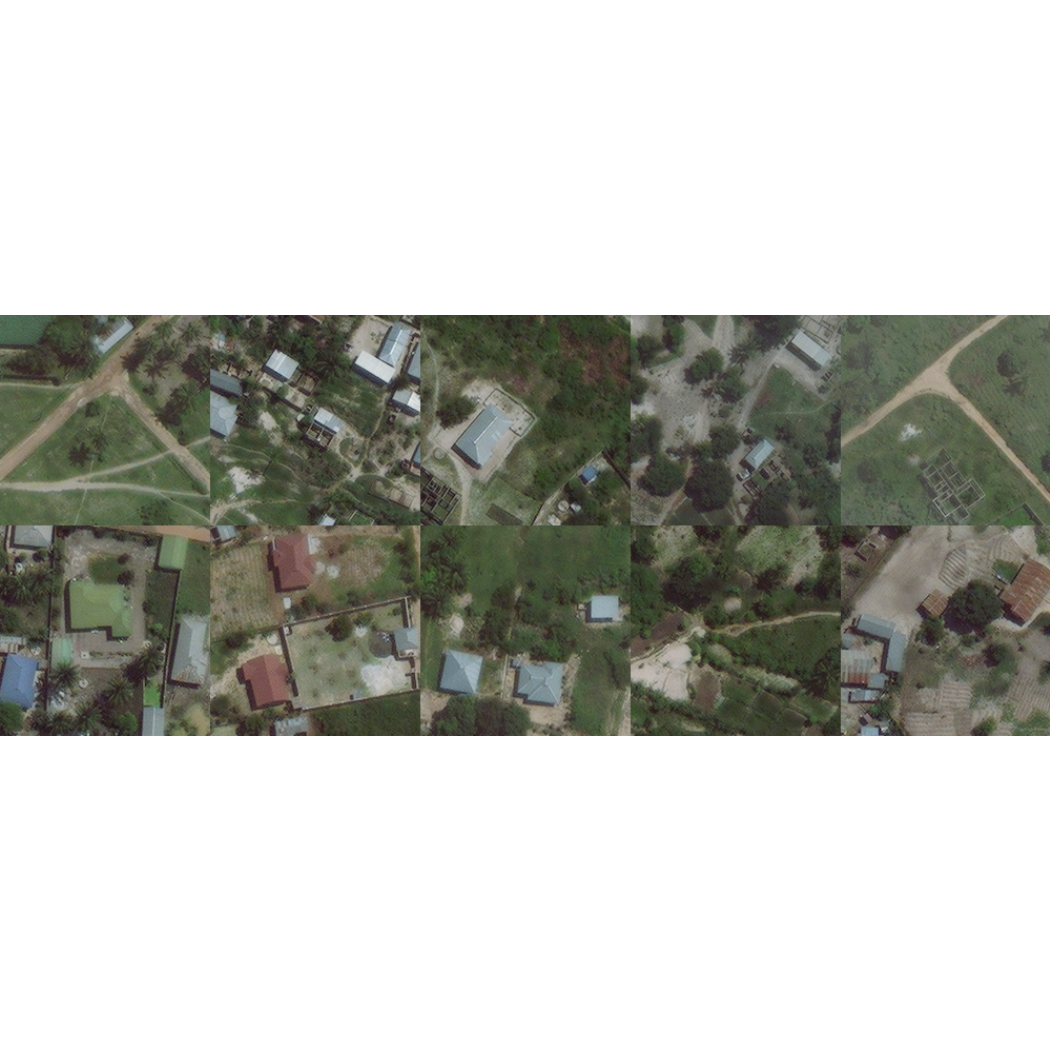}};
    \node (neg2) at (7,-6.3)
        {\includegraphics[width=0.4\linewidth]{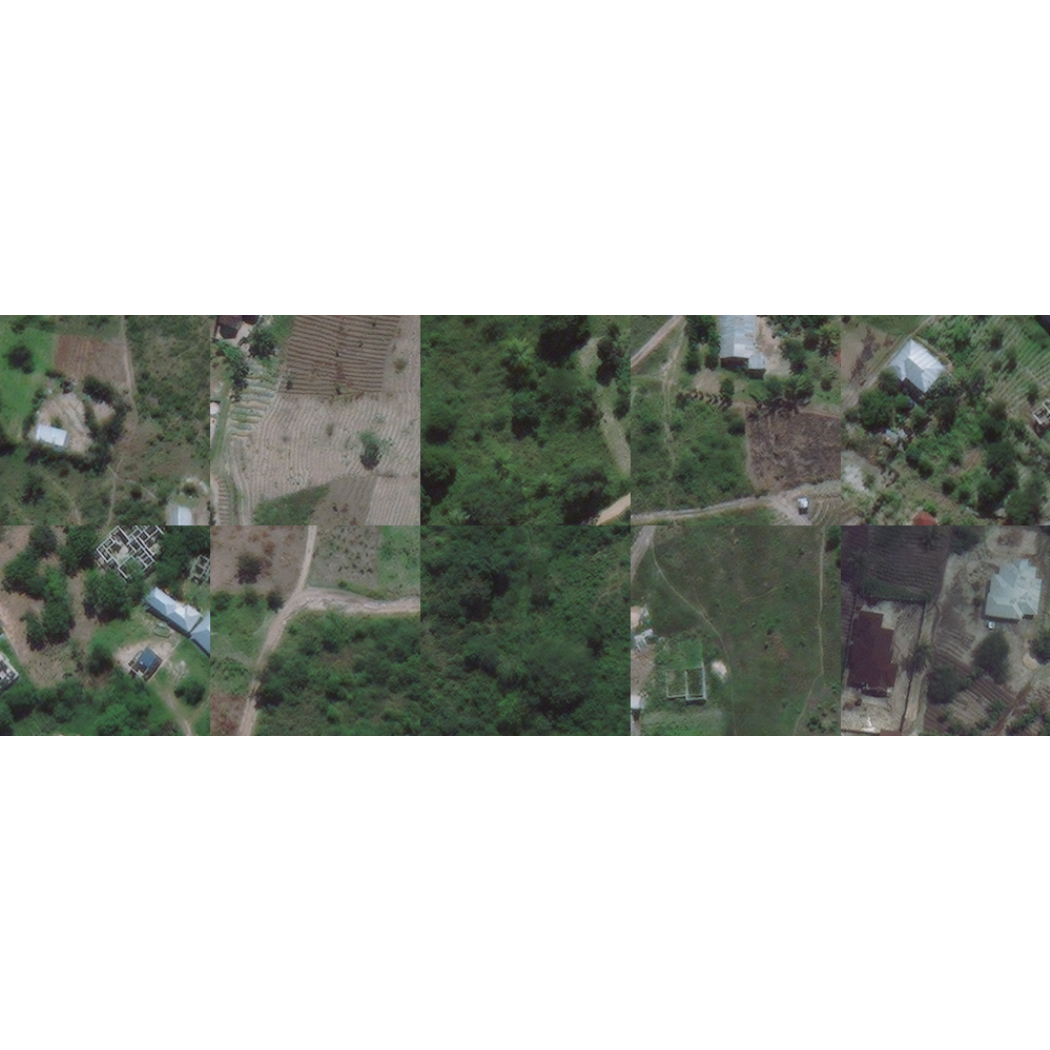}};
    \node (loadings) at (0,-7.5)
        {\includegraphics[width=0.4\linewidth]{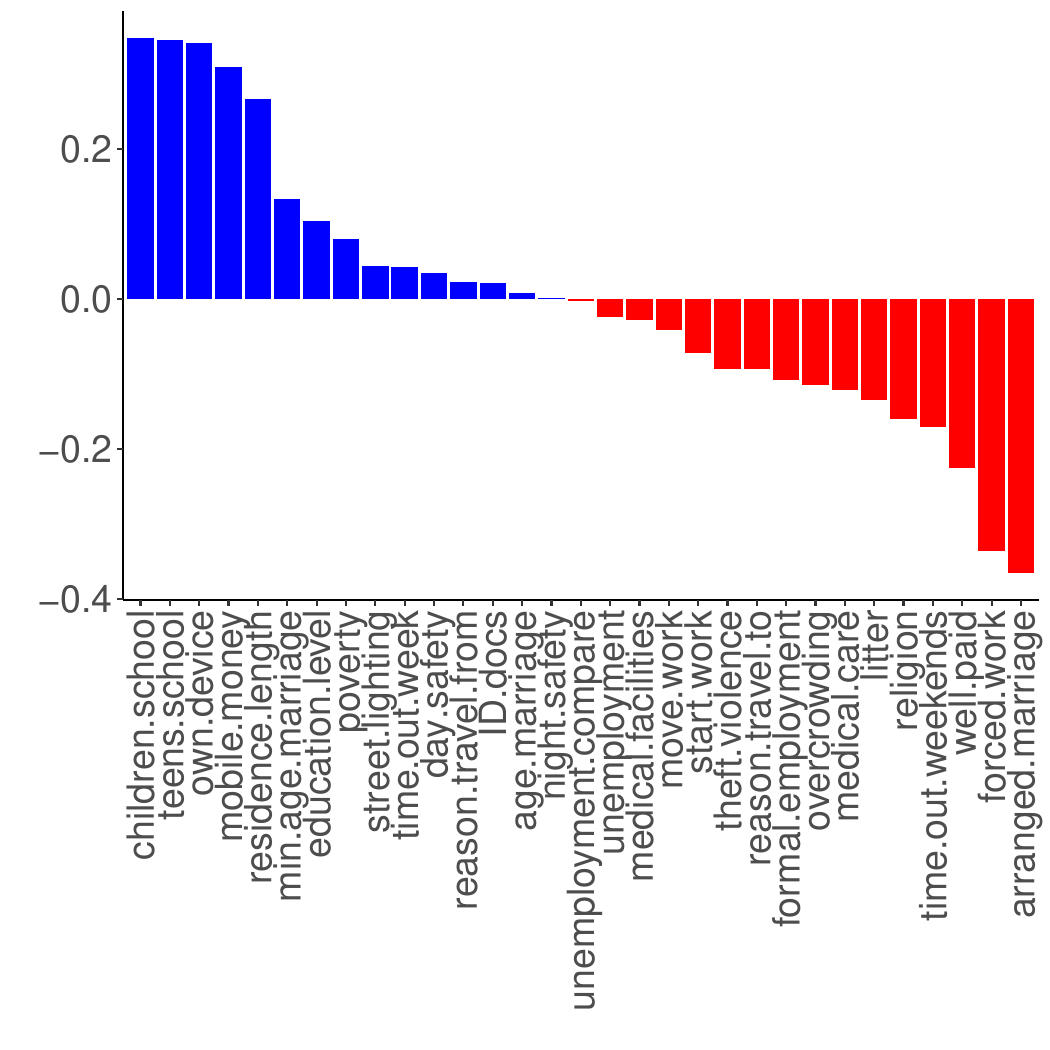}};
    \node (a) at (0,-3.7)
        {(a)};
    \node (b) at (0,-10.6)
        {(b)};
    \node (c) at (7,-8.2)
        {(c)};
    \draw [ultra thick, blue, ->] (-1.71,-1.6) to [bend left] (4,2.8);
    \draw [ultra thick, blue, ->] (-0.36,0.87) to [bend left] (4,0);
    \draw [ultra thick, red, ->] (-0.77,-1.62) to [bend left] (4,-3.5);
    \draw [ultra thick, red, ->] (-1.7,1.56) to [bend left] (4,-6.3);
    \node (t) at (-2.5,5)
        {\small{\textbf{Survey individual component}}};
    \node (p) at (7,4.2)
        {\footnotesize{\textbf{Patches from most positive subwards}}};
    \node (n) at (7,-2.1)
        {\footnotesize{\textbf{Patches from most negative subwards}}};
    \node (l) at (0,-4.5)
        {\footnotesize{\textbf{Survey feature loadings}}};
\end{tikzpicture}
    \caption{\small{(a) Subward scores for IC1\textsuperscript{Survey}, the survey individual component. Red corresponds to negative scores and blue to positive scores. Grey subwards are those for which we do not have image data. (b) Survey feature loadings, sorted from most positive to most negative. (c) Patches from the two most positive and two most negative subwards in IC1\textsuperscript{Survey}, with arrows showing to which subwards they correspond.}}
    \label{fig:survey}
\end{figure}

\begin{figure}
    \centering
    \includegraphics[width=0.9\linewidth]{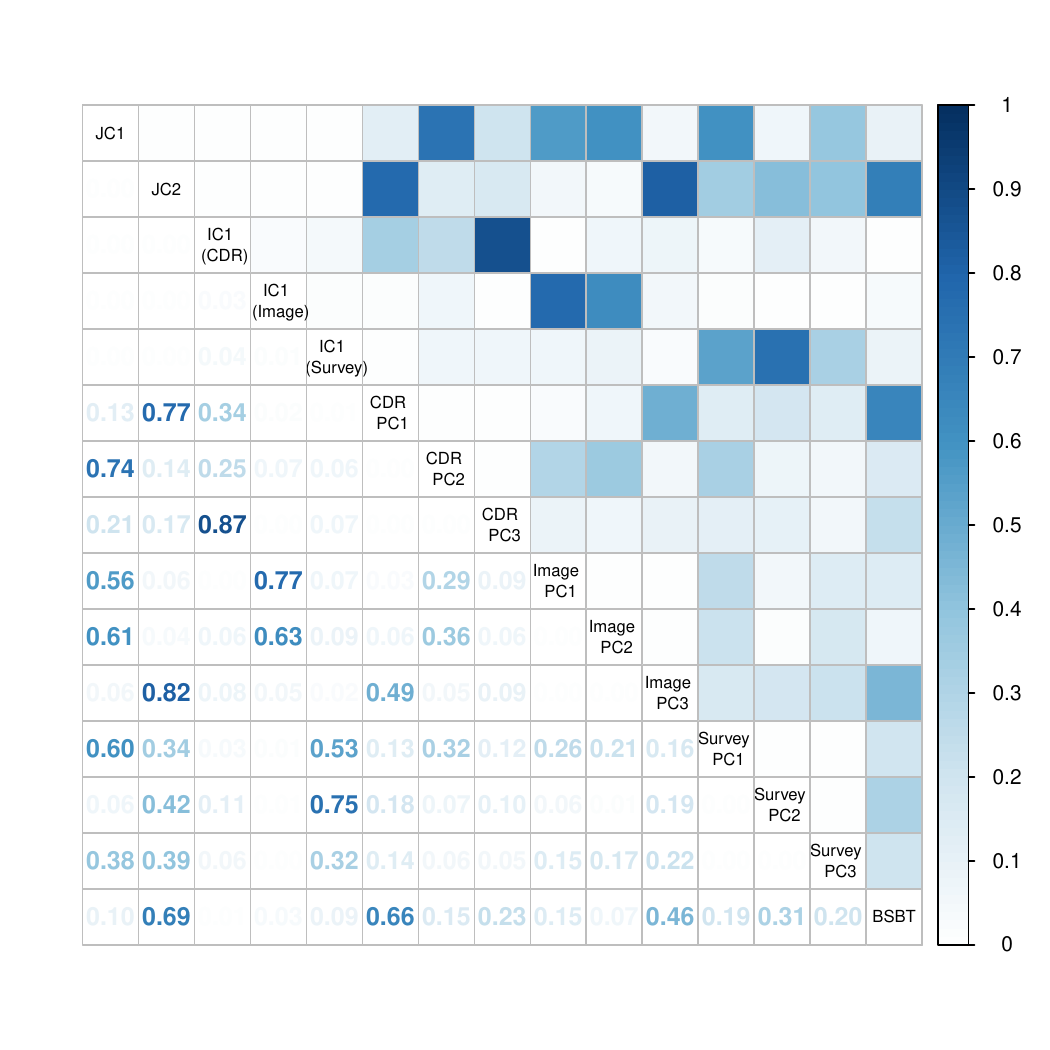}
    \caption{\small{Absolute values of correlations between joint, individual and PC scores for the CDR, image and survey data. (As previously, the signs of the score vectors are arbitrary, so the directions of the correlations between them are not important.) Deprivation scores estimated using the Bayesian Spatial Bradley-Terry model (see Section \ref{sec:data-md}) are also included for comparison.}}
    \label{fig:corrplot-survey}
\end{figure}

Figure \ref{fig:survey} shows a plot of the survey individual component (IC1\textsuperscript{Survey}) scores for each subward, and 5 patches from two most positive and most negative subwards. Corresponding plots for the other components (JC1, JC2, IC1\textsuperscript{CDR}, and IC1\textsuperscript{Image}) are shown in the supplementary material (Section \ref{sec:survey-2}): they are very similar to those we obtained using only the CDR and image data (Figures \ref{fig:joint-1} to \ref{fig:image}). Figure \ref{fig:corrplot-survey} shows correlations between AJIVE and PC components for the three data matrices. Comparing with Figure \ref{fig:svd-corrplot}, the relations between JC1, JC2, IC1\textsuperscript{CDR} and IC1\textsuperscript{Image}, and the CDR and image PCs are virtually unchanged, so incorporating the survey data does not have much effect on the decomposition. However, looking at the survey individual component allows us to see what information may be present in the survey data that we cannot get from the other data sets. From Figure \ref{fig:survey} (b), it appears that IC1\textsuperscript{Survey} highlights a risk of forced labour and exploitation: the variables with the most negative loadings are those relating to forced work and arranged marriage, whilst as the positive end we have higher rates of children and teenagers in education, and more use of technology, which may be associated with lower risks of forced labour.

\subsection{Regression modelling to predict deprivation} \label{sec:reg} 

We have investigated correlation between the dimension-reduced representations of the data (AJIVE and PC scores) and deprivation scores, but the approaches to dimension reduction were ``unsupervised" in the sense that they did not involve use of the deprivation data. In this section, we consider the ``supervised" approach of using deprivation as a response variable in a regression model geared towards prediction of deprivation from the CDR, image and survey data. We particularly investigate whether it aids prediction to use AJIVE and/or PCA as a step of dimension reduction to construct ``features" for the regression model.

A recently introduced approach \citep{ding2022cooperative} for incorporating multiple data ``views" into a regression model involves minimising the objective
\begin{equation}
    \frac{1}{2} \left\| \boldsymbol{y} - \sum_{m=1}^M \boldZ_m \boldsymbol{\beta}_m \right\|^2
        + \sum_{m=1}^M \lambda_m \| \boldsymbol \beta_m\|_1
        + \frac{\rho}{2} \sum_{m < m'}\left\| \boldZ_m \boldsymbol{\beta}_m - \boldZ_{m'} \boldsymbol{\beta}_{m'}\right \|^2,
\label{eqn:multiview:regression:objective}
\end{equation}
with respect to the regression parameters $\boldsymbol{\beta}_1 , \ldots , \boldsymbol{\beta}_M $, in which $\boldsymbol{\beta}_m $ is a $p_m \times 1$ vector of parameters corresponding to the $m$th view, $\boldZ_m$, and $\boldsymbol{y}$ is a response variable to be predicted. For the special case with parameters $\rho = 0$ and $\lambda_1 = \cdots = \lambda_M$, this is the objective for LASSO regression of $\boldsymbol{y}$ on the concatenated data $\boldZ = (\boldZ_1, \ldots, \boldZ_M)$. But when $\rho > 0$, the final term encourages ``cooperation" between the predictions from the different individual views, which in some circumstances leads to better predictive performance for out-of-sample data, i.e. new observations not used in fitting the model \citep{ding2022cooperative}. 

\begin{figure}
    \centering
    \includegraphics[width=0.7\linewidth]{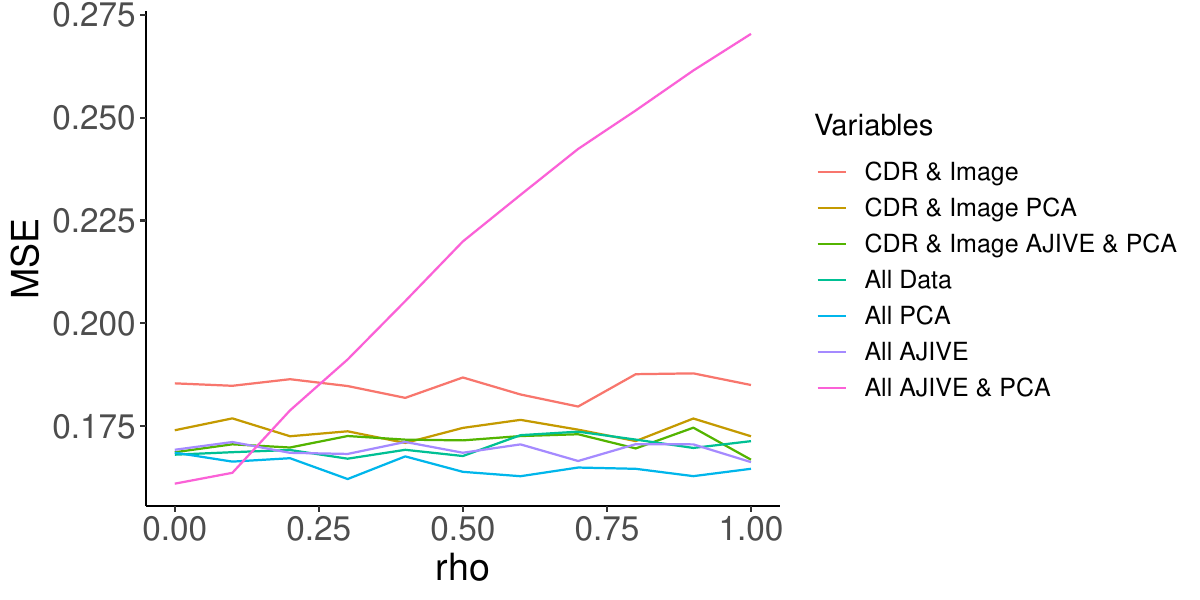}
    \caption{\small{Values of the MSE plotted against $\rho$, where we apply the regression model in \eqref{eqn:multiview:regression:objective} with cross-validation. Different lines correspond to different combinations of variables. For the case where we include all AJIVE and PC scores, increasing $\rho$ also increases the MSE, so the model performs worse. For other scenarios, the value of $\rho$ has little or no effect on performance.}}
    \label{fig:rho-plot}
\end{figure}

For a given choice of $\rho, \lambda_1, \ldots, \lambda_M$, the fitted parameters $\hat{\boldsymbol \beta} = (\hat{\boldsymbol{\beta}}_1^\top , \ldots , \hat{\boldsymbol{\beta}}_M^\top)^\top$ are determined by minimising \eqref{eqn:multiview:regression:objective}, and prediction of the response variable for a new observation with predictor vector $\boldsymbol{z} = (\boldsymbol{z}_1^\top , \ldots, \boldsymbol{z}_M^\top )^\top$ is $\hat{\boldsymbol{y}} = \boldsymbol{z}^\top \hat{\boldsymbol \beta}$. One way to select the hyper-parameters $\rho, \lambda_1, \ldots, \lambda_M$ is by cross-validation, i.e., repeatedly partitioning the observations into training and ``held-out'' test sets, finding $\hat{\boldsymbol \beta}$ based on the training data for various different values of the hyper-parameters, then selecting the hyper-parameter values that minimise mean-squared error in predictions for the held-out test data.

To compare the performance of different $\boldsymbol{Z}$'s, we select $\lambda_1, \ldots, \lambda_M$ using cross-validation with 20 folds, and calculate the average Mean Squared Error (MSE) across folds: we choose the value of $\boldsymbol{\lambda}$ which minimizes this. A lower MSE means the model is more accurate at predicting deprivation $\boldsymbol{y}$. In various numerical investigations exploring values $\rho \geq 0$ with different choices of predictor data $\boldZ$, we found no circumstances where $\rho > 0$ performed better than $\rho = 0$ --- see for example Figure \ref{fig:rho-plot} --- so we set $\rho = 0$ in all following experiments.

\begin{figure}
    \centering
    \includegraphics[width=0.9\linewidth]{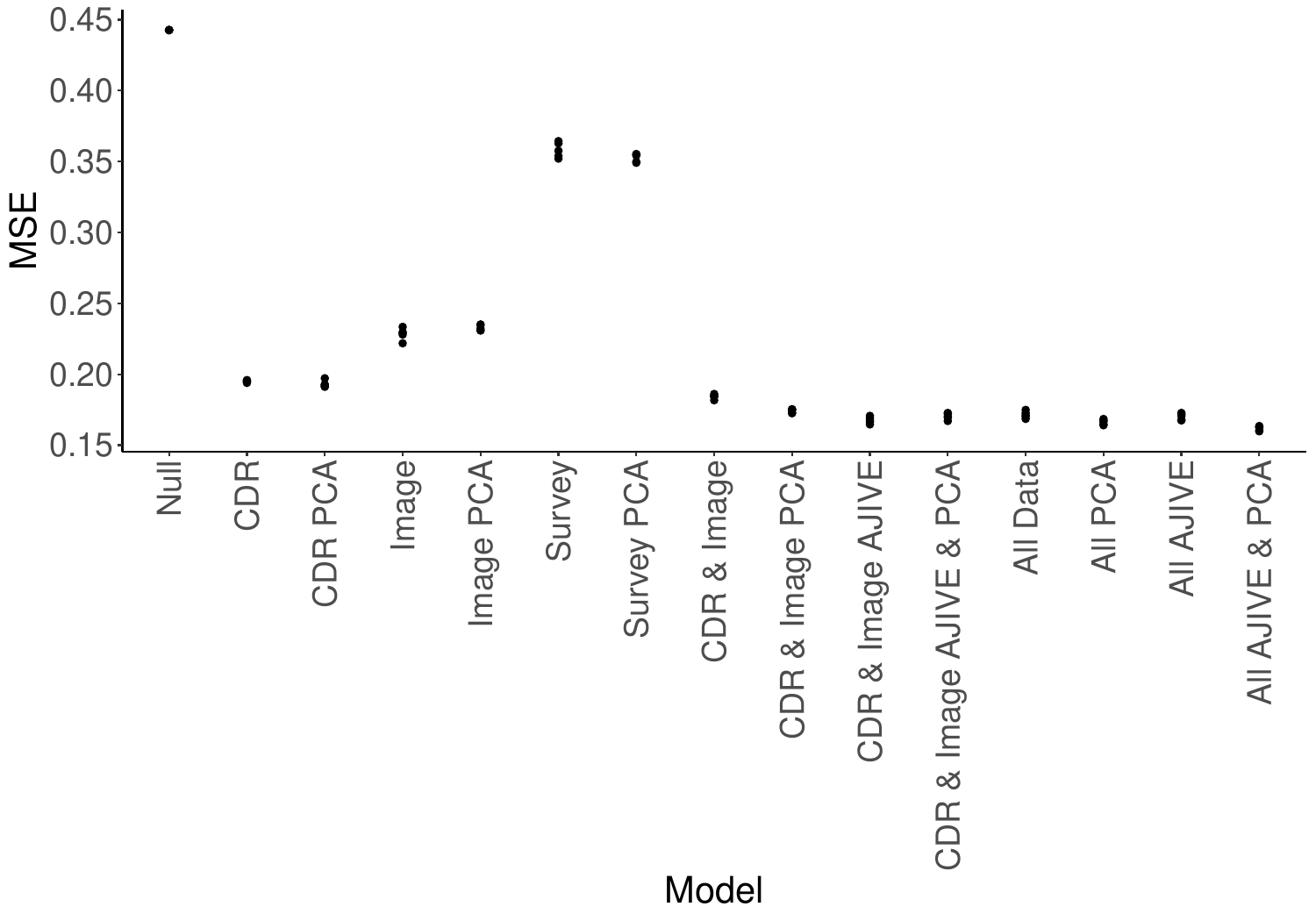}
    \caption{\small{Results of regression using different choices of $\boldZ$. In each case we implemented the model 5 times with data split randomly into 20 folds in each case. For the null model, $\boldsymbol{y}$ is modelled as a constant mean plus random noise.}}
    \label{fig:mse-regression}
\end{figure}

Figure \ref{fig:mse-regression} shows the results using $\rho = 0$ and different choices for $\boldZ$. Except for PCA on the CDR data (where we have a maximum of 20 components to work with), all AJIVE and PCA regressions are done with 30 total components, which are divided equally between data sets and, for AJIVE, between joint and individual components.

We first consider using each data set individually, setting $\boldZ$ to be either the entire data set or the matrix of principal components. The CDR data gives the lowest MSE, whilst the survey data does by far the worst. In each case using PCA gives similar results to using the entire data set. We then combine data sets: we consider using the CDR and image data, as we did for our main analysis, and then adding in the survey data. Here, we find that PCA regression does slightly better than using the entire data. The difference between PCA and AJIVE is less clear: when we have just the CDR and image data, AJIVE seems to do slightly better than PCA, but this is not the case once we also add in the survey data, although the differences in MSE are very small so it is difficult to draw firm conclusions. However, as mentioned above AJIVE with CDR, image and survey data gives the additional useful information about forced labour risk and so this is our preferred approach.

\begin{figure}
    \centering
    \includegraphics[width=0.7\linewidth]{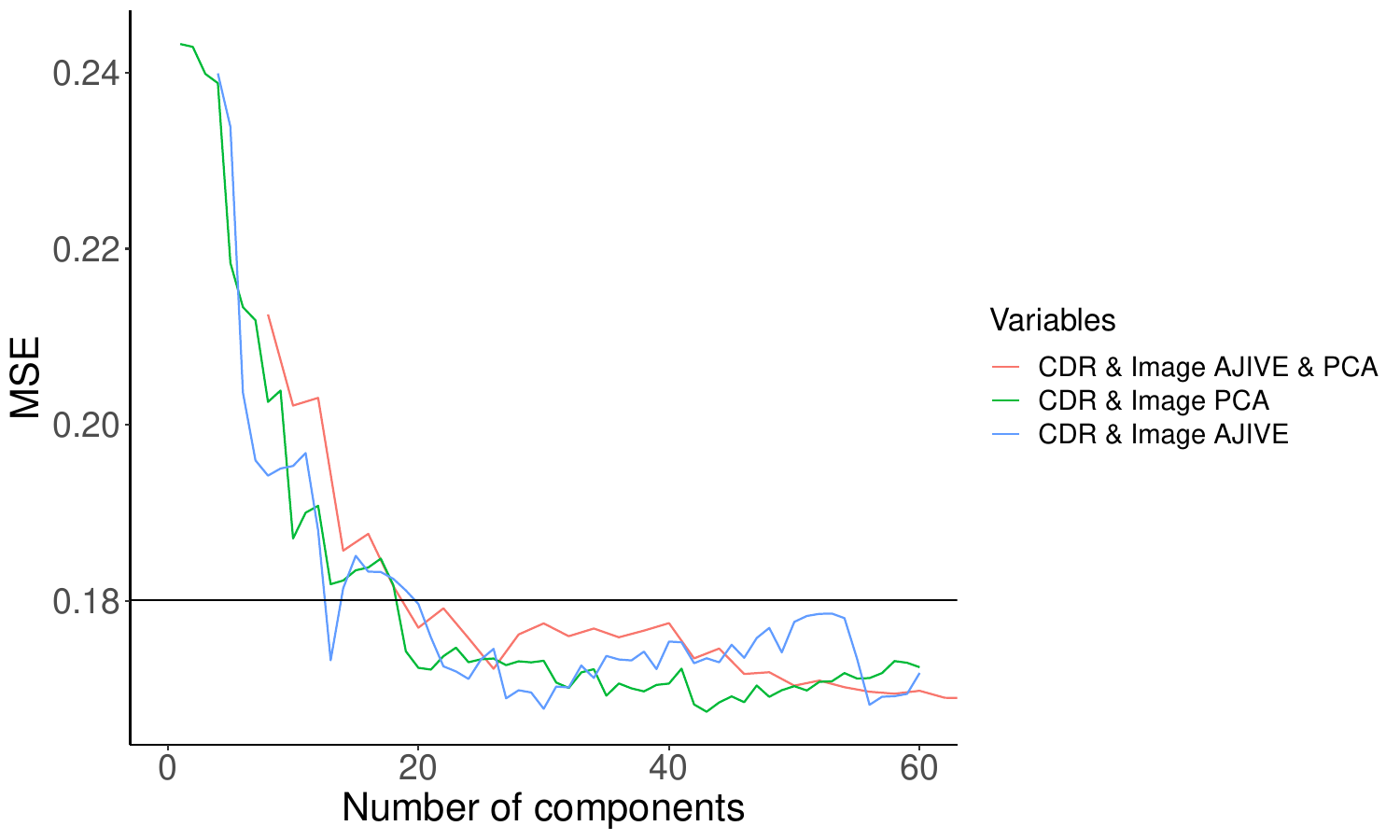}
    \caption{\small{Average MSE (across 10 runs) plotted against number of components for several different combinations of variables. The horizontal black line corresponds to the MSE for the model where the covariates are the CDR and image data, without applying AJIVE or PCA. Where relevant, components are divided equally (as far as possible) between joint and individual components, and between CDR and image components.}}
    \label{fig:regression-mse-vs-num-components}
\end{figure}

Figure \ref{fig:regression-mse-vs-num-components} shows a plot of the MSE versus the number of components, when we use several combinations of AJIVE and PCA components for the CDR and image data. The black horizontal line corresponds to model where we use the data as covariates, without applying AJIVE or PCA. In each case the MSE decreases until we have around 20 components, then levels off. There does not appear to be much difference in performance between the three choices of $\boldZ$ we use here.

\section{Discussion} \label{sec:conclusion}

As outlined in the introduction, the last decade has seen numerous applications of the kinds of data we utilise here (e.g. satellite imagery and phone CDR) to predict measures of poverty, deprivation and related constructs, in isolation; less frequently, other works (e.g. \cite{steele2017mapping}) have begun to bring such diverse data sources together. We have built upon this, using an approach which (i) extracts features from imagery in an unsupervised way, allowing us to learn which aspects of the images have the most variation and offer the most predictive power rather than imposing a structure on those features, and (ii) using AJIVE, which explicitly allows for (and quantifies) the extent to which different types of data (imagery, CDR, survey) vary together and independently, which facilitates better understanding of the resulting models predicting poverty/deprivation.

In the context of Dar es Salaam, we have shown how AJIVE allows identification of features that are common to both satellite imagery and CDR data, as well as features which are unique to these data sets. A key advantage over alternative approaches, such as working with PCA on individual and/or concatenated data, is the control AJIVE affords over the number of joint and different individual components of variation, which in turn allows predictive signal in lower-dimensional input data (CDR data in our case) to be found more easily and within a unified framework.

The finding that incorporating survey data in addition to the CDR and image data has little effect on the joint components of an AJIVE analysis or the ability to predict deprivation is, at first sight, surprising; but it is also potentially valuable, given the logistical costs of obtaining such data, and worthy of further exploration in several directions. Is this situation, for example, where the information in the survey data is already captured within other datasets such as imagery and CDR, common to other cities and/or countries? How sensitive is this finding to the (relative) timings of collection of the different data sets? Might there be other kinds of data which are relatively cheap to acquire which might add further predictive power and possibly render expensive surveys necessary less often? The task of understanding poverty and being able to predict it based on readily-available data is crucial, especially in rapidly changing settings like large cities in the developing world. AJIVE allows quantification of the benefit of these diverse modes of data, both individually and in combination, with their very different costs and ease of acquisition and updating.

It may perhaps be the case that our methods are less well suited to the structure of survey data, which consists mostly of categorical and ordinal variables, whereas PCA and AJIVE are designed for continuous data. We note, however, that the individual component from the survey data does seem to be correlated with another variable likely to be of interest: the risk of forced labour, and hence appears to be worthy of further investigation.

\section*{Acknowledgements}

This work was supported by the Engineering and Physical Sciences Research Council [grant reference EP/T003928/1].

\bibliography{references}
\bibliographystyle{apalike}

\newpage
\appendix

\section*{Supplementary material}

\section{List of CDR features}

Table \ref{tab:phone-features} lists CDR features with definitions.

\begin{table}[h]
    \centering
    \begin{tabular}{lll}
      \hline
        \textbf{Variable} & \textbf{Level} & \textbf{Description} \\
      \hline
        ratio\_call\_text & User & Number of calls made, divided by number of SMS sent \\
        initiated\_calls & User & Percent of calls which were outgoing \\
        percent\_initiated & User & Percent of calls and SMS which were outgoing \\
        percent\_pareto\_calls & User & Percentage of contacts that account for $80 \%$ of \\
            & & call interactions \\
        entropy\_contacts & User & Entropy of call contacts \\
        total\_sms & User & Total SMS sent or received \\
        norm\_entropy & User & Entropy of visited BTS (base transceiver stations) \\
        unique\_bts & User & Number of unique BTS at which calls/SMS sent or received \\
        frequent\_bts & User & Number of BTS that account for $80 \%$ of locations where \\ 
            & & the user sent or received calls or SMS \\
        interevent\_time\_sd* & User & Standard deviation of time between events (call or SMS) \\
        active\_days & User & Number of days on which the user sent or received a call \\
            & & or SMS \\
        interevent\_time\_call\_mean & User & Mean time between call initiations \\
        mean\_call\_interactions & User & Mean number of call interactions with each contact \\
        response\_delay\_mean & User & Mean response delay in seconds, for texts \\
        response\_delay\_sd & User & Standard deviation of response delay in seconds, for texts \\
      \hline
        active\_users & Tower & Number of users for whom this tower is their home tower \\
        day\_interactions\_overall & Tower & Total interactions between 8am and 7pm \\
        evening\_interactions\_overall & Tower & Total interactions between 7pm and 12pm \\
        night\_interactions\_overall & Tower & Total interactions between 12pm and 8am \\
        distance\_overall & Tower & Mean distance to call recipient \\
      \hline
     \end{tabular}
    \caption{\small{List of CDR features used in the analysis. The first 15 features are calculated for individual users, and averaged to get a feature vector for each tower. The last 5 features are calculated directly for each tower. Entropy is calculated using Shannon entropy \citep{shannon1948mathematical}: $H = -\sum_{l=1}^N P (c_l) \log P (c_l)$, where e.g. for entropy of contacts, $c_1, \ldots, c_N$ are the contacts of each user ($N$ is user dependent), and $P(c_l)$ is the proportion of a user's interactions which take place with contact $c_l$ (for $l = 1, \ldots, N$). The features are calculated for each subward (see below), each one then corresponding to a column of $\boldX_1$.}}
    \label{tab:phone-features}
\end{table}

\section{List of survey features}

Tables \ref{tab:survey-1} and \ref{tab:survey-2} list the survey features with definitions.

\begin{table}[h]
    \centering
    \begin{tabular}{ll}
      \hline
        \multicolumn{2}{l}{\textbf{How strongly do you agree with the following statement?} (1-5)} \\
      \hline
        \texttt{overcrowding} & Overcrowding is a problem in this subward. \\
        \texttt{litter} & The level of litter in this subward is a problem. \\
        \texttt{day.safety} & I would feel safe in this subward during the day. \\
        \texttt{night.safety} & I would feel safe in this subward during the night. \\
        \texttt{unemployment} & The level of unemployment in this subward is a problem. \\
        \texttt{unemployment.compare} & The level of unemployment in this subward is high compared \\
            & to the rest of Dar. \\
        \texttt{poverty} & Poverty is a problem in this subward. \\
        \texttt{medical.care} & There is a good availability to medical care in this subward. \\
        \texttt{theft.violence} & Theft or violence is a problem in this subward. \\
        \texttt{well.paid} & People are paid well in this subward compared to the rest of \\
            & Dar es Salaam. \\
      \hline
        \multicolumn{2}{l}{\textbf{Percentage scale} (0-20\%, 20-40\%, 40-60\%, 60-80\%, 80-100\%)} \\
      \hline
        \texttt{move.work} & In your opinion what percentage of people move to this subward \\
            & for work? \\
        \texttt{teens.school} & In your opinion what percentage of teenagers in this subward \\
            & aged 13-18 are in school? \\
        \texttt{children.school} & In your opinion what percentage of children in this subward \\
            & aged 12 and under are in school? \\
        \texttt{mobile.money} & What percentage of people in this subward use mobile money? \\
        \texttt{own.device} & What percentage of people in this subward own a mobile device? \\
    \end{tabular}
    \caption{\small{List of questions in the survey (part 1). The first group of questions are of the form ``How strongly do you agree with this statement?'' with answers given on a Likert scale: possible responses are ``Strongly Disagree", ``Disagree", ``Neither Agree nor Disagree", ``Agree", ``Strongly Agree." The second group of questions ask for percentages of the population where the respondent selects a range from one of 0-20\%, 20-40\%, 40-60\%, 60-80\%, 80-100\%.}}
    \label{tab:survey-1}
\end{table}

\begin{table}
    \centering
    \adjustbox{max width=\textwidth}{
    \begin{tabular}{lll}
      \hline
        Variable name & Question & Possible responses \\
      \hline
        \texttt{street.lighting} & Is there street lighting in this subward? & No / Yes, but it's limited \\
            & & and/or broken / Yes \\
        \texttt{formal.employment} & What is the most common type of employment \\
            & in this subward? & Informal / Formal \\
        \texttt{residence.length} & On average how long do people remain living in \\
            & this subward? & Under a year / 1-5 years / \\
            & & 5-10 years / 10+ years \\
        \texttt{reason.travel.to} & What is the most common reason people travel \\
            & to this subward? & Social / Mixture or Other / Work \\
        \texttt{reason.travel.from} & What is the most common reason people travel \\
            & out from this subward? & Social / Mixture or Other / Work \\
        \texttt{time.out.weekends} & How much of the day do residents spend outside & Less than half of the day / \\
            & of this subward during the weekends? & Half of the day / Most of the day \\
        \texttt{time.out.week} & How much of the day do residents spend outside & Less than half of the day / \\
            & of this subward during week days? & Half of the day / Most of the day \\
        \texttt{start.work} & In your opinion what age do people generally & 0-11 years / 12-14 years / \\
            & start paid work in this subward? & 15-17 years / 18-20 years / 21+ years \\
        \texttt{medical.facilities} & What medical facilities are available in this & Hospital / Small medical facility \\
            & subward? Select all that apply. & but not hospital / Doctors are working \\
            & & without a building, no small medical \\
            & & facility or hospital / None \\
        \texttt{age.marriage} & What age do people tend to get married in this \\
            & subward? \\
        \texttt{min.age.marriage} & What is the youngest age people get married in \\
            & this subward? \\
        \texttt{religion} & What religion are most people in this \\
            & subward? & Christian / Mixed or Other / Muslim \\
        \texttt{education.level} & What level of education do most people reach & Primary / O-Level / A-Level / \\
            & in this subward? & University \\
        \texttt{id.docs} & Do most people in this subward have \\
            & identification documentations (such as \\
            & passports, driving license)? & Yes / No \\
      \hline
    \end{tabular}
    }
    \caption{\small{Survey questions (continued): questions that do not fit into either of the first two groups (Table \ref{tab:survey-1}).}}
    \label{tab:survey-2}
\end{table}

\newpage
\section{AJIVE with survey data} \label{sec:survey-2}

\begin{figure}[h]
    \centering
    \includegraphics[width=0.48\linewidth]{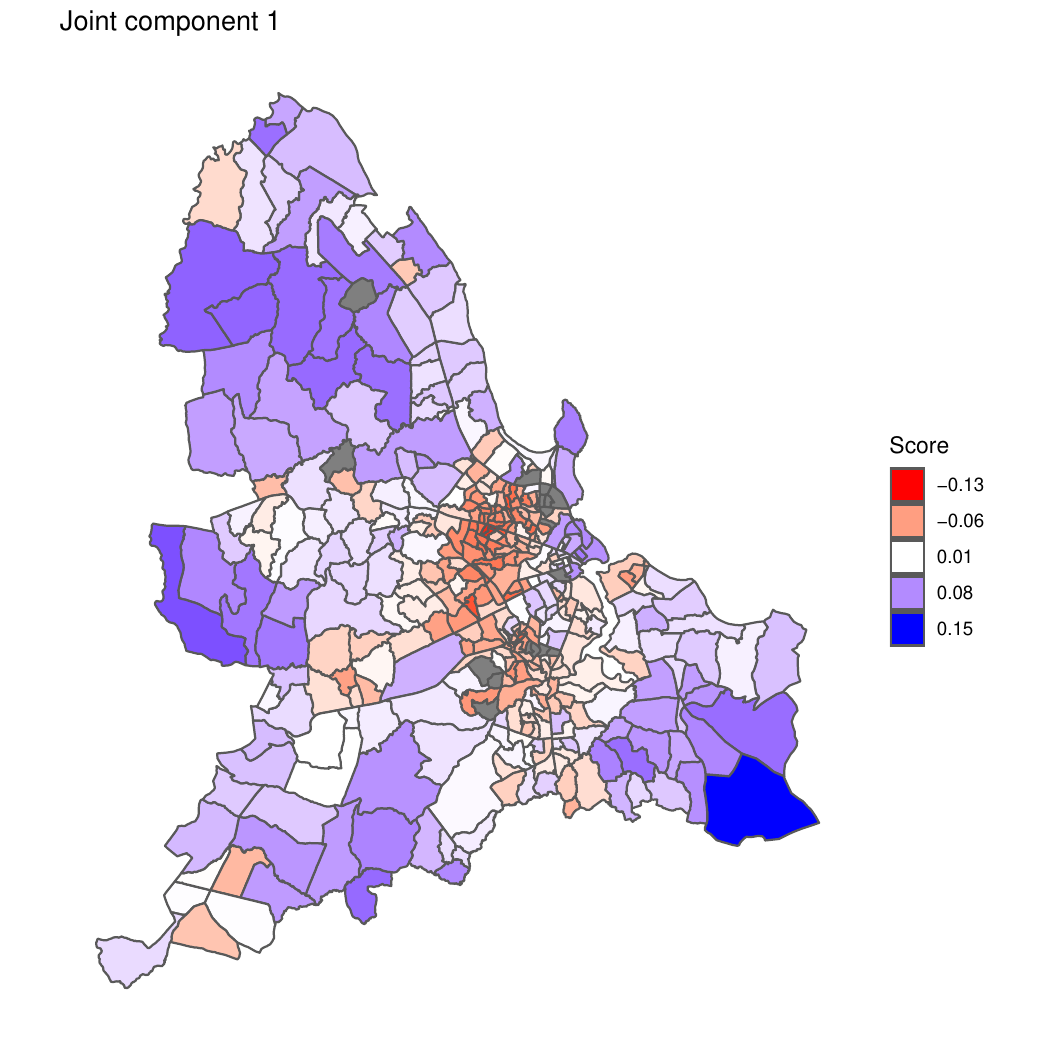}
    \includegraphics[width=0.48\linewidth]{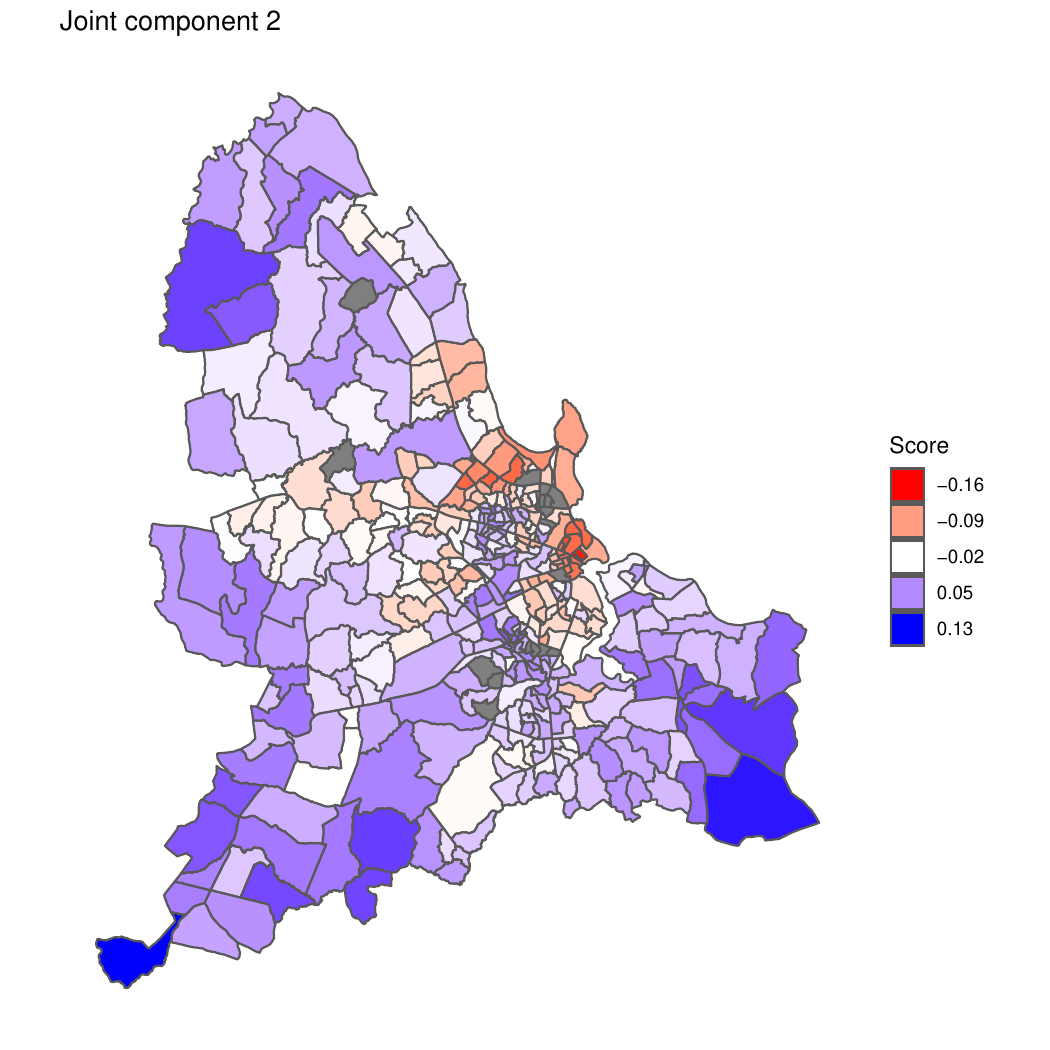}
    \includegraphics[width=0.48\linewidth]{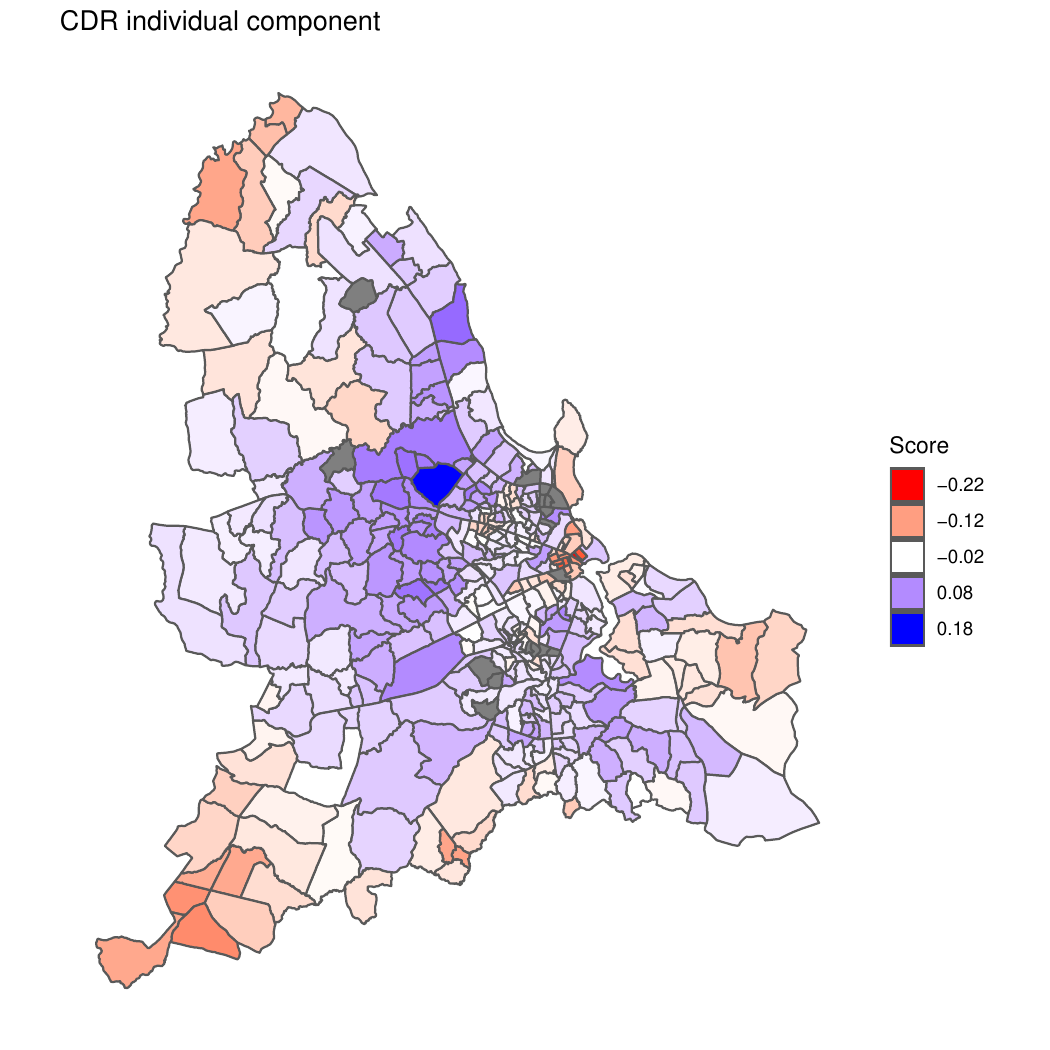}
    \includegraphics[width=0.48\linewidth]{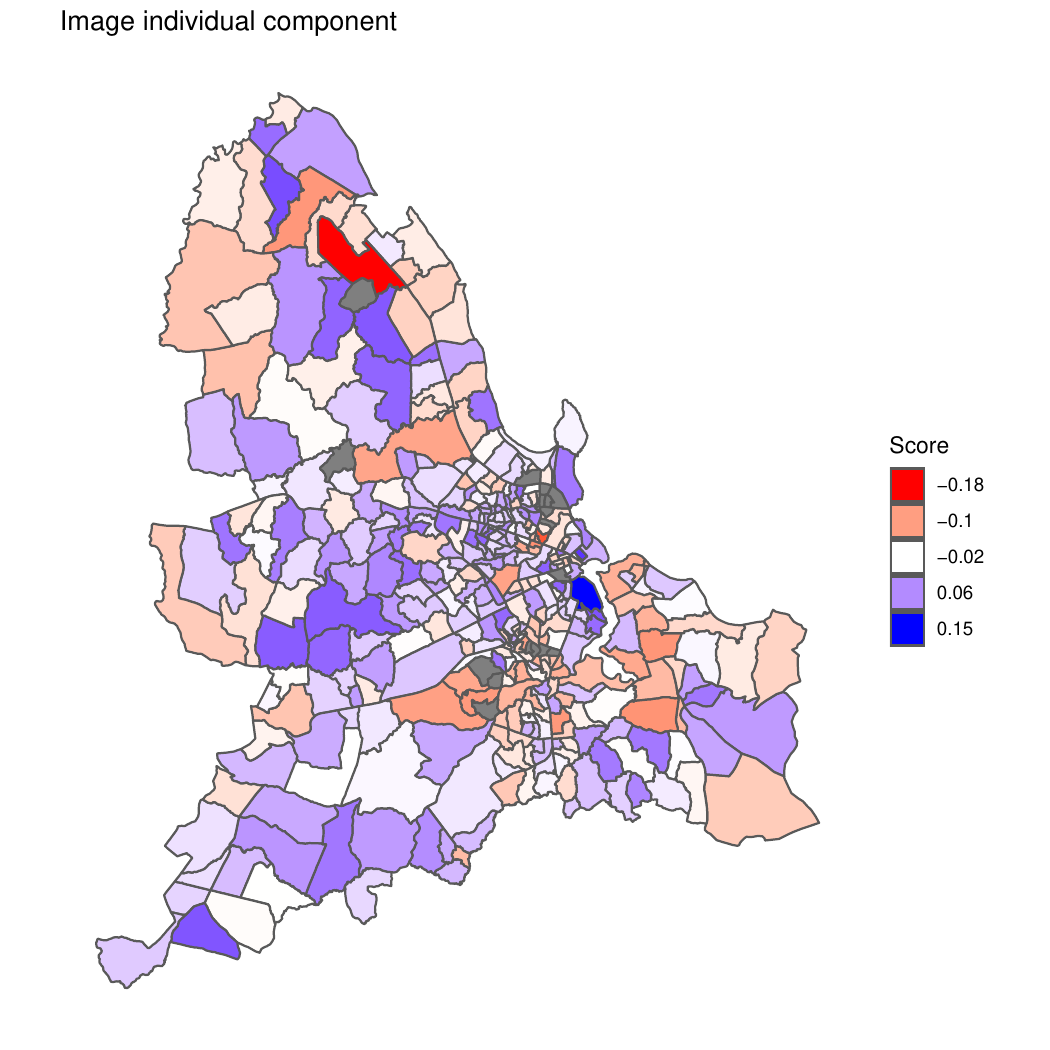}
    \caption{\small{Plots showing the subward scores for components JC1, JC2, IC1\textsuperscript{CDR} and IC1\textsuperscript{Image}, when we run AJIVE also including the survey data.}}
    \label{fig:survey-ajive-scores}
\end{figure}

\begin{figure}
    \centering
    \includegraphics[width=0.48\linewidth]{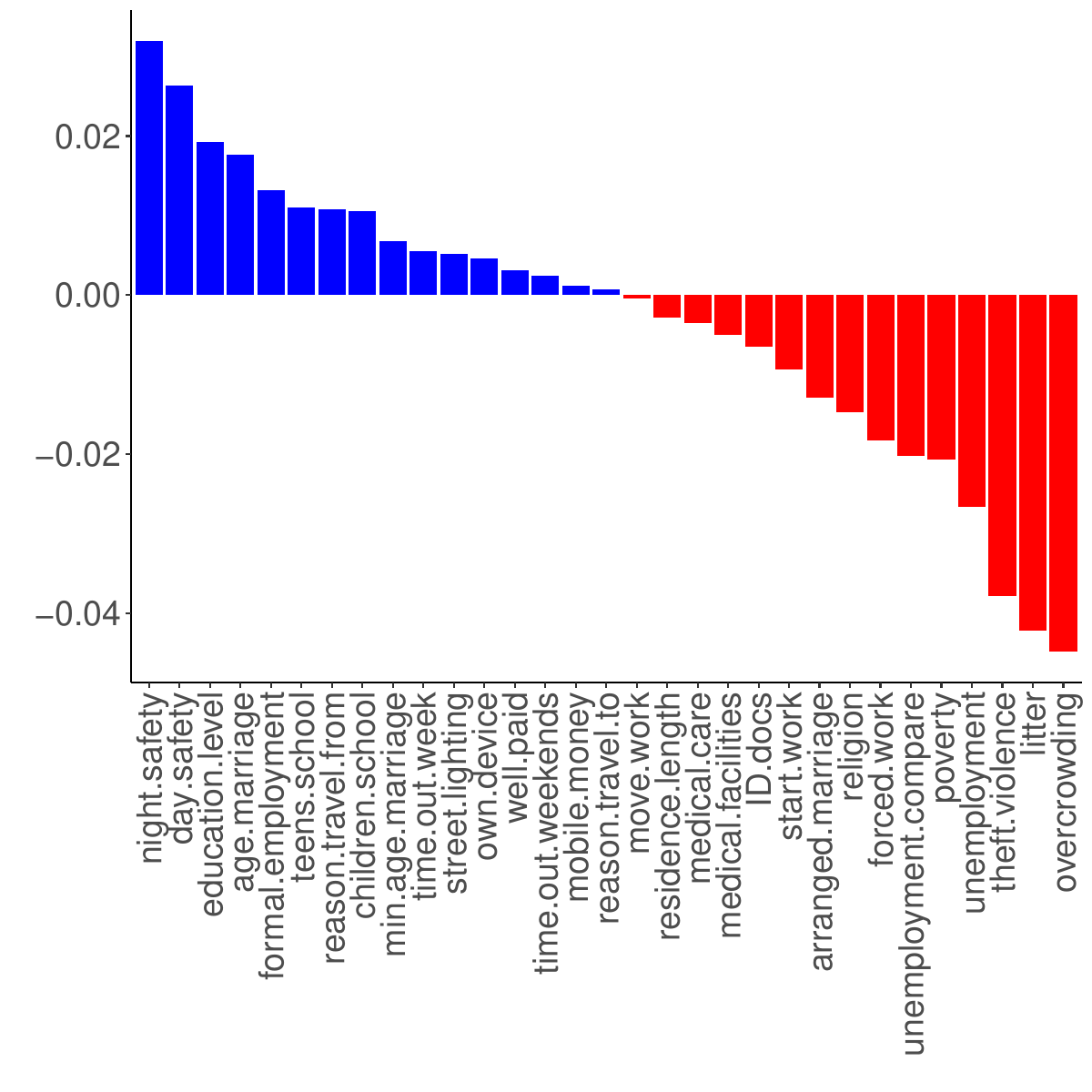}
        \put(-105,198){\footnotesize{JC1}}
    \includegraphics[width=0.48\linewidth]{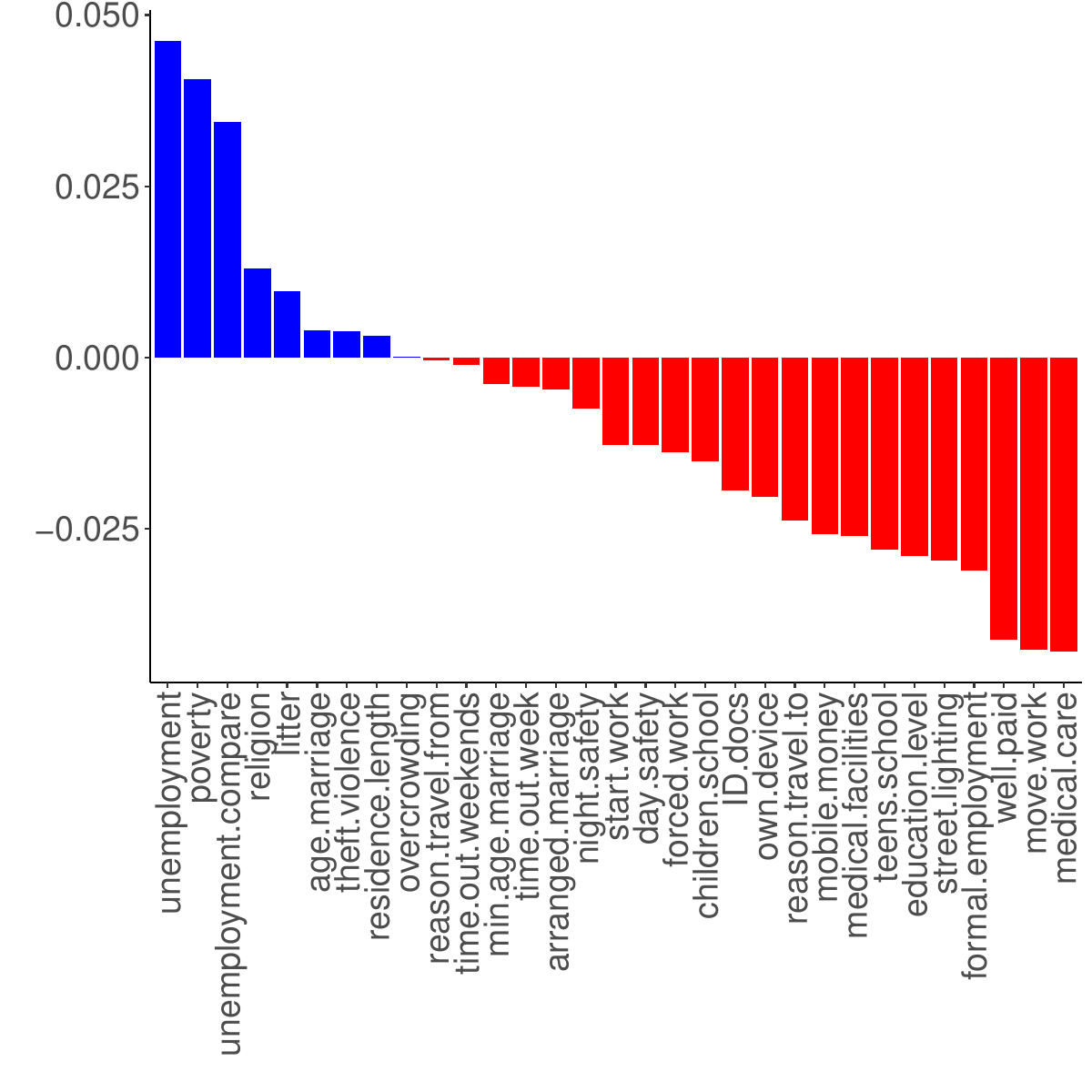}
        \put(-105,198){\footnotesize{JC2}}
    \caption{\small{Survey feature loadings for the joint components.}}
    \label{fig:survey-feature-loadings}
\end{figure}

\begin{figure}
    \centering
    \includegraphics[width=0.7\linewidth]{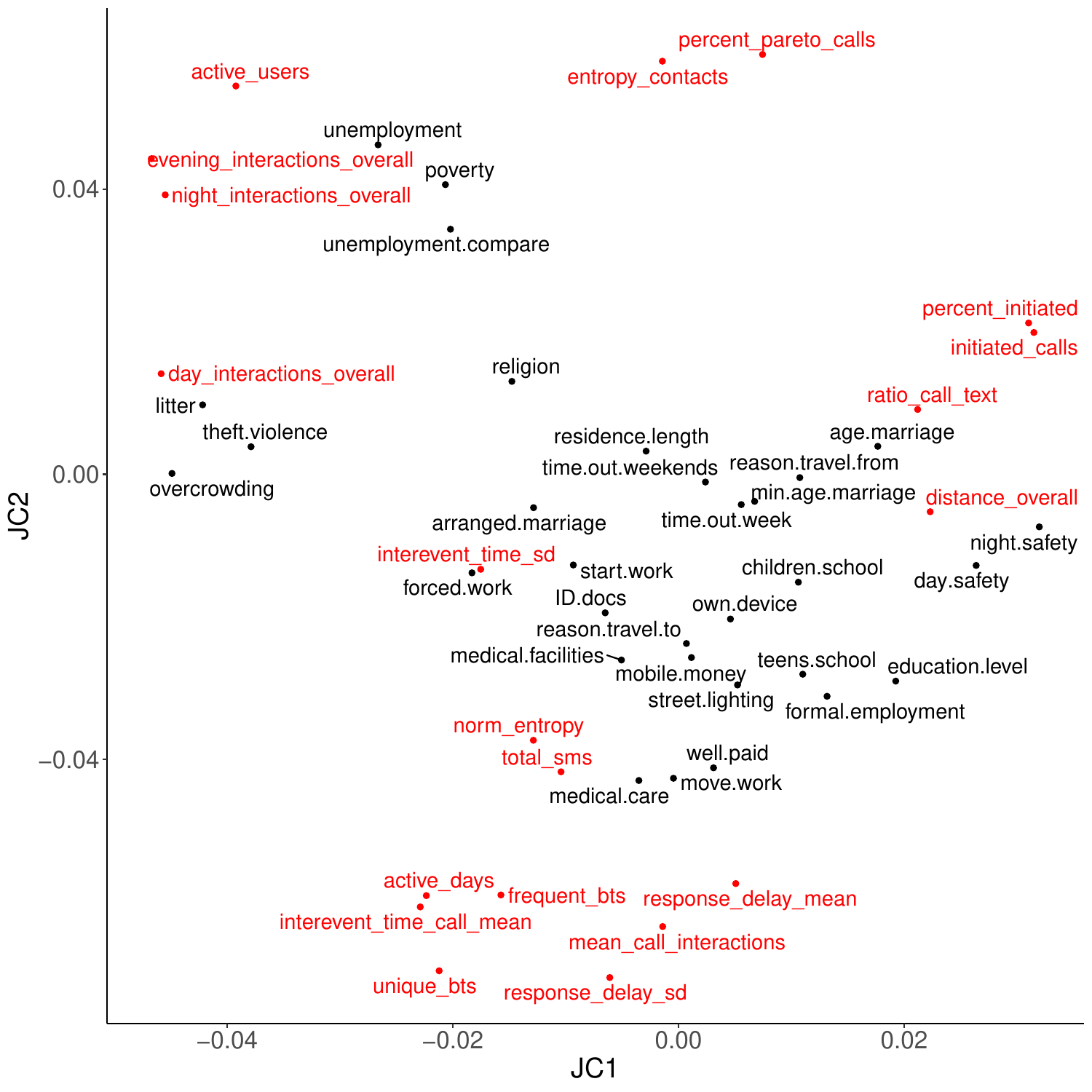}
    \caption{\small{Feature loadings for the joint components: survey features are in black, CDR features in red.}}
    \label{fig:jc-loadings}
\end{figure}

Figure \ref{fig:survey-ajive-scores} displays the subward scores for JC1, JC2, IC1\textsuperscript{CDR}, and IC1\textsuperscript{Image}, when the survey data is included in AJIVE. Figure \ref{fig:survey-feature-loadings} shows plots of the survey feature loadings for the two joint components. Figure \ref{fig:jc-loadings} shows a plot of the CDR and survey loadings for JC1 and JC2.

\section{Investigating the number of components}

In the main analysis we set the initial ranks for each data matrix to be 3. Here, we investigate the effect that using a different set of initial ranks has on the outcome.

\begin{figure}
    \centering
    \includegraphics[width=0.98\linewidth]{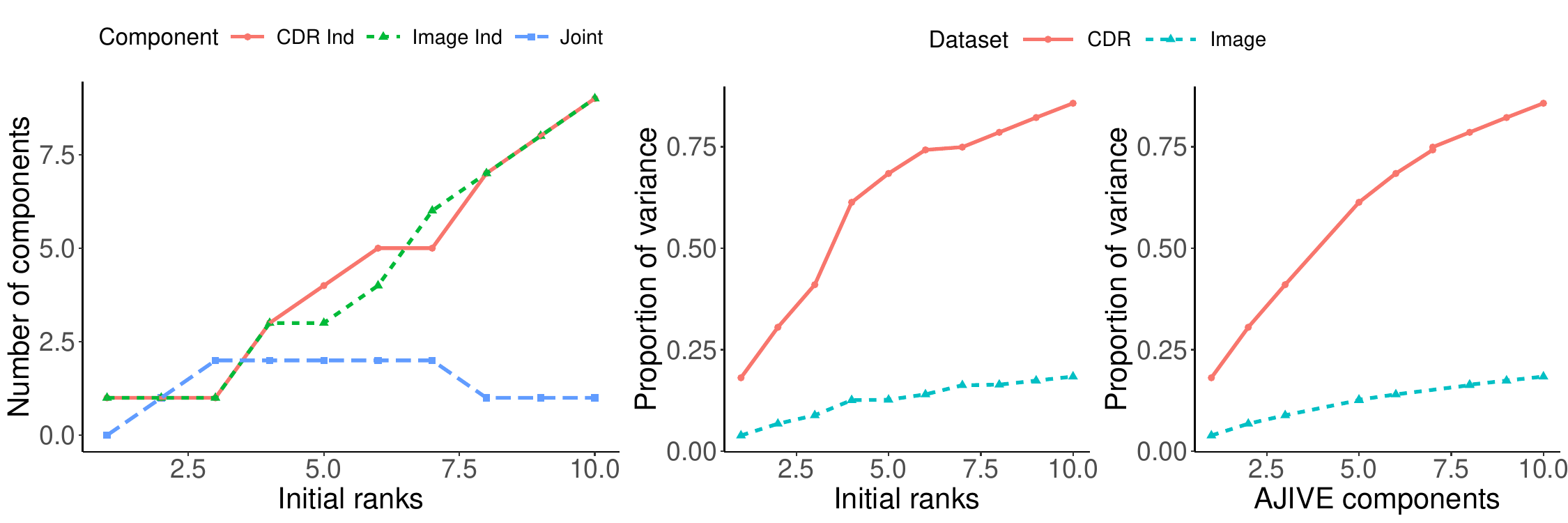}
        \put(-330, -7){\footnotesize{(a)}}
        \put(-180, -7){\footnotesize{(b)}}
        \put(-50, -7){\footnotesize{(c)}}
    \caption{\small{Panel (a) shows how the number of joint and individual components output by AJIVE changes with different initial ranks. (b) and (c) show how the proportion of variance explained by AJIVE increases with the initial ranks and total number of components.}}
    \label{fig:num-components}
\end{figure}

Figure \ref{fig:num-components} (a) shows a plot of the number of joint and individual components we obtain using AJIVE with the initial ranks for each data matrix set to be between 1 and 10 (in each case we use the same initial ranks for both datasets). The number of joint and individual ranks are estimated as described in Section \ref{sec:ranks}. The joint rank rises from 0 to 2, then returns to 3; the individual ranks remain at 1 until the initial ranks are greater than 3, then generally rise with the number of components. The loss of the second joint component when the initial ranks are large is unexpected, but it could be due to having a too strict bound to determine the number of joint components, as we discuss in Section \ref{sec:ranks}.

Figure \ref{fig:num-components} (b) and (c) show how the proportion of variance explained for each dataset rises as the initial ranks (in panel (b)) and the number of AJIVE components (panel (c)) rise. As panel (a) shows, an increase of 1 in the initial ranks can correspond to a different rise in the overall number of components in the AJIVE output, so panel (c) gives a fairer impression of how the proportion of variance explained changes with the number of components in the results. We see that as the number of components increases, adding further components has a smaller effect -- as we would expect, due to the way AJIVE (like PCA) orders the components. The much lower proportion of variance explained for the image dataset is due to the much larger dimension of this dataset ($p = 1536$ compared to $p = 20$ for the CDR data).

\end{document}